\documentclass[12pt,reqno]{article}

\usepackage{amsmath}
\usepackage{amssymb}
\usepackage{amsthm}
\usepackage{bm}
\usepackage[bf,footnotesize]{caption}
\usepackage{color}
\usepackage{eucal}
\usepackage{eurosym,enumitem}
\usepackage[top=2.4cm,left=2.7cm,right=2.7cm,bottom=3cm]{geometry}
\usepackage{graphicx}
\usepackage{multirow}
\usepackage{lineno}
\usepackage{mathpazo}
\usepackage{mathrsfs}
\usepackage{mathtools}
\usepackage[numbers,square,sort&compress]{natbib}
\usepackage{setspace}
\usepackage{subfig}
\usepackage{times}
\usepackage{titlesec}
\usepackage[normalem]{ulem}
\usepackage{verbatim}
\usepackage{xr}
\usepackage[all,cmtip]{xy}

\newcommand*\patchAmsMathEnvironmentForLineno[1]{%
	\expandafter\let\csname old#1\expandafter\endcsname\csname #1\endcsname
	\expandafter\let\csname oldend#1\expandafter\endcsname\csname end#1\endcsname
	\renewenvironment{#1}%
	{\linenomath\csname old#1\endcsname}%
	{\csname oldend#1\endcsname\endlinenomath}}%
\newcommand*\patchBothAmsMathEnvironmentsForLineno[1]{%
	\patchAmsMathEnvironmentForLineno{#1}%
	\patchAmsMathEnvironmentForLineno{#1*}}%
\AtBeginDocument{%
	\patchBothAmsMathEnvironmentsForLineno{equation}%
	\patchBothAmsMathEnvironmentsForLineno{align}%
	\patchBothAmsMathEnvironmentsForLineno{flalign}%
	\patchBothAmsMathEnvironmentsForLineno{alignat}%
	\patchBothAmsMathEnvironmentsForLineno{gather}%
	\patchBothAmsMathEnvironmentsForLineno{multline}%
}

\newcommand*{\citen}[1]{%
	\begingroup
	\romannumeral-`\x 
	\setcitestyle{numbers}%
	\cite{#1}%
	\endgroup   
}

\smallskip

\newtheoremstyle{plainCl1}
{9pt}
{9pt}
{\it}
{}
{\bfseries}
{}
{\newline}
{}

\newtheoremstyle{plainCl2}
{9pt}
{9pt}
{}
{}
{\bfseries}
{.}
{2mm}
{}

\theoremstyle{plainCl1}

\newtheorem{definition}{Definition}
\newtheorem{lemma}{Lemma}
\newtheorem{proposition}{Proposition}

\theoremstyle{plainCl2}
\newtheorem{remark}{Remark}
\newtheorem{example}{Example}

\newcommand{\p}{\mathbf{p}}
\newcommand{\q}{\mathbf{q}}
\newcommand{\eq}[1]{Eq.~\ref{eq:#1}}
\newcommand{\eqs}[2]{Eqs.~\ref{eq:#1}--\ref{eq:#2}}
\newcommand{\fig}[1]{Fig.~\ref{fig:#1}}
\newcommand{\sfig}[1]{Fig.~\ref{fig:#1}}
\newcommand{\sfigs}[2]{Figs.~\ref{fig:#1}--\ref{fig:#2}}
\newcommand{\FMTL}{\emph{FMTL}}
\newcommand{\lem}[1]{Lemma~\ref{lem:#1}}
\newcommand{\stab}[1]{Tab.~\ref{tab:#1}}
\newcommand{\vid}[1]{Vid.~\ref{vid:#1}}
\newcommand{\meth}{Methods}
\newcommand{\si}{Supporting Information}

\title{\begin{center} \bfseries \Large \singlespacing ~\\[-0.8cm]
		Evolutionary instability of selfish learning in repeated games
\end{center}}
\author{\parbox[c]{16cm}{\onehalfspacing \normalsize \centering ~\\[-0.5cm] Alex McAvoy$^{1,2}$, Julian Kates-Harbeck$^{3}$, Krishnendu Chatterjee$^{4}$, and Christian Hilbe$^{5}$ \\ \quad\\ \footnotesize
		$^{1}$Department of Mathematics, University of Pennsylvania, Philadelphia, PA, USA; \\
		$^{2}$Center for Mathematical Biology, University of Pennsylvania, Philadelphia, PA, USA; \\
		$^{3}$Department of Physics, Harvard University, Cambridge, MA, USA; \\
		$^{4}$Institute of Science and Technology Austria, Klosterneuburg, Austria; \\
		$^{5}$Max Planck Research Group: Dynamics of Social Behavior, Max Planck Institute for Evolutionary Biology, Pl\"{o}n, Germany}
	\date{}
}

\begin{document}
	
\allowdisplaybreaks

\maketitle

\begin{abstract}
\noindent
Across many domains of interaction, both natural and artificial, individuals use past experience to shape future behaviors. The results of such learning processes depend on what individuals wish to maximize. A natural objective is one's own success. However, when two such ``selfish'' learners interact with each other, the outcome can be detrimental to both, especially when there are conflicts of interest. Here, we explore how a learner can align incentives with a selfish opponent. Moreover, we consider the dynamics that arise when learning rules themselves are subject to evolutionary pressure. By combining extensive simulations and analytical techniques, we demonstrate that selfish learning is unstable in most classical two-player repeated games. If evolution operates on the level of long-run payoffs, selection instead favors learning rules that incorporate social (other-regarding) preferences. To further corroborate these results, we analyze data from a repeated prisoner's dilemma experiment. We find that selfish learning is insufficient to explain human behavior when there is a trade-off between payoff maximization and fairness.
\end{abstract}

\section{Introduction}
Individuals naturally adapt to their environment, either by modifying their existing behaviors or by considering alternative ones when necessary \citep{traulsen:PNAS:2010,rand:TCS:2013}. The study of behavioral adaptation is far-reaching, with applications ranging from microbial dynamics \citep{vulic:G:2001,zomorrodi:NC:2017} to social preferences in humans \citep{fehr:QJE:1999,charness:QJE:2002,fischbacher:AER:2010} to learning algorithms in multi-agent systems \citep{bloembergen:JAIR:2015,zhang:HRLC:2021}. Evolutionary game theory, a tool for modeling behavioral adaptations \citep{hofbauer:CUP:1988,friedman:E:1991,weibull:MIT:1995,sigmund:PUP:2010,mcnamara:JRSI:2013,tanimoto:S:2015,javarone:S:2018,newton:Games:2018}, has been used to describe how people learn to engage in reciprocity \citep{trivers:TQRB:1971,axelrod:Science:1981,press:PNAS:2012,hilbe:NHB:2018,stewart:PNAS:2012,van-segbroeck:prl:2012,fischer:PNAS:2013,stewart:PNAS:2013,pinheiro:PLoSCB:2014,akin:Games:2015,Yi:JTB:2017,hilbe:pnas:2017,mcavoy:PRSA:2019}, how social norms evolve over time \citep{ohtsuki:JTB:2004a,Santos:Nature:2018}, how groups are formed \citep{javarone:PLOSONE:2017}, how thriving communities can suddenly be undermined by corruption \citep{abdallah:JRSI:2014,Lee:PNAS:2019}, and how artificial agents behave ``in silico'' \citep{nowak:Nature:1993,zhong:IEEE:2002,batut:BMCB:2013,kiourt:AB:2016}. 
A key assumption in evolutionary game theory is that individuals might not act optimally from the outset. Rather, adaptation happens over time through either cultural or genetic mechanisms \citep{hofbauer:CUP:1988,friedman:E:1991,weibull:MIT:1995,sigmund:PUP:2010,mcnamara:JRSI:2013,tanimoto:S:2015}.

To describe how people learn new behaviors, the respective literature has considered various cognitive processes. Some models stipulate that individuals learn by imitating their peers \citep{szabo:PRE:1998,traulsen:JTB:2007b,amaral:PRE:2018}, whereas others assume that learning is based on aspiration levels \citep{oechssler:JEBO:2002,du:2013}, reinforcement \citep{sandholm:BioSys:1996,masuda:BMB:2009}, or introspection \citep{Hauser:Nature:2019,couto:NJP:2022}. Crucially, however, learning rules are frequently based on the assumption that individuals strive to increase their own immediate payoffs. Although errors, mutations, and chance events may temporarily lead individuals to adopt inferior strategies, better performing strategies are favored on average. We refer to learning rules with this property as ``selfish learning.''

In reality, many models ask neither whether individuals actually learn based on strict payoff maximization nor whether they have a long-run incentive to do so \citep{szabo:PRE:1998,traulsen:JTB:2007b,oechssler:JEBO:2002,du:2013,sandholm:BioSys:1996,masuda:BMB:2009,amaral:PRE:2018,Hauser:Nature:2019,couto:NJP:2022}. The assumption of selfish learning might be justified when interactions lack any strategic component (as in single-player optimization). However, the rationale for selfish learning is less clear in social dilemmas, where there are conflicts of interest between the individual and the group \citep{dawes:ARP:1980,kerr:TREE:2004,nowak:JTB:2012}. When social dilemmas give rise to multiple equilibria, selfish optimization may easily result in detrimental outcomes that are socially inefficient. This drawback is already well-recognized in the field of multi-agent reinforcement learning, where selfish learners are considered ``na\"{i}ve'' and serve as a benchmark for other learning rules \citep{foerster2018learning}. If selection acts upon the learning rules that determine how players choose strategies, then it may select for different learning rules altogether.

The problem of optimal learning is best illustrated with a key model of evolutionary game theory: the repeated prisoner's dilemma \citep{axelrod:Science:1981}. In each round of the game, two players independently choose whether to cooperate or defect. Each player has an incentive to defect even though mutual cooperation is in everybody's interest. For repeated games with sufficiently long time horizons, there is a wide range of possible equilibria \citep{friedman:RES:1971}. In particular, players may always cooperate, always defect, or alternate between cooperation and defection \citep{stewart:pnas:2014}. After decades of research, much is known about particular strategies that sustain cooperation, such as ``tit-for-tat'' \citep{axelrod:BB:1984} or ``win-stay, lose-shift'' \citep{nowak:Nature:1993}. Much less is known how such desirable strategies can be learned in the first place, especially when players are learning concurrently. If strategy updating is described by imitation, for example, players may end up defecting for a substantial amount of time \citep{stewart:scirep:2016}. Such detrimental outcomes become increasingly likely when the game involves only a few rounds, offers a small benefit of cooperation, or when players commit errors \citep{hilbe:NHB:2018}.

In this study, we ask whether there is a learning rule that helps individuals find more profitable equilibria, even when the opponent is self-interested. To this end, we imagine two individuals interacting in a repeated game. Each individual has its own learning rule for adapting its strategy, with the ultimate goal of achieving a high payoff. As a baseline, we consider a variant of selfish learning. At regular time intervals, a selfish learner compares the performance of its present strategy with the (hypothetical) performance of a slightly perturbed version of its strategy. The learner adopts the perturbed strategy if it yields a higher payoff, regardless of its effects on others.

We contrast selfish learning with a learning rule we term ``fairness-mediated team learning''~(\FMTL{}). Players with this learning rule balance two objectives, efficiency and fairness. To promote efficiency, \FMTL{} favors strategies that increase the total payoff of the group (i.e., the ``team'' payoff). In this way, \FMTL{} aims to avoid equilibria that leave everybody worse off. To prevent themselves from getting exploited, however, \FMTL{} players simultaneously aim to minimize payoff differences within their group (i.e., promote ``fairness''). The respective weights assigned to efficiency and fairness are dynamically adjusted based on the players' current payoffs. Increasing efficiency is the primary objective when payoff inequalities are negligible.

The definition of \FMTL{} has a natural connection to the field of multi-agent learning \citep{bowling:AI:2002,tuyls:AAMAS:2005,shoham:AI:2007,stone:AI:2007,tuyls:AI:2012}. Two of the most well-studied areas involve so-called ``fully-cooperative'' and ``fully-competitive'' interactions. In fully-cooperative interactions, players have identical payoff functions; what is good for one player is equally good for the other. In fully-competitive settings, the players' incentives are perfectly opposed. The iterated prisoner's dilemma falls somewhere in between and is often referred to as a ``mixed'' or ``general-sum'' game \citep{hu:JMLR:2003,hoen:LAMAS:2006}. As such, it presents a more difficult learning problem. In striving for fairness, \FMTL{} forces the players to have approximately equal payoffs, which is reminiscent of the fully-cooperative setting. Once payoffs are sufficiently close, \FMTL{} views itself and the opponent as a team and attempts to optimize the total payoff. The approach of optimizing a team score is common in cooperative settings, especially when individuals have mostly (but not completely) aligned incentives \citep{zhang:HRLC:2021}.

When {\it all} learners are driven by fairness and efficiency, it may not be surprising that the learning dynamics favor socially beneficial outcomes. Remarkably, however, such outcomes already arise when only {\it one} learner has these objectives. For a wide range of two-player games, we show that \FMTL{} players tend to settle at equilibria that are both individually optimal and socially efficient even when interacting with selfish opponents. Based on these observations, we explore the dynamics that arise when the two learning rules themselves are subject to evolution. We consider a process with two timescales. On a short timescale, individuals have a fixed learning rule that they use to guide their strategic choices. On a longer timescale, players can switch their learning rule, based on how successful it proved to be. The resulting evolutionary dynamics depend on which game is played, and on the relative pace at which learning rules are updated (compared to how often strategies are updated). When learning rules and strategies evolve at a similar timescale, selfish learning is favored. However, when learning rules evolve at a slower rate, selfish learning is routinely invaded and ceases to be stable in many classical games.

\section{Results}

\subsection{A model of learning in repeated games}
\begin{figure}[p]
	\centering
	\includegraphics[width=0.8\linewidth]{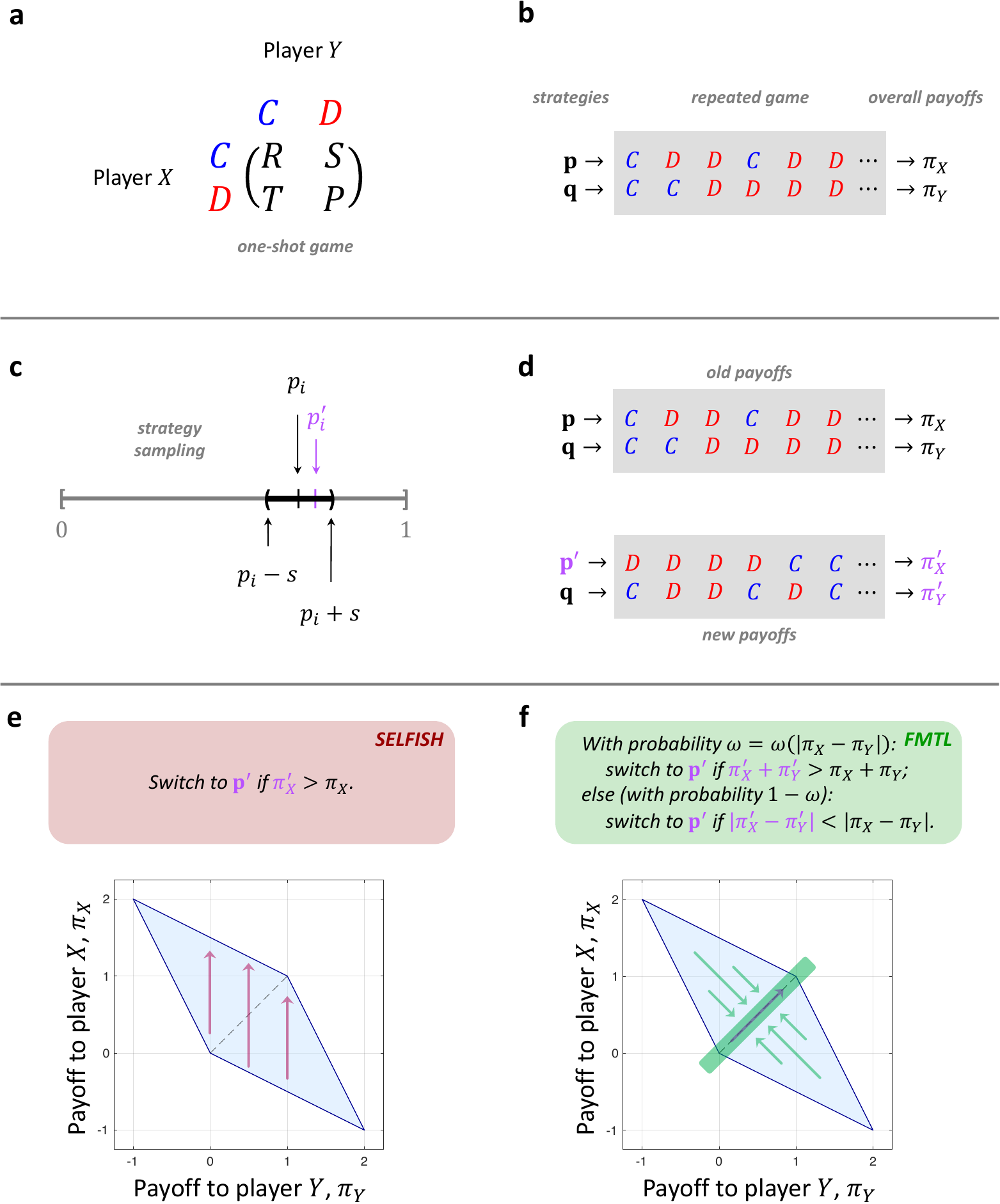}
	\caption{\textbf{Learning in repeated games and \FMTL{}.} {\bf a,} In each encounter of a repeated game, players engage in a ``one-shot'' game. Both choose actions, $C$ or $D$, and receive payoffs, $R$, $S$, $T$, or $P$. {\bf b,} A repeated game consists of a sequence of one-shot games. Both players choose strategies, $\p$ and $\q$, and receive overall payoffs, $\pi_{X}$ and $\pi_{Y}$, respectively. \textbf{c,} To update strategies, players occasionally sample a new, nearby strategy $\mathbf{p'}$ (``local search''). \textbf{d}, When $X$ uses strategy $\mathbf{p'}$ against $Y$'s strategy, $\q$, the players get new payoffs, $\pi_{X}'$ and $\pi_{Y}'$. This new strategy is then evaluated based on a player's learning rule. \textbf{e}, If $X$ is a selfish learner, $\mathbf{p'}$ is accepted only if it improves $X$'s payoff, i.e. if $\pi_{X}'>\pi_{X}$. \textbf{f}, If $X$ uses \FMTL{}, then with probability $1-\omega$ she takes $\mathbf{p'}$ only if it brings the two players' payoffs closer together (fairness). Otherwise, with probability $\omega$, she takes $\mathbf{p'}$ only if it improves the sum of the two players' payoffs (efficiency). The probability $\omega$ is a decreasing function of the payoff difference so that fairness becomes more important to \FMTL{} as one player starts to do better than the other.\label{fig:modelSetup}}
\end{figure}

To compare different learning rules, we consider two players who interact in a repeated game. In each round, players independently decide whether to cooperate or defect. As a result, each player obtains a payoff (\fig{modelSetup}a). After every round, players interact for another round with probability~$\lambda$. They decide whether to cooperate based on their strategies. A strategy is a rule that tells the player what to do in the next round, given the history of previous play. In the simplest version of the model, players use memory-one strategies \citep{sigmund:PUP:2010}. These strategies are contingent on only the last round of play, and they reasonably approximate human behavior in economic experiments \citep{englewarnick:ET:2006,dalbo:AER:2011,bruttel:TD:2011,dalbo:JEL:2018}. A memory-one strategy for player $X$ consists of an initial probability of cooperation, $p_{0}$, together with a four-tuple of conditional cooperation probabilities, $\p =\left(p_{CC},p_{CD},p_{DC},p_{DD}\right)\in\left[0,1\right]^{4}$. Here, $p_{xy}$ is the probability that $X$ cooperates in the next round when $X$ played $x$ and $Y$ played $y$ in the previous round. For example, an unconditional defector is represented by a memory-one strategy with $p_{0}=0$ and $\p =\left(0,0,0,0\right)$. Tit-for-tat takes the form $p_{0}=1$ and $\p =\left(1,0,1,0\right)$. Given strategies $\mathbf{p}$ and $\mathbf{q}$ of the two players, we can compute how likely they are to cooperate over the course of the game, and which overall payoffs $\pi_{X}\left(\p,\q\right)$ and $\pi_{Y}\left(\p,\q\right)$ they obtain (\fig{modelSetup}b and \meth{}).

After each repeated game, players may update their strategies based on their experience with the opponent. They do so by implementing a learning rule. We consider learning rules that consist of four elements: a distribution~$\mathcal{D}$ over initial strategies, a sampling procedure~$\mathcal{S}$, a set of objective functions~$\mathcal{V}$, and a priority assignment~$\Omega$. The first element, the distribution over initial strategies, determines the player's default strategy that is used before any learning takes place. The second element, the sampling procedure, determines how to generate an alternative strategy in each learning step. Throughout the main text, we assume that a player $X$ with current strategy~$\p$ generates an alternative strategy~$\mathbf{p'}$ using local random search \citep{solis:MOR:1981} (\fig{modelSetup}c). After generating an alternative strategy, the player decides whether to accept it. To this end, the set of objective functions~$\mathcal{V}$ specifies what the player's strategy ought to maximize. If $X$'s objective is to maximize $V\in\mathcal{V}$, the player switches to $\mathbf{p'}$ if and only if $V\left(\mathbf{p'},\q\right) >V\left(\p,\q\right)$. This decision may involve, for example, a comparison between the player's current payoff, $\pi_{X}\left(\p,\q\right)$, and the payoff $\pi_{X}\left(\mathbf{p'},\q\right)$ the player could have obtained using the alternative strategy~(\fig{modelSetup}d). 
However, a player's priorities over her objectives may change over time. The priority assignment~$\Omega\left(\p,\q\right)$ determines with which probability each objective~$V\in\mathcal{V}$ is chosen.

We compare the performance of two learning rules. According to selfish learning, a player switches to the alternative strategy if and only if it increases the player's payoff. Within our framework, selfish learning can be represented by the objective function $V_{S}\left(\p,\q\right) =\pi_{X}\left(\p,\q\right)$, which the player strives to maximize~(\fig{modelSetup}e). We refer to such a player as a selfish learner. The other learning rule is \FMTL{} (\fig{modelSetup}f). \FMTL{} has two objective functions, $\mathcal{V}=\left\{V_{E},V_{F}\right\}$. The first objective is to achieve efficiency. With the objective function $V_{E}\left(\p,\q\right) =\pi_{X}\left(\p,\q\right) +\pi_{Y}\left(\p,\q\right)$, the player aims to maximize the group's total payoff. The other objective is fairness. Here, a player aims to minimize payoff differences, which is equivalent to maximizing the objective $V_{F}\left(\p,\q\right) =-\left|\pi_{X}\left(\p,\q\right) -\pi_{Y}\left(\p,\q\right)\right|$. \FMTL{} prioritizes efficiency if the players' current payoffs are sufficiently close, and it prioritizes fairness if there is substantial inequality. For a precise description of the priority assignment, see \meth{}.

\subsection{Learning dynamics across different repeated games}
To compare the two learning rules, we first assume each player's learning rule is fixed. Across a range of different two-player games, we explore how the players' learning rules affect their payoffs. As the baseline scenario, we consider two selfish learners. We then contrast this scenario with groups where either one or both players use \FMTL{}.

\subsubsection{Prisoner's dilemma}
\begin{figure}[p]
	\centering
	\includegraphics[width=0.9\linewidth]{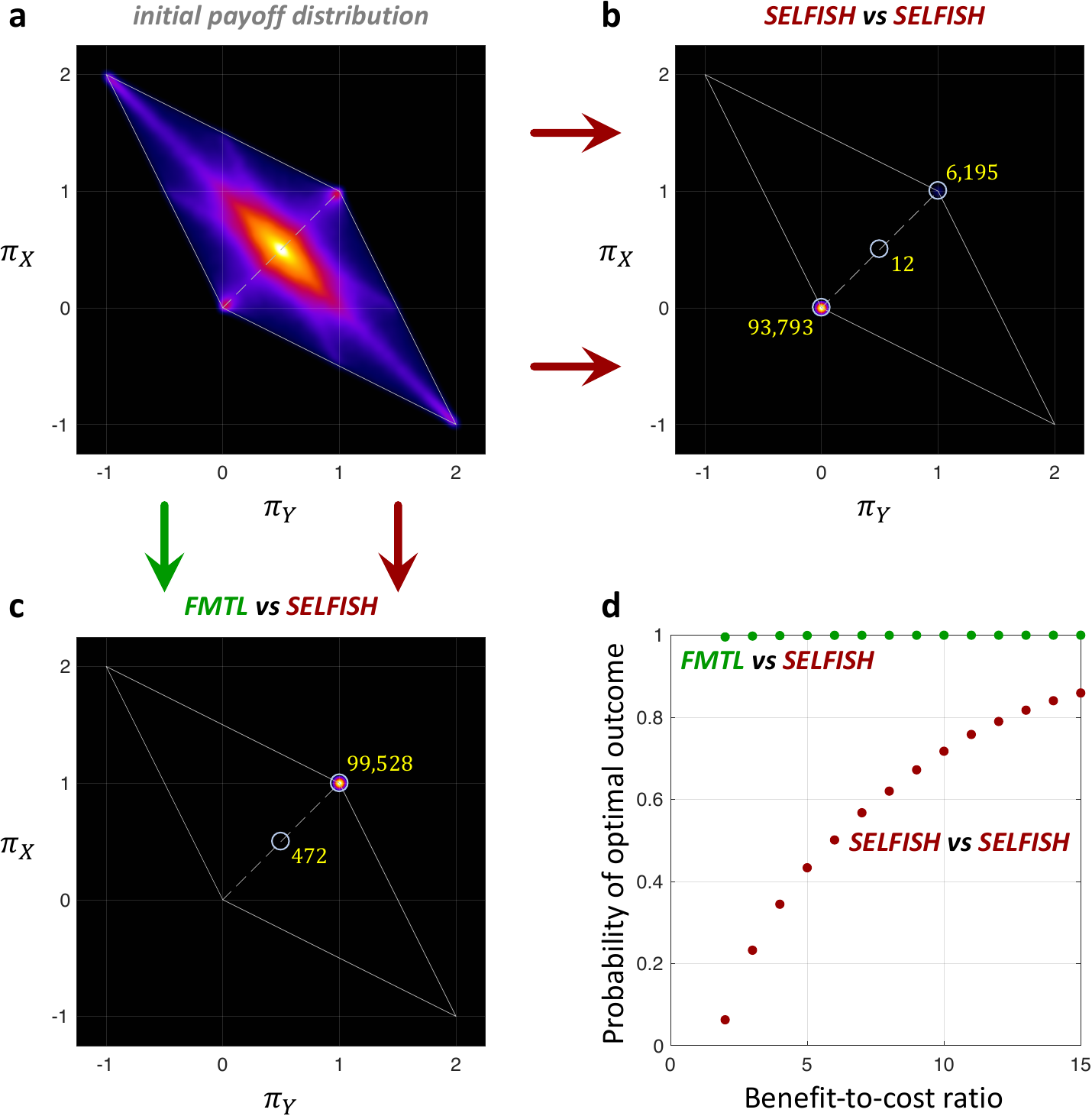}
	\caption{\textbf{\FMTL{} versus a selfish learner in the repeated donation game.} \textbf{a}, $X$ and $Y$ initially choose random strategies and receive payoffs that fall within the feasible region. \textbf{b}, When $X$ and $Y$ are both selfish learners, we let $X$ and $Y$ update until neither one has accepted a new strategy for $10^{4}$ update steps. This process is repeated for $10^{5}$ iterations. The resulting distribution of payoffs is concentrated around the mutual defection payoff. Only a small number of runs (approximately $6\%$) result in mutual cooperation, and an even smaller number of runs settle at alternating cooperation. \textbf{c,} When $X$ plays \FMTL{} instead, final payoffs are concentrated around the payoff for mutual cooperation. This is also the fair and socially optimal outcome. \textbf{d,} As one may expect, two selfish learners require a substantial benefit-to-cost ratio to coordinate on the optimal outcome ($b-c$) with high probability. In contrast, \FMTL{} against a selfish learner gives excellent outcomes even when $b/c$ is small. The endpoints in \textbf{b} and \textbf{c} are based on $10^{5}$ random initial strategy pairs, and for $b=2$ and $c=1$.\label{fig:donationGame}}
\end{figure}

We start by exploring how players fare in one of the most basic and well-studied social dilemmas, the donation game \citep{sigmund:PUP:2010}. Here, cooperation means paying a cost $c>0$ to deliver a benefit of $b>c$ to the co-player. This results in a prisoner's dilemma: players individually prefer to defect, yet mutual cooperation yields a better payoff than mutual defection. Players start out with random \mbox{memory-one} strategies~(\fig{donationGame}a) and the game is repeated for many rounds (see \meth{}). Previous work shows there are four types of symmetric equilibria among memory-one players \citep{stewart:pnas:2014}: players either both cooperate, both defect, alternate, or they use so-called equalizers \citep{boerlijst:AMM:1997,press:PNAS:2012}.

We use simulations to explore which of these equilibrium outcomes is eventually realized (if any), depending on which learning rules the players apply. When two selfish learners interact, most of the time they either end up in mutual cooperation or mutual defection~(\fig{donationGame}b). Although cooperation is socially optimal, an overwhelming majority of simulations give rise to an all-defection equilibrium with low payoffs. The observed learning dynamics change completely if one (or both) of the two learners switches to \FMTL{}. In that case, most simulated pairs of learners end up cooperating (\fig{donationGame}c; see \sfig{final_strategies} for a depiction of the final strategies). Remarkably, players yield almost full cooperation already for low benefit-to-cost ratios, for which cooperation is usually difficult to establish \citep{hilbe:NHB:2018} (\fig{donationGame}d). 

\begin{figure}[p]
	\centering
	\includegraphics[width=0.9\linewidth]{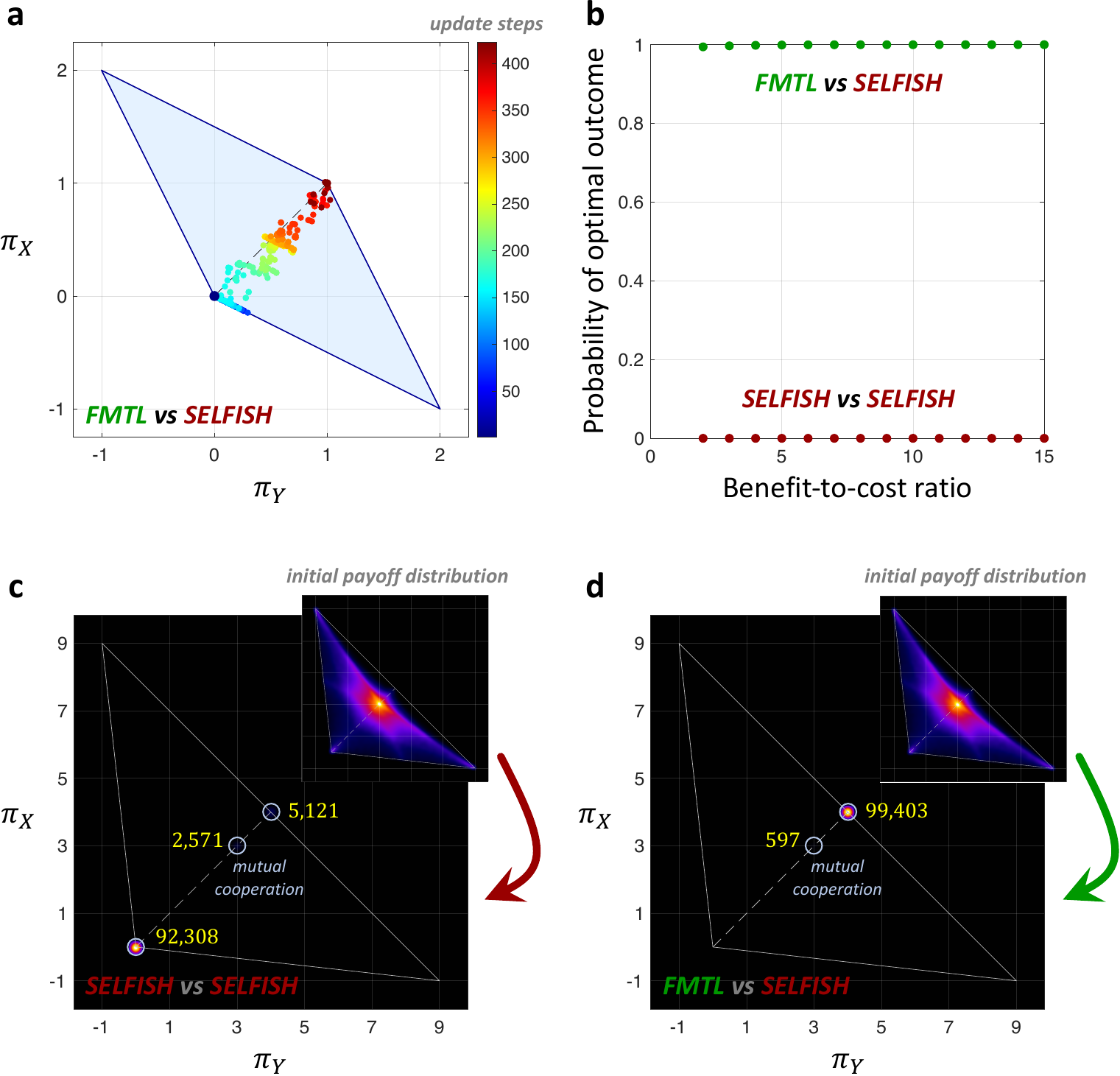}
	\caption{\textbf{Robustness of \FMTL{}.} \textbf{a}, To illustrate how \FMTL{} can help players to escape from mutual defection, we consider a scenario where $X$ uses \FMTL{} and $Y$ is a selfish learner. Initially, both $X$ and $Y$ play ALLD (``always defect''). Even under such hostile initial conditions, the process moves toward mutual cooperation. \textbf{b}, Across all benefit-to cost-ratios, pairs with an \FMTL{} learner reliably escape mutual defection. In contrast, two selfish learners never escape. {\bf c, d,} In addition to the donation game, we also consider simulations for a prisoner's dilemma with $\left(S+T\right) /2>R$. In this game, optimal play requires alternating cooperation and defection. When \FMTL{} is matched with a selfish learner, players are indeed most likely to settle at this outcome, although they have no explicit information about $R$ relative to $\left(S+T\right) /2$. Game parameters: \textbf{a, b,} $R=1$,~$S=-1$, ~$T=2$, ~and $P=0$; \textbf{c, d,} $R=3$,~$S=-1$,~$T=9$,~and $P=0$.\label{fig:escapeAndST}}
\end{figure}

To explore which mechanism allows \FMTL{} to evade inefficient equilibria against selfish learners, we consider simulations in which initially both players defect (\fig{escapeAndST}a). In this initial state, the players' payoffs are equal but inefficient. As a result, \FMTL{} prioritizes efficiency over fairness, leading the respective learner to cooperate occasionally. While this reduces the payoff to \FMTL{}, it also provides some strategic leverage. By cooperating conditionally, the learner can affect how profitable it is for the selfish opponent to cooperate. Once the \FMTL{} player adopts a strategy that makes cooperation a best response, even a selfish opponent has an incentive to adapt (\sfig{escapeALLD} and \vid{trajectory}). In this way, \FMTL{} triggers dynamics of ever-increasing cooperation rates. Eventually, players reach an equilibrium in which both players cooperate. Different simulated trajectories vary due to stochasticity, but they almost always lead from defection to cooperation, independent of the magnitude of the benefit-to-cost ratio (\fig{escapeAndST}b). We observe similar dynamics when players start out with random strategies (see \sfig{average_trajectories}).

Donation games may favor successful learning because efficiency requires only a simple pattern of behavior: both players merely need to cooperate each round. Instead, additional coordination problems might arise if efficiency requires the players to cooperate in turns. This problem occurs, for example, in a prisoner's dilemma with $R<\left(S+T\right) /2$. Here, it is socially optimal to agree on a policy of alternation: $X$ cooperates and $Y$ defects in even rounds, whereas $X$ defects and $Y$ cooperates in odd rounds. Again, we explore the dynamics of this game when players use either selfish learning or \FMTL{}. When they both use selfish learning, they typically fail to cooperate altogether (\fig{escapeAndST}c). But if one of them adopts \FMTL{}, they overwhelmingly discover the optimal policy of alternation (\fig{escapeAndST}d). We observe a similar pattern in the context of a harmony game (here, $R>T$ and $S>P$, such that cooperation is dominant~\citep{martinez-vaquero:plosone:2012}). When $R<\left(S+T\right) /2$, two selfish learners typically coordinate on mutual cooperation. But if one of the two players switches to \FMTL{}, they both achieve the superior alternation outcome. Specifically, when two selfish learners interact in a game with payoffs $R=1$, $S=2$, $T=0.5$, and $P=0$, we find that over $99\%$ of runs give a payoff of $\left(R,R\right) =\left(1,1\right)$. Once one of the learners switches to \FMTL{}, more than $99\%$ of runs end up at $\left(\left(S+T\right) /2,\left(S+T\right) /2\right) =\left(1.25,1.25\right)$. 

\subsubsection{Generalized social dilemmas: stag hunt and snowdrift games}
While the prisoner's dilemma has been instrumental in modeling human cooperation, there are other natural rankings of the game's payoffs. These alternative rankings result in weaker forms of conflict. To further explore the performance of \FMTL{}, we consider generalized social dilemmas, defined by three properties \citep{dawes:ARP:1980,kerr:TREE:2004,nowak:JTB:2012}: \emph{(i)} the payoff for mutual cooperation exceeds the payoff for mutual defection,~$R>P$; \emph{(ii)} when players choose different actions, the defector obtains a larger payoff than the cooperator, $T>S$; and \emph{(iii)} irrespective of their own action, players prefer their opponent to cooperate, $R>S$ and $T>P$. In addition to the prisoner's dilemma and the harmony game, there are two more generalized social dilemmas \citep{hauert:JTB:2006a}: the stag hunt \citep{skyrms:CUP:2003} and the snowdrift \citep{sugden1986economics,hauert:Nature:2004} game.

The stag hunt game has the payoff ranking $R>T>P>S$. In particular, mutual defection is an equilibrium of the one-shot game, but so is mutual cooperation. For repeated stag hunt games, we find that players do not need to settle at socially optimal outcomes, even if they both adopt \FMTL{} (\sfig{others}a-f). Such inefficiencies may arise more generally. They can occur in all games with equilibria that are fair but inefficient, and where one player alone is unable to raise the group payoff (see \si{} for analytical results). Nevertheless, we find that \FMTL{} makes players less likely to settle at such inefficient equilibria in the first place. As a result, each player performs better on average if at least one of them adopts \FMTL{}.

In the other generalized social dilemma, the snowdrift game, the payoff ranking is $T>R>S>P$. Mutual defection is no longer an equilibrium because unilateral cooperation is a better outcome for \emph{both} players. When two selfish learners engage in a repeated snowdrift game, they often approach one of these two pure equilibria. Eventually, one player cooperates each round and the other defects (\sfig{others}g and \sfig{others_neutral}a). Which of the two players ends up cooperating depends on their initial strategies and on chance. In contrast, if one of them switches to \FMTL{}, players most likely coordinate on a pattern of play that is both fair and efficient. Similar to the prisoner's dilemma, this pattern requires players to either mutually cooperate (if $\left(S+T\right) /2 <R$, \sfig{others}h), or to cooperate in an alternating fashion (if $\left(S+T\right) /2 > R$, \sfig{others_neutral}b). In the latter case, already two selfish learners tend to achieve an efficient (albeit possibly unfair) outcome. The role of \FMTL{} here, relative to selfish learning, is to eliminate inequality.

\subsubsection{Alternative forms of conflict: hero game}
Finally, we consider an example that does not meet the conditions of a social dilemma: the hero game \citep{tanimoto:S:2015}. This game, sometimes referred to as a (symmetric) battle of the sexes \citep{maynardsmith:CUP:1982,luce:D:1989}, satisfies $S>T>R\geqslant P$. Here, mutual $C$ is preferred to mutual $D$, but when both players use $C$, a single player can act as a ``hero'' and improve both players' payoffs by switching to $D$ \citep{rapoport:BS:1967}. The one-shot game has two pure equilibria, but players disagree on which equilibrium they prefer. When two selfish learners engage in the repeated game, they frequently converge toward one of these pure one-shot equilibria (\sfig{others_neutral}d). In contrast, groups with at least one \FMTL{}~player reliably learn to alternate (\sfig{others_neutral}e,f). Selfish learning is able to generate efficient outcomes, but only \FMTL{} makes sure the realized outcome is fair, irrespective of the learning rule of the opponent.
\stab{SummaryFixedLearningRules} summarizes these results across the different games we study.

\subsection{Evolutionary dynamics of learning rules}
After analyzing how different learning rules affect adaptation, we study how the learning rules themselves evolve over time. We consider two timescales. In the short run, the players' learning rules are fixed. Players use their learning rule to adapt to their opponent. In the long run, learning rules reproduce, based on how well players with the respective rule perform. This process may reflect cultural or biological evolution (i.e. successful learners are either imitated more often, or they have more offspring). Just as repeated games can be thought of as a ``supergame'' layered over a one-shot game \citep{friedman:RES:1971} (\fig{modelSetup}a,b), the process describing the evolution of learning rules can be thought of as a supergame layered over the repeated game (\fig{modelSetup}c--f).

\subsubsection{Description of the evolutionary process}
To describe this supergame formally, we consider a population of learners who use either selfish learning ($\textrm{S}$) or \FMTL{} ($\textrm{F}$). In the short run, players are randomly matched to engage in repeated games with a fixed partner. Over the course of their interactions, they update their strategies according to their learning rules, as in the previous section. As a result, they receive a payoff that depends on the game being considered, the players' learning rules, and on the time that has passed for learning to unfold. For a given game, let $a_{ij}\left(n\right)$ denote the expected payoff of a learner~$i$ against another learner~$j$ after $n$ learning steps (i.e. after the players had $n$ opportunities to revise their strategies). We estimate these payoffs using numerical simulations.

For a given number of learning steps $n$, the four payoffs $a_{\textrm{SS}}\left(n\right)$, $a_{\textrm{SF}}\left(n\right)$, $a_{\textrm{FS}}\left(n\right)$, $a_{\textrm{FF}}\left(n\right)$ can be assembled in a $2\times 2$ payoff matrix. We interpret the entries of this matrix as the payoff of each learning rule, and we interpret $n$ as the players' learning horizon. Payoff matrices for different values of~$n$ reflect different assumptions on how patient players are. For small $n$, players are impatient. They assess the quality of their learning rule by how well they perform after only a few learning steps. In contrast, for large $n$, players assess the quality of their learning rule according to how well it performs eventually (even if it may be ineffective in the short run). 

For a given payoff matrix, we use the replicator equation to model the evolutionary dynamics among learning rules \citep{taylor:MB:1978}. The equation tracks the frequencies of each learning rule over time, and it favors those rules with higher average payoffs (see \meth{}). For $2\times 2$ games, as in our case, replicator dynamics can result in four different scenarios \citep{hofbauer:CUP:1988,sigmund:PUP:2010}. Either one learning rule is globally stable (dominance), there is a mixed population that is globally stable (coexistence), each learning rule is locally stable (bistability), or any mixed population is stable (neutrality). 

\subsubsection{Evolutionary dynamics across different games}
The resulting dynamics between the two learning rules depend on the game and on the players' learning horizon. \fig{replicatorDynamics}a shows the possible scenarios for the donation game. If the learning horizon is short, then learning rules are selected according to whether they result in an immediate advantage. As a result, we find that selfish learning is dominant, as one might expect. However, as players become increasingly patient, the dynamics first take the form of bistable competition, and then \FMTL{} becomes globally stable. In this parameter regime, which starts after $n\approx 60$ learning steps, already a small initial fraction of \FMTL{} learners is sufficient to drive selfish learning to extinction (\fig{replicatorDynamics}b). Since clusters of \FMTL{} learners perform better than clusters of selfish learners, the evolutionary advantage of \FMTL{} is even more pronounced in structured populations \citep{nowak:PTRSB:2010,perc:JRSI:2013} (see \meth{} for details).

\begin{figure}[p]
	\centering
	\includegraphics[width=1.0\linewidth]{Fig4.pdf}
	\caption{\textbf{Evolutionary dynamics of \FMTL{} in the donation game.} \textbf{a}, In general, the players' payoffs depend on their learning rule and on how many learning steps they had to update their strategies. Here, we show these (supergame) payoffs for the donation game with $b=2$ and $c=1$. If the number of learning steps is small (fewer than $\approx 20$ in this example), the payoffs of \FMTL{} and selfish learning take the form of a prisoner's dilemma: \FMTL{} yields the better payoff when adopted by everyone, but it is dominated by selfish learning. With slightly more learning steps, the dynamics transition into bistable competition. Moreover, after only $\approx 60$ learning steps, \FMTL{} globally dominates selfish learning. \textbf{b}, Starting from an initial frequency of $x_{0}=10^{-3}$, \FMTL{} quickly spreads in a population of selfish learners under replicator dynamics. Depicted here are evolutionary trajectories for the donation game when $c=1$ and $b$ varies from $2$ to $15$. The dynamics in \textbf{b} are shown for the supergame payoffs after convergence of the learning process; but due to \textbf{a}, similar results hold when the timescale of learning is much shorter.\label{fig:replicatorDynamics}}
\end{figure}

These patterns generalize to other games. In each case, selfish learning is dominant when players have a very short learning horizon. As players become sufficiently patient, \FMTL{} becomes dominant in all variants of the prisoner's dilemma and the stag-hunt game, as well as in the snowdrift game when $\left(S+T\right) /2<R$ (\fig{other_pds} and \sfig{supergame_others}). Only for the snowdrift game with $\left(S+T\right)/2 >R$ and the hero game may selfish learning prevail for long learning horizons. \stab{SummaryEvolution} gives a summary of these results. Interestingly, the games in which \FMTL{} becomes globally stable are exactly those in which two selfish learners are at a considerable risk of settling at inefficient equilibria (\stab{SummaryFixedLearningRules}). Conversely, the games in which selfish learning can prevail are those in which selfish learners tend to achieve efficient outcomes (\sfig{others_neutral}a,d). For example, in all simulated cases of the hero game, selfish learners settle at equilibria with maximum average payoffs (even though payoffs may be shared unfairly). Because replicator dynamics depend on only the average payoffs (not on the distribution of payoffs), and because selfish learning achieves the maximum average payoff against itself, \FMTL{} cannot invade. This conclusion does not require replicator dynamics. Instead, it remains true for all evolutionary dynamics that depend on only a trait's average payoff, including stochastic models of weak selection \citep{mcavoy:NHB:2020}.

Importantly, for \FMTL{} to be successful, each of its components, fairness and efficiency, is vital. To illustrate this point, we repeat the previous simulations for the donation game with alternative learning rules (\fig{alternative_rules}): players either value only fairness, only efficiency, or they combine fairness with selfishness. Against a selfish learner, we find that each alternative rule performs worse than~\FMTL{}. When the focal player values only fairness, payoffs tend to be equal but inefficient (\fig{alternative_rules}a). When the focal player values only efficiency, the focal player tends to cooperate unconditionally, whereas the selfish co-player defects (\fig{alternative_rules}b). Finally, when the focal player combines fairness and selfish learning, the overall performance is similar to the case of two selfish learners (\fig{alternative_rules}c). These results highlight that \FMTL{}'s two components are effective only when combined. Learners who only aim for efficiency are subject to exploitation; learners who only value fairness obtain equal payoffs, but at the price of obtaining low payoffs.

\subsection{Selfishness versus fairness in a repeated prisoner's dilemma experiment}

The results presented herein cast doubt on the assumption that selfish learning can fully explain human adaptation processes in repeated games. While it has been suggested that selfish learning reasonably approximates cooperative behavior in linear multiplayer games \citep{Burton-Chellew:PRSB:2015,Burton-Chellew:NHB:2021}, such games differ in crucial aspects from the two-player interactions studied herein. For example, in multiplayer games, each individual has less of an impact on other group members. Moreover, each player's defection can only be punished collectively (by withholding cooperation from {\it all} other group members). In contrast, pairwise interactions provide individuals with more immediate possibilities to affect their co-player's future behavior. It is pairwise games where learning rules have the strongest strategic impact.

\begin{figure}[p]
	\centering
	\includegraphics[width=1.0\linewidth]{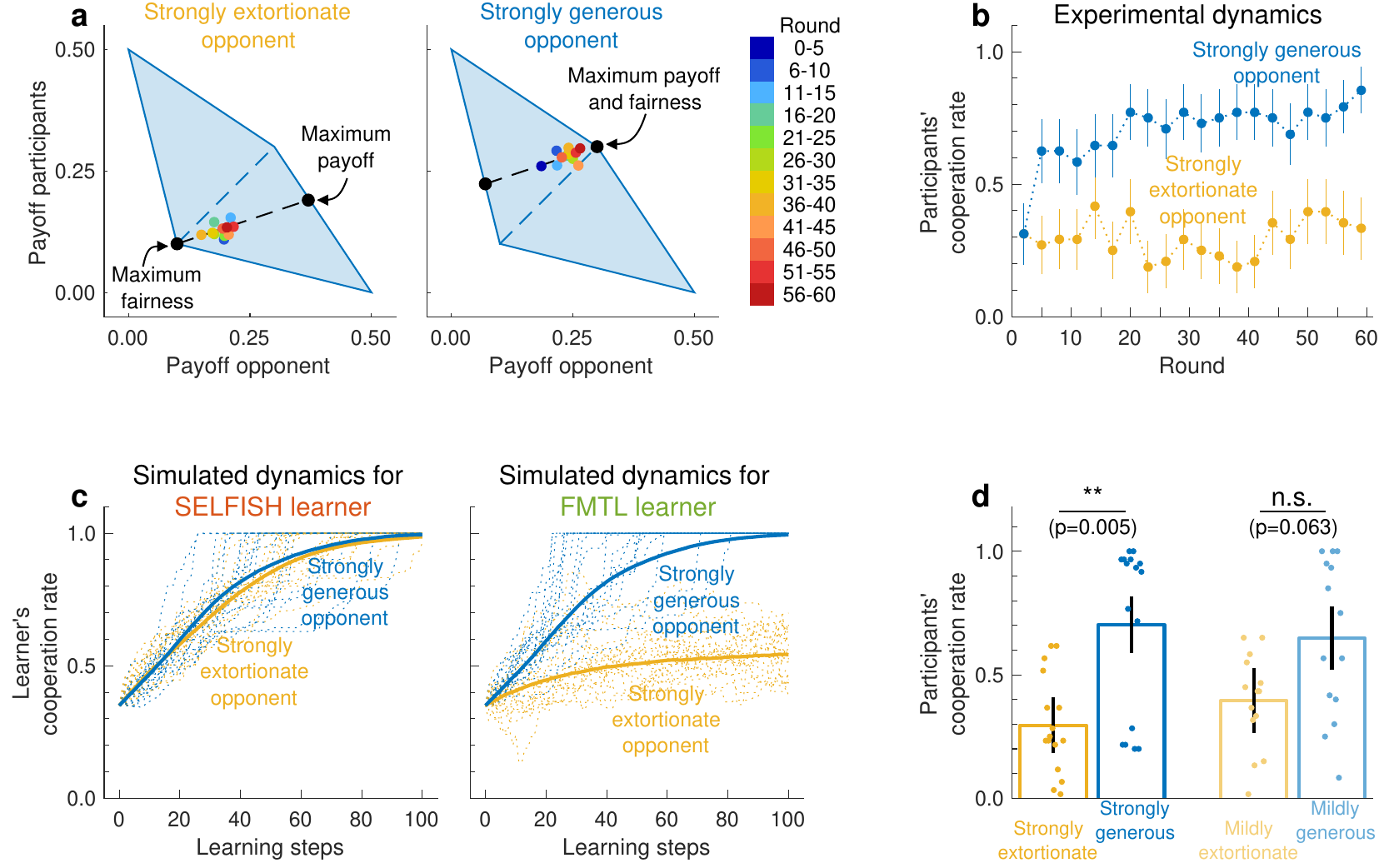}
	\caption{\textbf{Selfish learning fails to explain human behavior when there is a tension between payoff maximization and fairness.} To empirically distinguish between selfish learning and fairness-mediated learning, we re-analyze data from a repeated prisoner's dilemma experiment~\citep{hilbe:ncomms:2014}. In this experiment, human participants interact with a computerized opponent who implements a fixed strategy. Human participants are not informed about the nature of their opponent. {\bf a,}~The computer implements either an extortionate \citep{press:PNAS:2012} or a generous \citep{stewart:PNAS:2013} zero-determinant strategy. Such strategies ensure that there is a linear relationship between the payoffs of the two players (indicated by the black dashed line). If human participants wish to maximize their payoffs, they should learn to cooperate in either case. If participants wish to enhance fairness, they should cooperate against generous strategies but not against extortioners. {\bf b,} The experiment finds that humans become more cooperative against generous strategies. There is no trend toward cooperation against extortioners. {\bf c,}~To shed further light on these observations, we simulate the possible learning dynamics. We consider a single learner, who either adopts selfish learning or \FMTL{}. Only for \FMTL{} do we recover that subjects fail to fully cooperate against extortioners. {\bf d,}~The previous results are based on two treatments in which the trade-off between fairness and payoff maximization is strongest. When the computer instead implements so-called mildly generous and extortionate strategies, this trade-off is weaker. As a result, while generous co-players still tend to induce more cooperation, the effect is now smaller. Error bars indicate standard errors, and any statistical statements are based on non-parametric tests (Mann-Whitney U test, Wilcoxon test). For details, see \meth{}.\label{fig:experiment}
	}
\end{figure}

To explore the relevance of selfish learning for pairwise games, we re-analyze data from a prisoner's dilemma experiment \citep{hilbe:ncomms:2014}. In this experiment, human participants play against a computer program (unbeknownst to the human subjects). The computer program either implements an extortionate \citep{press:PNAS:2012} or a generous \citep{stewart:PNAS:2013} ``zero-determinant'' strategy \citep{press:PNAS:2012,hilbe:NHB:2018}. Against both classes of strategies, the payoff-maximizing choice for human participants is to cooperate in every round. However, while full cooperation is also the fairness-maximizing choice against generous opponents, it leads to maximally unequal outcomes against extortioners (\fig{experiment}a).

There are two reasons why this experimental paradigm allows for a clean comparison between selfish learning and other learning rules based on other-regarding preference: ({\it i}) Because the computer's strategy is fixed (and known to the researcher), each human participant's learning behavior can be studied in isolation. ({\it ii}) The two competing learning rules make opposing predictions for this experiment: If human participants use selfish learning, they should equally learn to cooperate against either computer strategy. In contrast, if behavior is better described by \FMTL{}, we expect more cooperation against the generous strategy (\fig{experiment}c; see \meth{} for a detailed description of the experimental setup and our predictions).

In \fig{experiment}b, we compare experimental data for the ``strongly extortionate'' and the ``strongly generous'' computer strategy (for which fairness considerations are most likely to impede cooperation in the extortion treatment). For both treatments, human cooperation rates are similar in the beginning ($31.3$\% during the first three rounds). Against the generous program, humans increase their cooperation rate to $85.4$\% by the end of the experiment (during the last three rounds). In contrast, against the extortionate program, overall cooperation rates are largely unchanged ($33.3$\%), although the monetary incentives for cooperation are identical. These experimental results are consistent with learning behavior that is shaped by fairness considerations. In line with this view, when the computer implements a strategy that is only mildly extortionate or generous, there is less of a difference in human cooperation rates (\fig{experiment}d). Moreover, the difference disappears altogether if individuals are informed ahead of the experiment that they are matched with a computerized opponent, in which case fairness considerations can be expected to be absent \citep{Xu:NComms:2016}. 
Overall, these results suggest that fairness motives are of crucial importance to describe human behavior in pairwise interactions, even though they may be less salient in multiplayer games \citep{Burton-Chellew:PRSB:2015,Burton-Chellew:NHB:2021}.

\section{Discussion}
Given the extensive effort to explore strategies that sustain cooperation in repeated social dilemmas \citep{trivers:TQRB:1971,axelrod:Science:1981,press:PNAS:2012,hilbe:NHB:2018,stewart:PNAS:2012,van-segbroeck:prl:2012,fischer:PNAS:2013,stewart:PNAS:2013,pinheiro:PLoSCB:2014,akin:Games:2015,Yi:JTB:2017,hilbe:pnas:2017,mcavoy:PRSA:2019}, it seems quite remarkable that relatively little is known about how strategies with desirable properties can be learned most effectively. Instead, much of the existing work tends to take the way in which individuals learn as given. While details vary between studies \citep{sandholm:BioSys:1996,masuda:BMB:2009,szabo:PRE:1998,oechssler:JEBO:2002,traulsen:JTB:2007b,du:2013,amaral:PRE:2018,Hauser:Nature:2019}, most often it is assumed that individuals adopt strategies that enhance their own payoff, abandoning strategies that are personally disadvantageous. This modeling assumption could be justified on theoretical grounds if selfish payoff maximization were indeed an optimal learning policy. Here, we have explored under which conditions selfish learning can be expected to succeed. We show theoretically that selfish learning performs well when individuals wish to optimize their short-run performance. However, if individuals are motivated by how well they fare eventually, selfish learning can be of limited use in navigating conflicts of interest.

To assess the performance of selfish learning, we contrast it with an alternative rule termed \FMTL{}. Rather than maximizing just one's own payoff, a learner who adopts \FMTL{} strives to enhance either the efficiency of the resulting game outcome or its fairness. By striving to increase efficiency, \FMTL{} attempts to evade inferior equilibria that leave all group members worse off. By striving to increase fairness, \FMTL{} avoids exploitation when other group members continue to learn selfishly. Using individual-based simulations, we show that \FMTL{} can help individuals to settle at better equilibria. These equilibria are either more equitable or more efficient. Moreover, when players select their learning rules according to how well they perform eventually, \FMTL{} outcompetes selfish learning for most of the games we study. Qualitatively, the ability of \FMTL{} to draw out better outcomes also extends to learning based on imitation as opposed to introspection (\S\ref{sec:imitation} of \si{} and \sfigs{imitation}{imitation_trend}) as well as asymmetric games (\S\ref{sec:asymmetric_games} of \si{} and \sfig{FMTL_asymmetric}).

Our theoretical results are further corroborated by an analysis of human cooperation in a repeated prisoner's dilemma with fixed opponent strategies \citep{hilbe:ncomms:2014,Xu:NComms:2016}. When interactions entail a trade-off between fairness and payoff maximization, selfish learning fails to explain crucial patterns of human behavior (\fig{experiment}). There is a considerable literature within behavioral economics seeking to describe human behavior that deviates from models of pure self-interest \citep{fehr:Handbook:2006}. In particular, there is ample empirical evidence that fairness and efficiency are important drivers of human behavior. Humans value fairness starting from a young age \citep{fehr:Nature:2008,mcauliffe:NHB:2017}, and they are often willing to accept substantial reductions in their own income to achieve more egalitarian outcomes \citep{Dawes:Nature:2007}. Such a demand for fairness can have substantial economic consequences, as it constrains a firm's profit seeking behavior \citep{kahneman:AEA:1986} and market prices \citep{fischbacher:JEBO:2009}. There is similar experimental evidence about the importance of efficiency \citep{charness:QJE:2002, engelmann:AER:2004}. In dictator games, human participants often give up some of their own payoff in order to increase that of the pair \citep{andreoni:E:2002,fehr:Handbook:2006}. At the same time, however, human decisions to increase efficiency via ``gifts'' are constrained by fairness considerations \citep{guth:JEBO:2003}, consistent with the constraint built-in to \FMTL{}.

Our results are related to the ``indirect evolutionary approach,'' which explores the evolution of preferences with game-theoretic methods \citep{guth:IJGT:1995,guth:RS:1998,huck:GEB:1999,heifetz:ET:2007}. The respective literature distinguishes between objective and subjective payoffs. Objective payoffs include monetary rewards and reproductive success (fitness). Subjective payoffs capture how individuals experience certain outcomes and what they strive to maximize when making decisions. Similar to our model, this approach is ``indirect'' because preferences that guide behaviors do not need to align with objective payoffs. However, while this literature explicitly models the evolutionary dynamics of preferences \citep{akcay:PNAS:2009,heifetz:ET:2007}, it usually does not describe how preferences affect the way individuals learn (see \si{} for a more detailed discussion). The learning process is the main focus of our study. We explore how different learning heuristics influence the way in which subjects navigate between equilibria of differing efficiency. To this end, our framework requires a notion of learning rules that is slightly more general than the notion of preferences considered before. Learning rules do not only specify the objectives that players wish to maximize. Instead, they also determine how players choose between different objectives and how they generate alternative strategies. In this way, our framework also applies to learning on shorter time scales, wherein subjects update their learning rules even before the learning dynamics reach an equilibrium (\stab{SummaryEvolution}).

Our results also have implications for objective design in multi-agent learning. Previous work has shown that objectives based on a convex combination of the players' payoffs can improve outcomes relative to selfish learning in stag hunt games \citep{peysakhovich:AAMAS:2018}. An alternative approach is to implement a look-ahead into the opponent's learning process in order to shape their future behavior \citep{foerster2018learning} (however, this forward-looking approach requires substantial information about the opponent; see \si{}). While the space of possible objective functions is vast, we introduce a  learning rule (one of many, perhaps) that can outcompete na\"{i}ve selfish learning. Compared to the rules considered previously, \FMTL{} has the advantage of being comparably simple, and its components are natural for both humans and machines to implement. Much in the same way that the most rudimentary strategy (tit-for-tat) won Axelrod's tournaments \citep{axelrod:BB:1984}, here too a simple learning rule is able to align incentives with a selfish learner.

Even if individuals are ultimately driven by their own advantage, optimal learning rules may require them to take into account other considerations, such as the well-being of others. Of course, the repeated two-player games studied herein cannot capture all realistic interactions in which learning is relevant. However, these simple baseline models can help to understand the general principles at work in more complex settings \citep{smaldino:CSP:2017}. Delineating which aspects are crucial for successful learning is, in our view, one of the most exciting directions for future research.

\section{Methods}

\subsection{Strategies and payoffs in repeated games}
All of the strategies we consider for repeated games are ``memory-one'' strategies, which means that they consist of a five-tuple of probabilities, $\left(p_{0},p_{CC},p_{CD},p_{DC},p_{DD}\right)$, where $p_{0}$ is the probability of cooperating (action $C$) in the initial round and $p_{xy}$ is the probability a player cooperates after using $x\in\left\{C,D\right\}$ in the previous round against an opponent using $y\in\left\{C,D\right\}$. The initial distribution over the outcomes $\left(CC,CD,DC,DD\right)$, where the first action is that of $X$ and the second is that of $Y$, is
\begin{align}
\nu_{0}\left(\p ,\q\right) &= \left(p_{0}q_{0},p_{0}\left(1-q_{0}\right) ,\left(1-p_{0}\right) q_{0},\left(1-p_{0}\right)\left(1-q_{0}\right)\right) .
\end{align}
Following the initial round, transitions between states are described by the stochastic matrix
\begin{align}
M\left(\p ,\q\right) &= 
\begin{pmatrix}
p_{CC}q_{CC} & p_{CC}\left(1-q_{CC}\right) & \left(1-p_{CC}\right) q_{CC} & \left(1-p_{CC}\right)\left(1-q_{CC}\right) \\
p_{CD}q_{DC} & p_{CD}\left(1-q_{DC}\right) & \left(1-p_{CD}\right) q_{DC} & \left(1-p_{CD}\right)\left(1-q_{DC}\right) \\
p_{DC}q_{CD} & p_{DC}\left(1-q_{CD}\right) & \left(1-p_{DC}\right) q_{CD} & \left(1-p_{DC}\right)\left(1-q_{CD}\right) \\
p_{DD}q_{DD} & p_{DD}\left(1-q_{DD}\right) & \left(1-p_{DD}\right) q_{DD} & \left(1-p_{DD}\right)\left(1-q_{DD}\right)
\end{pmatrix} .
\end{align}

After $t\geqslant 0$ rounds, the distribution over states is $\nu_{t}\left(\p ,\q\right) =\nu_{0}\left(\p ,\q\right) M\left(\p ,\q\right)^{t}$. With discounting factor $\lambda\in\left[0,1\right)$, the mean payoffs to $X$ and $Y$ when $X$ plays $\p$ and $Y$ plays $\q$ are
\begin{subequations}\label{eq:discountedPayoff}
\begin{align}
\pi_{X}\left(\p ,\q\right) &= \left(1-\lambda\right)\sum_{t=0}^{\infty}\lambda^{t} \left< \nu_{t}\left(\p ,\q\right) , \left(R,S,T,P\right) \right> \nonumber \\
&= \left< \left(1-\lambda\right) \nu_{0}\left(\p ,\q\right)\left(I-\lambda M\left(\p ,\q\right)\right)^{-1} , \left(R,S,T,P\right)\right> ; \\
\pi_{Y}\left(\p ,\q\right) &= \left(1-\lambda\right)\sum_{t=0}^{\infty}\lambda^{t} \left< \nu_{t}\left(\p ,\q\right) , \left(R,T,S,P\right) \right> \nonumber \\
&= \left< \left(1-\lambda\right) \nu_{0}\left(\p ,\q\right)\left(I-\lambda M\left(\p ,\q\right)\right)^{-1} , \left(R,T,S,P\right)\right> ,
\end{align}
\end{subequations}
respectively, where $\left<\cdot ,\cdot\right>$ denotes the standard inner (dot) product on $\mathbb{R}^{4}$. 
To approximate the results of an infinite-horizon game, we use a discounted game with a small discounting rate, $\lambda =1-10^{-3}$. In this way, we ensure that the limiting payoffs always exist, even if the players adopt strategies that allow for multiple absorbing states, such as tit-for-tat \citep{sigmund:PUP:2010}.

\subsection{Learning rules}
In the following, we describe the four components of a learning rule (the initial strategy distribution, the sampling procedure, the set of objective functions, and the priority assignment) for both selfish learning and \FMTL{}.

\subsubsection{Distribution over initial strategies}
When two individuals are first paired with one another, they each choose an initial strategy for their first interaction, which is to be subsequently revised during the learning process. Two of the most natural choices are \emph{(i)} to choose each coordinate of the strategy independently from a uniform distribution on $\left[0,1\right]$, and \emph{(ii)} to choose each coordinate independently from an arcsine ($\textrm{Beta}\left(1/2,1/2\right)$) distribution on $\left[0,1\right]$. For the figures presented herein, we use the latter distribution because it is more effective in exploring the corners of the space $\left[0,1\right]^{5}$ of memory-one strategies \citep{nowak:Nature:1993}. However, we obtain similar qualitative results for a uniform initial distribution. In addition, we also explore the learning dynamics that arise when players initially defect unconditionally (\fig{escapeAndST}a,b).

\subsubsection{Sampling procedure}
We assume strategy sampling to be local in the following sense (see also \fig{modelSetup}c). Let $s\in\left[0,1\right]$ and suppose that $z_{i}$ is uniformly distributed on $\left[-s,s\right]$ (with $z_{i}$ independent of $z_{j}$ for $j\neq i$). If $p_{i}$ is the coordinate being ``mutated'' at a given time step, then the candidate sample of this coordinate in the next step is $p_{i}'=\min\left\{ \max\left\{ p_{i} + z_{i} , 0 \right\} , 1 \right\}$. We use $s=0.1$ in our examples, which allows for exploration while ensuring that the candidate strategy is not too distant from the current strategy (so that desirable properties of the current strategy are not immediately discarded). Relatively small values of $s$ make the trajectories of the learning process more interpretable (e.g. \fig{escapeAndST}a), but they also slow down the learning process. While taking a different value of $s$ can change the overall performance of each learning rule, we did not find a scenario in which it reverses the relative ranking of selfish learning compared to \FMTL{}.

\subsubsection{Objective functions}
Once a candidate strategy is sampled, the respective player decides whether to accept it based on the player's objectives. To this end, each learning rule specifies a set of objective functions, 
\begin{align}
\mathcal{V} &= \Big\{V\ \mid\ V:\left[0,1\right]^{5}\times\left[0,1\right]^{5} \rightarrow \mathbb{R}\Big\} .
\end{align}
Each objective function $V$ takes the players' memory-one strategies $\mathbf{p}$ and $\mathbf{q}$ as an input, and returns a value that indicates to which extent the players' objectives are met. Throughout our study, we consider objective functions that depend on only the players' payoffs, $\pi_{X}\left(\mathbf{p},\mathbf{q}\right)$ and $\pi_{Y}\left(\mathbf{p},\mathbf{q}\right)$, but more general formulations are possible. A candidate strategy $\mathbf{p'}$ is accepted if $V\left(\mathbf{p'},\mathbf{q}\right) >V\left(\mathbf{p},\mathbf{q}\right)$, and it is discarded otherwise. 

For the simulations, we assume the candidate strategy is accepted if and only if $V\left(\mathbf{p'},\mathbf{q}\right) >V\left(\mathbf{p},\mathbf{q}\right) +\varepsilon$, where $0<\varepsilon\ll 1$ is a small threshold. This assumption prevents floating point errors from resulting in faulty decisions, particularly when $V\left(\mathbf{p},\mathbf{q}\right)$ is extremely close to $V\left(\mathbf{p'},\mathbf{q}\right)$. While any one such faulty decision might have negligible effects on the learning process, these mistakes can accumulate over many time steps. In all of our numerical examples, the threshold we use is $\varepsilon =10^{-12}$. The use of such a threshold can also be interpreted in terms of bounded rationality \citep{simon:W:1957}. Due to limitations on cognition or information, the learner may not be able to distinguish two values that are sufficiently close together. Similarly, in the presence of noise perturbing the observed payoffs, $\varepsilon$ may be thought of parametrizing the confidence an agent has that there will truly be an improvement in $V$ by switching to the new strategy.

We note that objective functions take both players' strategies as an input. One interpretation of this assumption is that the learner needs to have precise knowledge of the co-player's strategy in order to forecast whether a given strategy change is profitable. However, we note that this rather stringent assumption is in fact not necessary. Instead, we only need to assume that individuals can reliably assess the sign of $V\left(\mathbf{p}',\mathbf{q}\right) -V\left(\mathbf{p},\mathbf{q}\right) -\varepsilon$. That is, players only need to be able to make qualitative assessments.

For selfish learners, the set of objective functions contains a single element, $\mathcal{V}=\left\{V_{S}\right\}$ with $V_{S}\left(\mathbf{p},\mathbf{q}\right) =\pi_{X}\left(\mathbf{p},\mathbf{q}\right)$. For \FMTL{}, the set of objective functions is $\mathcal{V}=\left\{V_{E},V_{F}\right\}$. Here, $V_{E}\left(\mathbf{p},\mathbf{q}\right) = \pi_{X}\left(\mathbf{p},\mathbf{q}\right) + \pi_{Y}\left(\mathbf{p},\mathbf{q}\right)$ reflects the objective to increase efficiency, whereas $V_{F}\left(\mathbf{p},\mathbf{q}\right) = -\left|\pi_{X}\left(\mathbf{p},\mathbf{q}\right) -\pi_{Y}\left(\mathbf{p},\mathbf{q}\right)\right|$ corresponds to the objective of enhancing fairness.

\subsubsection{Priority assignments}
A learning rule's priority assignment $\Omega$ determines which of the players' different objectives is currently maximized. Formally, a priority assignment is a map $\Omega :\left[0,1\right]^{5}\times\left[0,1\right]^{5}\rightarrow\Delta\left(\mathcal{V}\right)$, where $\Delta\left(\mathcal{V}\right)$ denotes the space of probability distributions on $\mathcal{V}$. For each of the players' current memory-one strategies $\mathbf{p},\mathbf{q}\in\left[0,1\right]^{5}$ it determines the probability with which each possible objective $V\in\mathcal{V}$ is chosen for maximization. 

In the case of selfish learning, the priority assignment is trivial, because there is only one possible objective to choose from. In the case of \FMTL{}, the priority assignment takes the form $\Omega\left(\mathbf{p},\mathbf{q}\right)=\left(\omega ,1-\omega\right)$, where $\omega =\omega\left(\mathbf{p},\mathbf{q}\right)$ is the weight assigned to efficiency. In our formulation of \FMTL{}, $\omega$ depends on the current magnitude of the payoff difference (see \sfig{w_function}a),
\begin{align}
\omega\left(\mathbf{p},\mathbf{q}\right) &= \exp\left( -\frac{\left(\pi_{X}\left(\mathbf{p},\mathbf{q}\right) -\pi_{Y}\left(\mathbf{p},\mathbf{q}\right)\right)^{2}}{2\sigma^{2}} \right). \label{eq:gaussian_function}
\end{align}
The parameter $\sigma >0$ reflects a player's tolerance with respect to inequality. In the limit $\sigma\rightarrow 0$ a player always aims to enhance fairness, whereas in the opposite limit $\sigma\rightarrow\infty$, the player always aims to improve efficiency. In general, neither efficiency nor fairness alone are sufficient for establishing good outcomes against selfish learners (\fig{alternative_rules}). Finding a proper balance between these two objectives depends on the nature of the interaction, and as a result the optimal value of $\sigma$ can vary from game to game.

We choose $\sigma$ by taking the average payoff for \FMTL{} versus a selfish learner over a small number of runs ($10^{3}$), for each $\sigma\in\left\{0.01,0.02,\dots ,1.00\right\}$. We then select $\sigma$ based on which value maximizes the average payoff of the \FMTL{} individual, except for when this value is comparable for all such $\sigma$, in which case we choose the value that minimizes the runtime. For the donation game (\fig{donationGame}c,d, \fig{alternative_rules}c, \fig{escapeAndST}a,b, \fig{replicatorDynamics}, \sfig{average_trajectories}, \sfig{final_strategies}, \sfig{FMTL_asymmetric}, and \vid{trajectory}), we use $\sigma =0.1$; for the prisoner's dilemma with $\left(S+T\right) /2>R$ (\fig{escapeAndST}c,d and \fig{other_pds}b) and with $\left(S+T\right) /2<P$ (\fig{other_pds}a), we use $\sigma =1$; for the stag hunt game with $\left(S+T\right) /2>P$ (\sfig{others}a,b,c and \sfig{supergame_others}a), we use $\sigma =0.01$; for the stag hunt game with $\left(S+T\right) /2<P$ (\sfig{others}d,e,f and \sfig{supergame_others}b), we use $\sigma =1$; for the snowdrift game with $\left(S+T\right) /2<R$ (\sfig{others}g,h,i and \sfig{supergame_others}c), we use $\sigma =0.1$; for the snowdrift game with $\left(S+T\right) /2>R$ (\sfig{others_neutral}a,b,c and \sfig{supergame_others_neutral}a), we use $\sigma =1$; and for the hero game (\sfig{others_neutral}d,e,f and \sfig{supergame_others_neutral}b), we use $\sigma =0.1$. The use of different values of $\sigma$ in different games is a result of treating $\sigma$ as a hyperparameter \citep{bergstra:JMLR:2012} that can be fine-tuned according to the nature of a given repeated game. However, we note that we get qualitatively similar (even if not completely optimized) results when we use a single value of $\sigma$ across all our main examples (e.g. $\sigma =0.25$).

Instead of a bell curve, $\omega$ could be a bump function such as
\begin{align}
\omega\left(\mathbf{p},\mathbf{q}\right) &= 
\begin{cases}
\displaystyle e^{-\frac{\left(\pi_{X}\left(\p,\q\right) -\pi_{Y}\left(\p,\q\right)\right)^{2}}{\alpha^{2}\sigma^{2}\left(\alpha^{2}-\left(\pi_{X}\left(\p,\q\right) -\pi_{Y}\left(\p,\q\right)\right)^{2}\right)}} & \left|\pi_{X}\left(\p,\q\right) -\pi_{Y}\left(\p,\q\right)\right| <\alpha , \\[.8cm]
0 & \left|\pi_{X}\left(\p,\q\right) -\pi_{Y}\left(\p,\q\right)\right|\geqslant\alpha
\end{cases} \label{eq:bump_function}
\end{align}
(see \sfig{w_function}b). In practice, we do not see significant qualitative differences between these two functions provided the parameters are chosen properly. However, relative to \eq{bump_function}, the function in \eq{gaussian_function} does have the advantage of depending on only one shape parameter, $\sigma$.

\subsubsection{Implementation of learning rules}
For two players with given learning rules, we explore the resulting learning dynamics with simulations. The code is available (see {\bf Code availability statement}). For these simulations, the learning process is terminated after neither learner has accepted a new candidate strategy in a fixed threshold number of update steps (here,~$10^{4}$). All of the examples we consider terminate. 

In addition, \fig{replicatorDynamics}a, \fig{other_pds}, \sfig{supergame_others}, and \sfig{supergame_others_neutral} illustrate the expected payoffs as a function of the number of learning steps. In these graphs, the horizontal axis shows how many opportunities the two players had to revise their strategies. The vertical axis represents the average payoff over sufficiently many simulations. This payoff depends on the learning rules of the focal player and the opponent.

\subsection{Evolutionary dynamics of learning rules}
The learning rules considered here may be viewed as strategies for a ``supergame.'' The payoffs of this supergame are given by the players' expected payoffs after $n$ updating steps. Here, we consider $n$ as a fixed parameter of the model, and we refer to it as the player's learning horizon. When $n$ is small, individuals evaluate the performance of their learning rule based on how well they fare after a few learning steps. In contrast, when $n$ is large, learning rules are evaluated according to how well they fare eventually.

Because we consider the competition between two learning rules, this supergame can be represented as a $2\times 2$ matrix. To this end, we fix a base game $\textrm{G}\in\left\{\textrm{HE},\textrm{PD},\textrm{SH},\textrm{SD}\right\}$, where $\textrm{HE}$ is the hero game, $\textrm{PD}$ is the prisoner's dilemma, $\textrm{SH}$ is the stag hunt game, and $\textrm{SD}$ is the snowdrift game. For a given learning horizon $n$, we can write the resulting $2\times 2$ matrix as
\begin{align}
\bordermatrix{%
& \textit{FMTL} & \textit{SELFISH} \cr
\textit{FMTL} &\ a_{\textrm{FF}}^{\textrm{G}}\left(n\right) & \ a_{\textrm{FS}}^{\textrm{G}}\left(n\right) \cr
\textit{SELFISH} &\ a_{\textrm{SF}}^{\textrm{G}}\left(n\right) & \ a_{\textrm{SS}}^{\textrm{G}}\left(n\right) \cr
}\ . \label{eq:supergameMatrix}
\end{align}
When the game and the learning horizon is clear (or irrelevant), we sometimes drop the indices $G$ and $n$ for better readability.

\subsubsection{Evolutionary dynamics in well-mixed populations}
Given the payoff matrix, we explore the evolutionary dynamics between learning rules using the replicator equation \citep{taylor:MB:1978}. The replicator equation describes deterministic evolution in infinite populations. Let $x\in\left[0,1\right]$ denote the frequency of \FMTL{} in the population. The frequency of selfish learners is $1-x$. Using the payoffs of \eq{supergameMatrix}, the fitness values of the two types are
\begin{subequations}
\begin{align}
f_{\textrm{F}}\left(x\right) &= x a_{\textrm{FF}} + \left(1-x\right) a_{\textrm{FS}} ; \\
f_{\textrm{S}}\left(x\right) &= x a_{\textrm{SF}} + \left(1-x\right) a_{\textrm{SS}} ,
\end{align}
\end{subequations}
respectively. The average fitness in the population is $\overline{f}\left(x\right) =xf_{\textrm{F}}\left(x\right) +\left(1-x\right) f_{\textrm{S}}\left(x\right)$. Under replicator dynamics, the frequency of \FMTL{} satisfies the ordinary differential equation
\begin{align}
\frac{dx}{dt} &= x \left( f_{\textrm{F}}\left(x\right) - \overline{f}\left(x\right) \right) .
\end{align}
When $a_{\textrm{FF}} \geqslant a_{\textrm{SF}}$ and $a_{\textrm{FS}} \geqslant a_{\textrm{SS}}$, we have $f_{\textrm{F}}\left(x\right) \geqslant f_{\textrm{S}}\left(x\right)$, and thus $f_{\textrm{F}}\left(x\right) \geqslant \overline{f}\left(x\right)$ for all $x\in\left(0,1\right)$. 
Moreover, if one of the inequalities is strict, $a_{\textrm{FF}}>a_{\textrm{SF}}$ or $a_{\textrm{FS}}>a_{\textrm{SS}}$, then $f_{\textrm{F}}\left(x\right) > \overline{f}\left(x\right)$. This payoff relationship holds in prisoner's dilemmas and stag hunt games, as well as in the snowdrift game with $\left(S+T\right) /2<R$, when $n$ is sufficiently large. In that case, there are two equilibria. The equilibrium $x=0$ is unstable, whereas $x=1$ is globally stable. It follows that \FMTL{} evolves from all initial populations with $x>0$. \fig{replicatorDynamics}b illustrates these dynamics for the donation game as the benefit of cooperation varies.
It is worth pointing out that the relative value of $a_{\textrm{SF}}$ compared to $a_{\textrm{FS}}$ does not affect replicator dynamics. More precisely, even though a selfish learner always gets at least the co-player's payoff in any interaction with \FMTL{} (i.e. even though $a_{\textrm{SF}}\geqslant a_{\textrm{FS}}$ for all games we studied), selfish learning may still go extinct. 

\subsubsection{Evolutionary dynamics in structured populations}
The classical replicator equation describes populations in which all individuals are equally likely to interact with each other. To explore how population structure is expected to affect our evolutionary results, we use the approach of Ohtsuki and Nowak \citep{ohtsuki:JTB:2006b}. They show that under weak selection, various stochastic evolutionary processes on regular graphs can be approximated by a replicator equation with modified payoffs. Instead of $a_{ij}$, the payoff of strategy $i$ against strategy $j$ is now given by $\widetilde{a}_{ij} := a_{ij}+b_{ij}$. Here, $b_{ij}$ depends on the game, the evolutionary process under consideration, and the degree $k>2$ of the network. For example, for death-birth updating,
\begin{equation}
b_{ij} = \frac{\left(k+1\right) a_{ii}+a_{ij}-a_{ji}-\left(k+1\right) a_{jj}}{\left(k+1\right)\left(k-2\right)} .
\end{equation}
When we apply this formula to the four payoffs $a_\textrm{FF}$, $a_\textrm{FS}$, $a_\textrm{SF}$, and $a_\textrm{SS}$, the modified payoffs are 
\begin{subequations}
\begin{align}
\widetilde{a}_\textrm{FF} &= a_\textrm{FF} ; \\
\displaystyle \widetilde{a}_\textrm{FS} &= a_\textrm{FS}+\frac{\left(k+1\right) a_\textrm{FF}+a_\textrm{FS}-a_\textrm{SF}-\left(k+1\right) a_\textrm{SS}}{\left(k+1\right)\left(k-2\right)} ; \\
\displaystyle \widetilde{a}_\textrm{SF} &= a_\textrm{SF}+\frac{\left(k+1\right) a_\textrm{SS}+a_\textrm{SF}-a_\textrm{FS}-\left(k+1\right) a_\textrm{FF}}{\left(k+1\right)\left(k-2\right)} ; \\
\widetilde{a}_\textrm{SS} &= a_\textrm{SS} .
\end{align}
\end{subequations}
For replicator dynamics, only payoff differences matter. These can be written as follows
\begin{subequations} \label{Eq:ModifiedReplicatorCondition}
\begin{align}
\widetilde{a}_\textrm{FF}-\widetilde{a}_\textrm{SF} &= \left(1+\frac{1}{\left(k+1\right)\left(k-2\right)}\right)\left(a_\textrm{FF}-a_\textrm{SF}\right) \nonumber \\
&\quad + \frac{1}{\left(k+1\right)\left(k-2\right)}\left(a_\textrm{FS}-a_\textrm{SS}\right) +\frac{k}{\left(k+1\right)\left(k-2\right)}\left(a_\textrm{FF}-a_\textrm{SS}\right) ; \\
\widetilde{a}_\textrm{FS}-\widetilde{a}_\textrm{SS} &= \left(1+\frac{1}{\left(k+1\right)\left(k-2\right)}\right)\left(a_\textrm{FS}-a_\textrm{SS}\right) \nonumber \\
&\quad + \frac{1}{\left(k+1\right)\left(k-2\right)}\left(a_\textrm{FF}-a_\textrm{SF}\right) +\frac{k}{\left(k+1\right)\left(k-2\right)}\left(a_\textrm{FF}-a_\textrm{SS}\right) .
\end{align}
\end{subequations}
In a well-mixed population, the condition for \FMTL{} to be globally stable is $a_{\textrm{FF}} \geqslant a_{\textrm{SF}}$ and $a_{\textrm{FS}} \geqslant a_{\textrm{SS}}$, with at least one inequality being strict. In the examples where we observe \FMTL{} to globally stable, the inequality $a_\textrm{FF}>a_\textrm{SS}$ is also satisfied (\fig{replicatorDynamics}a, \fig{other_pds}, and \sfig{supergame_others}). For those games, Equation~\ref{Eq:ModifiedReplicatorCondition} allows us to conclude that if \FMTL{} is globally stable in a well-mixed population, then it is also stable in any regular network. 
Moreover, especially if the degree $k$ of the network is small, regular networks may require a smaller learning horizon for \FMTL{} to become globally stable.

\subsection{Empirical analysis of a repeated prisoner's dilemma experiment}

\subsubsection{Experimental methods}
To compare our theoretical predictions to actual human behavior, we re-analyze the experimental data collected by \citet{hilbe:ncomms:2014}. In this experiment, human subjects play sixty rounds of a repeated prisoner's dilemma against a computer program. Humans are not told the nature of their opponent; instead, they only learn that they are ``matched with an opponent with whom they will interact for many rounds.'' In each round, participants can either cooperate or defect. The payoffs per round are derived from the payoffs used in Axelrod's tournament \citep{axelrod:Science:1981}:
\begin{equation} \label{eq:Axelrod}
R=0.30~\text{EUR};~~~~S=0.00~\text{EUR};~~~~T=0.50~\text{EUR};~~~~P=0.10~\text{EUR}.
\end{equation} 
In addition, all participants get a show-up fee (independent of their performance) of $10$~EUR. All participants are first-year biology students recruited from the universities of Kiel and Hamburg, Germany. 

The experiment consists of four treatments, which only differ in the memory-one strategies implemented by the computer program. The four strategies are referred to as being ``strongly extortionate,'' ``mildly extortionate,'' ``mildly generous,'' and ``strongly generous,'' and they are specified as follows: 
\begin{equation} \label{eq:zdstrategies}
\begin{array}{lll}
\text{Strongly extortionate:}   &p_0=0, &\mathbf{p}=\left(0.692,~0,~0.538,~0\right) ; \\
\text{Mildly extortionate:} &p_0=0, &\mathbf{p}=\left(0.857,~0,~0.786,~0\right) ; \\
\text{Mildly generous:} &p_0=1, &\mathbf{p}=\left(1,~0.077,~1,~0.154\right) ; \\
\text{Strongly generous:} &p_0=1,   &\mathbf{p}=\left(1,~0.182,~1,~0.364\right) .
\end{array}
\end{equation}
For the given payoff values, these four strategies represent so-called zero-determinant strategies \citep{press:PNAS:2012,hilbe:NHB:2018}. By using one of these strategies, the computer program ensures that there is an approximately linear relationship between the payoff of the human participant $\pi_H$ and the payoff of the computer program $\pi_C$. The respective linear relationships are \citep{hilbe:ncomms:2014}:
\begin{equation} \label{eq:zdrelations}
\begin{array}{lc}
\text{Strongly extortionate:}   &\pi_H - P = \frac{1}{3} \left(\pi_C - P\right) ; \\
\text{Mildly extortionate:}     &\pi_H - P = \frac{2}{3} \left(\pi_C - P\right) ; \\
\text{Mildly generous:}         &\pi_H - R = \frac{2}{3} \left(\pi_C - R\right) ; \\
\text{Strongly generous:}       &\pi_H - R = \frac{1}{3} \left(\pi_C - R\right) .
\end{array}
\end{equation}
The interpretation of these payoff relationships is as follows. If the program is extortionate, the human co-player's surplus (over the mutual defection payoff) is only $1/3$ or $2/3$ of the computer's surplus. On the other hand, if the program is generous, the human co-player's loss (compared to the payoff for mutual cooperation) is only $1/3$ or $2/3$ of the computer's loss. In particular, an extortionate computer program always obtains at least as much as the human co-player, whereas a generous program only obtains at most the payoff of the human co-player \citep{hilbe:NHB:2018}. In~\fig{experiment}a, the payoff relationships for the strong strategy variants are indicated by a black dashed line. In the extortionate case, the black dashed line is always on or below the main diagonal. In the generous case, the black dashed line is always on or above the main diagonal. The colored dots represent experimental data, averaged over all participants of the respective treatment. Each dot indicates the payoff of the human participant and the computer opponent for consecutive $5$-round intervals. Overall, the study has been conducted with $60$ participants ($16$ participants in each of the two strong treatments, $14$ participants in each of the two mild treatments). 

\subsubsection{Theoretical predictions}

For the interpretation of the experimental results with respect to our theoretical framework, three aspects of the strategies are crucial. 
\begin{enumerate}[itemsep=0pt]
\item According to Eqs.~\eqref{eq:zdrelations}, there is a positive linear relationship between the payoffs of the computer program and the human participant. In particular, if humans wish to maximize their own payoff, they should strive to maximize their opponent's payoff. That is, they should cooperate in every single round.
\item In contrast, if humans wish to have equal payoffs $\pi_H=\pi_C$ against the extortionate program, they should defect in every round (in which case $\pi_H=\pi_C=P$). If they wish to have equal payoffs against the generous program, humans should cooperate in every round  (in which case $\pi_H=\pi_C=R$). 
\item Importantly, the monetary incentives for humans to become more cooperative are the same in the strongly extortionate and in the strongly generous treatment. In either case, for every cent that they increase the co-player's payoff by being more cooperative, their own payoff is increased by $1/3$ of a cent. Similarly, the monetary incentives for increasing cooperation in the two mild treatments are also identical. 
\end{enumerate}

Because of the first and the third property, we would predict that participants who wish to increase their own payoffs cooperate equally often, independent of whether they are matched with a strongly extortionate or with a strongly generous opponent. An analogous prediction applies to the two mild treatments. In contrast, because of the second property, we would predict that participants who wish to enhance fairness are more likely to cooperate in the generous treatments. These predictions are general; they do not depend on the exact implementation of the participants' learning rules. 

In addition to these qualitative predictions, we have also run simulations for the learning rules studied herein. In \fig{experiment}c, we show the average of $1{,}000$ simulations (bold solid lines) as well as a sample of $20$ representative simulation runs (thin dotted lines). For each simulation, there is one learner who either implements selfish learning or \FMTL{}. Consistent with our theoretical analysis, the learner is restricted to memory-one strategies. Initially, the learner's strategy is unconditional, with $p_0=0.35$ and $\mathbf{p}=\left(0.35,~0.35,~0.35,~0.35\right)$. In each simulation, the learner is given 100 opportunities to revise its memory-one strategy. New strategies are sampled within an $s=0.1$-neighborhood of the parent strategy. For \FMTL{}, we use a sensitivity parameter of $\sigma =0.1$, as in our main text analysis of the standard prisoner's dilemma. The learner's opponent applies a fixed memory-one strategy that is either strongly extortionate or strongly generous, as specified by Eqs.~\eqref{eq:zdstrategies}. Consistent with the qualitative predictions above, we find that selfish learners equally learn to cooperate no matter whether the opponent is extortionate or generous. In contrast, the \FMTL{} player only learns to fully cooperate against a generous opponent. 

\subsubsection{Statistical analysis}
To test these predictions, we first compare the cooperation dynamics against the strongly extortionate strategy and the strongly generous strategy (participants matched with the strongly extortionate strategy face the strongest trade-off between payoff maximization and fairness). In \fig{experiment}b, each dot represents the human participants' cooperation rates, averaged over three rounds. In particular, the panel shows that initially participants are equally likely to cooperate against both computer programs ($31.3$\% during the first three rounds for both). However, only for the strong generosity treatment is there a significant increase in cooperation rates (to $85.4$\% during the last three rounds; Wilcoxon matched-pairs signed-rank test: $Z=2.9341, p=0.003$, all tests are two-tailed). In contrast, in the strong extortion treatment, there is no such increase ($33.3$\%; Wilcoxon matched-pairs signed-rank test: $Z=0.3494, p=0.726$.).

We obtain similar results if we compare overall cooperation rates over all $60$ rounds~(\fig{experiment}d). Against the strongly extortionate program, this average cooperation rate is $29.6$\%, compared to $70.3$\% against the strongly generous program (Mann-Whitney U test, $Z=2.789, p=0.005$). For the two mild treatments, the difference in cooperation rates is smaller, and fails to be significant; however, players still tend to cooperate more against the generous program (the cooperation rates are $39.5$\% and $64.8$\%, Mann-Whitney U test, $Z=1.860, p=0.063$). Overall, these results suggest that fairness considerations affect human cooperation rates. Moreover, this effect is more pronounced the stronger the trade-off between payoff maximization and fairness.

\section*{Acknowledgments}
The authors are grateful to J\"{o}rg Oechssler for many helpful comments. A.M. was supported by a Simons Postdoctoral Fellowship (Math+X) at the University of Pennsylvania; K.C. was supported by European Research Council Consolidator Grant 863818 (ForM-SMArt); and C.H. was supported by European Research Council Starting Grant 850529 (E-DIRECT).

\section*{Code availability statement}
Supporting code is available at https://github.com/alexmcavoy/fmtl/.

\clearpage
\newpage

\makeatletter
\@fpsep\textheight
\makeatother

\setcounter{figure}{0}
\setcounter{table}{0}
\renewcommand{\thefigure}{S\arabic{figure}}
\renewcommand{\thetable}{S\arabic{table}}

\begin{figure}[p]
	\centering
	\includegraphics[width=0.6\linewidth]{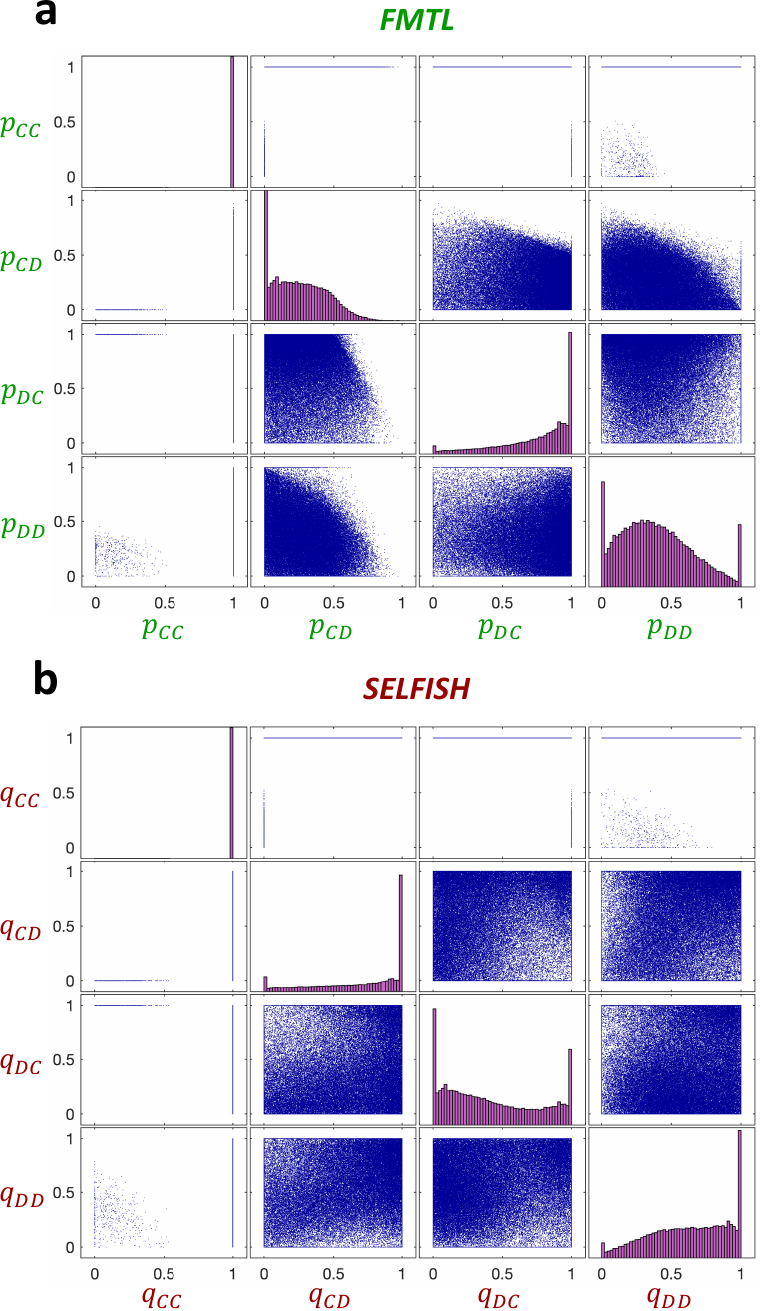}
	\caption{\textbf{Memory-one strategies resulting from \FMTL{} against a selfish learner.} In the donation game with $b=2$ and $c=1$, we let the learning process unfold for \FMTL{} paired with a selfish learner for $10^{5}$ randomly-chosen initial conditions. The conditional probabilities making up the resulting memory-one strategies are shown in \textbf{a} for the \FMTL{} learner and in \textbf{b} for the selfish learner. Notably, the strategies that result from this learning process are quite diverse, with the simplest common feature being that $p_{CC}=q_{CC}=1$ in most runs.\label{fig:final_strategies}}
\end{figure}

\clearpage
\newpage

\begin{figure}[p]
	\centering
	\includegraphics[width=1\linewidth]{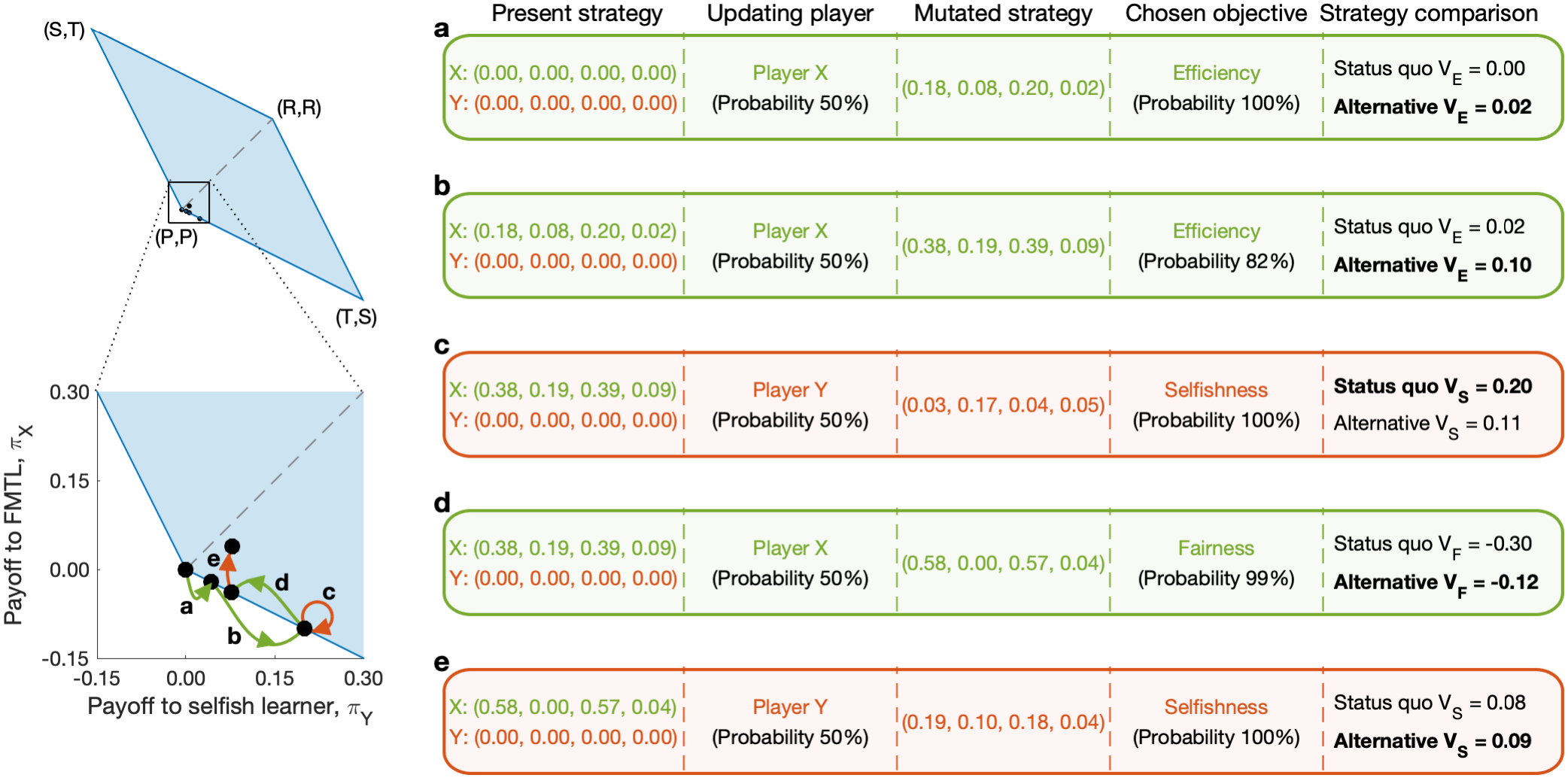}
	\caption{\textbf{\FMTL{} can escape from mutual defection in the repeated donation game.} We consider an interaction between an \FMTL{} player (player $X$) and a selfish learner (player $Y$). Learning occurs as described in the main text. In each time step, one player is randomly chosen to revise its strategy. To this end, the respective player compares its present strategy with a nearby strategy. The players' objectives are either efficiency and fairness (for \FMTL{}), or selfishness (for the selfish learner). The mutated strategy is adopted if and only if it has a better performance with respect to the objective (marked in bold). {\bf a,b,}~Initially, both players defect unconditionally. Because payoffs are equal but inefficient, the \FMTL{} player consecutively adopts strategies that increase efficiency. {\bf c,}~In the meanwhile, the selfish player has no incentive yet to adapt. Unconditional defection is in the player's best interest. {\bf d,} To avoid exploitation by the selfish player, \FMTL{} regularly revises its strategy to ensure a fairer outcome. {\bf e,} Once \FMTL{} adopts a strategy that rewards cooperation sufficiently, it is also in the selfish player's interest to become more cooperative. In the end, both players are better off than initially, even if \FMTL{} has a reduced payoff in the short run (during steps {\bf a-d}). We note that the same sequence of updating players and mutated strategies would have had no effect if both players were selfish learners. Parameters: We consider a donation game with $b=2$ and $c=1$ without discounting ($\lambda =1$). The depicted memory-one strategies are of the form $\left(p_{CC},p_{CD},p_{DC},p_{DD}\right)$; the player's initial cooperation probability has been dropped as it is irrelevant in undiscounted games \citep{sigmund:PUP:2010}. To reduce the number of learning steps required for this illustration, we use a slightly larger mutation kernel: each entry of the mutated strategy can differ by at most $s=0.2$ from the present strategy. All other parameters are the same as in \fig{escapeAndST}a.\label{fig:escapeALLD}}
\end{figure}

\clearpage
\newpage

\begin{figure}[p]
	\centering
	\includegraphics[width=1.0\linewidth]{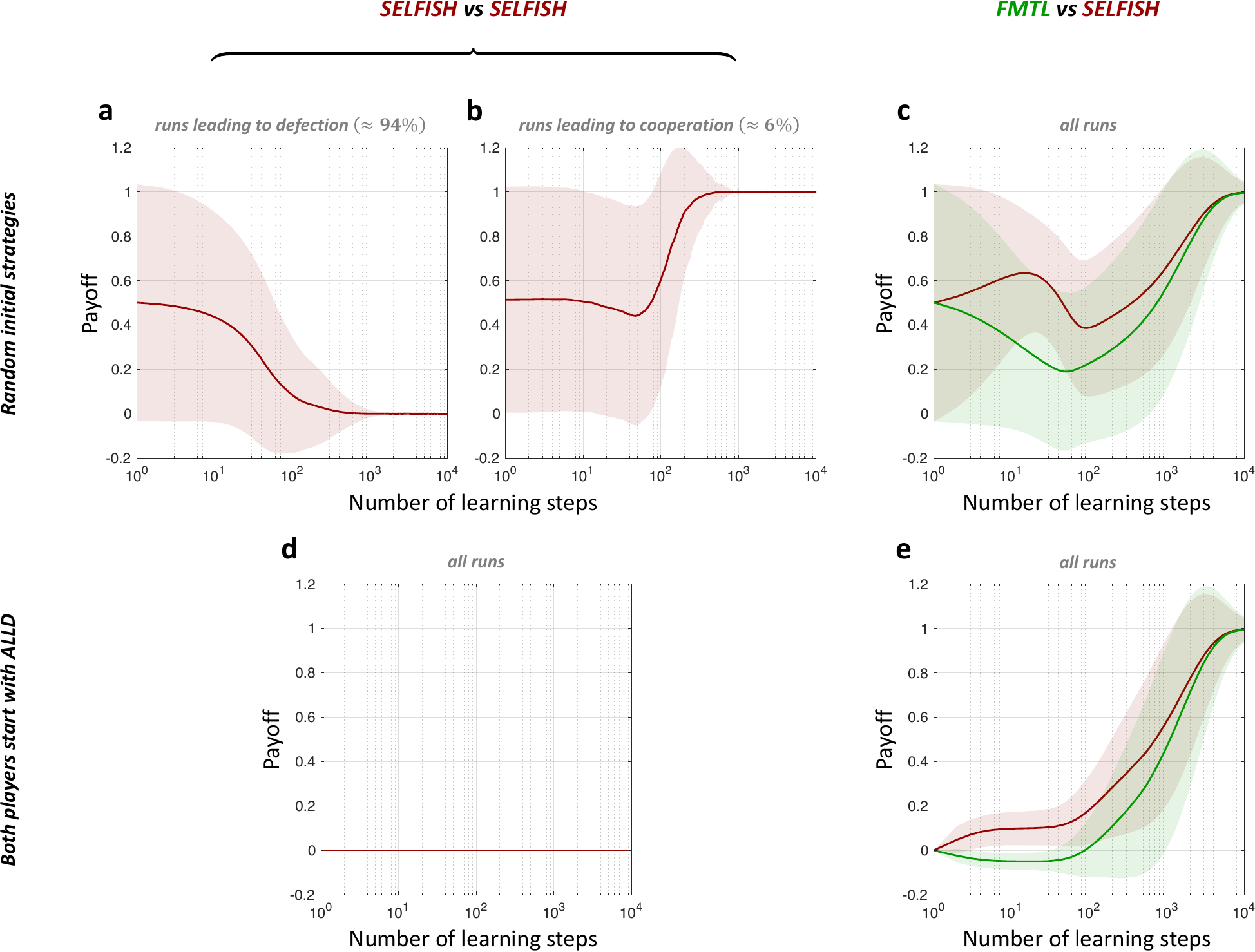}
	\caption{\textbf{Comparing trajectories of selfish learning and \FMTL{}.} Starting with random strategies, with each coordinate chosen independently from an arcsine distribution (\textbf{a,b,c}), or ``always defect'' (ALLD) (\textbf{d,e}), we observe the payoff trajectories for both learners over $10^{4}$ time steps in the donation game with a benefit of $b=2$ and a cost of $c=1$. These trajectories are averaged over $10^{5}$ initial conditions (solid lines), with shaded regions representing one standard deviation from the mean. With random initial conditions, a selfish learner against another selfish learner leads to outcomes of defection approximately 94\% of the time (\textbf{a}) and outcomes of cooperation approximately 6\% of the time (\textbf{b}). When \FMTL{} is paired with a selfish learner, the outcome is nearly always cooperative, as \FMTL{} can incentivize a selfish learner to avoid detrimental equilibria (\textbf{c}). These effects are amplified when the initial conditions are ``bad,'' such as when both players start off at ALLD. Here, a selfish learner paired with a selfish learner can never escape defection (\textbf{d}), but \FMTL{} can still destabilize defection and lead a selfish partner to a cooperative outcome (\textbf{e}).\label{fig:average_trajectories}}
\end{figure}

\clearpage
\newpage

\begin{figure}[p]
	\centering
	\includegraphics[width=1.0\linewidth]{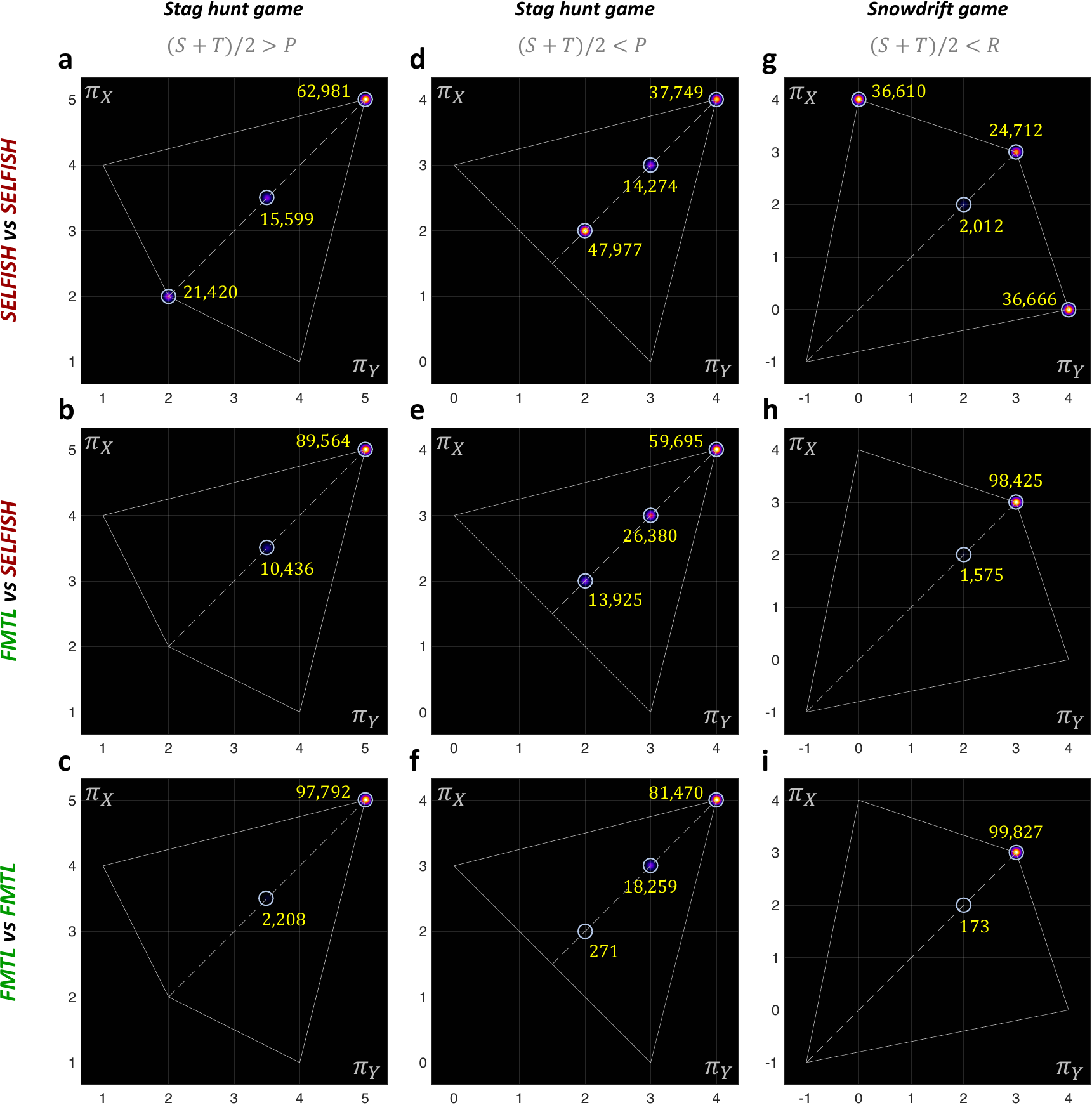}
	\caption{\textbf{Other conflicts for which selfish learning is suboptimal.} In addition to the prisoner's dilemmas depicted in the main text, we consider three other classes of conflicts for which selfish learning leads to inefficient outcomes. The first two are stag hunt games, one with $\left(S+T\right) /2>P$ (\textbf{a,b,c}) and one with $\left(S+T\right) /2<P$ (\textbf{d,e,f}). The third is the snowdrift game with $\left(S+T\right) /2<R$ (\textbf{g,h,i}). In both kinds of stag hunt games, \FMTL{} is able to significantly improve the average payoffs of the individuals, despite not being able to ensure that all players receive an optimal payoff with certainty. In the snowdrift game, \FMTL{} not only significantly increases the average payoffs; it also ensures that the two players obtain equal payoffs in all runs. The points that correspond to alternating cooperation in \textbf{h} and \textbf{i} are not actual rest points; however, escaping from these points requires more time than allowed by the termination condition (no update for $10^{4}$ steps). Game parameters: (\textbf{a},\textbf{b},\textbf{c}) $R=5$, $S=1$, $T=4$, and $P=2$;~~ (\textbf{d},\textbf{e},\textbf{f}) $R=4$, $S=0$, $T=3$, and $P=2$;~~(\textbf{g},\textbf{h},\textbf{i}) $R=3$, $S=0$, $T=4$, and $P=-1$.\label{fig:others}}
\end{figure}

\clearpage
\newpage

\begin{figure}[p]
	\centering
	\includegraphics[width=0.7\linewidth]{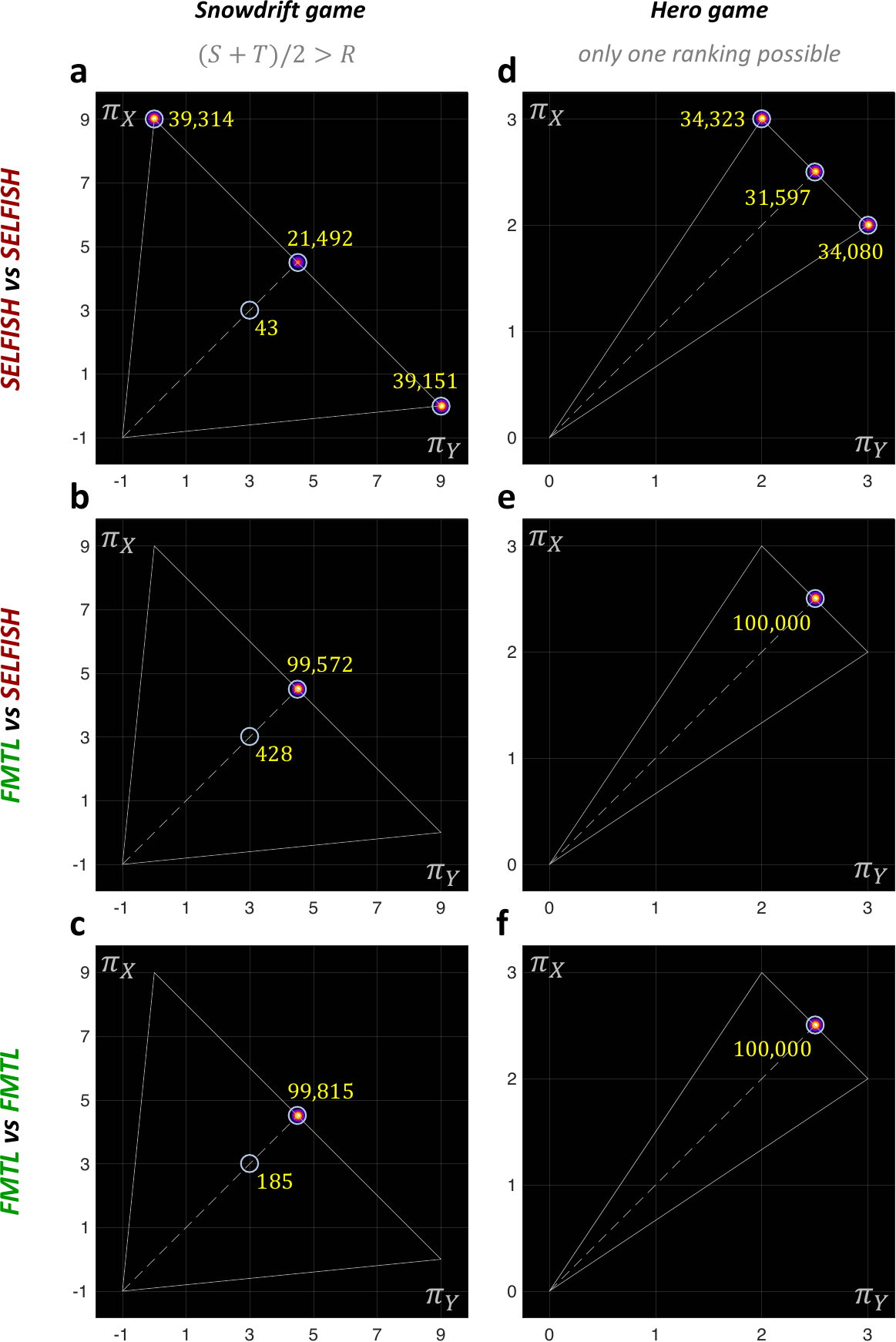}
	\caption{\textbf{Performance of \FMTL{} in conflicts for which selfish learning is nearly optimal.} In the snowdrift game with $\left(S+ T\right) /2>R$ (\textbf{a,b,c}) and the hero game (\textbf{g,h,i}), a selfish learner against a selfish learner leads to nearly optimal outcomes on average (\textbf{a} and \textbf{d}). For the snowdrift game, \FMTL{} versus a selfish learner actually leads to slightly lower average payoffs due to the insistence on fairness, which can lead (by chance) to a small amount of additional runs ending at mutual cooperation. For the hero game, the average payoffs in \textbf{d,e,f} are the same since the game is undiscounted. (Otherwise, the initial few rounds needed to coordinate a policy of alternation can affect payoffs slightly, but even in this case the average payoffs in \textbf{d,e,f} are approximately equal.) Game parameters: (\textbf{a},\textbf{b},\textbf{c}) $R=3$, $S=0$, $T=9$, and $P=-1$; (\textbf{d},\textbf{e},\textbf{f}) $R=1$, $S=3$, $T=2$, and $P=0$.\label{fig:others_neutral}}
\end{figure}

\clearpage
\newpage

\begin{figure}[p]
	\centering
	\includegraphics[width=0.9\linewidth]{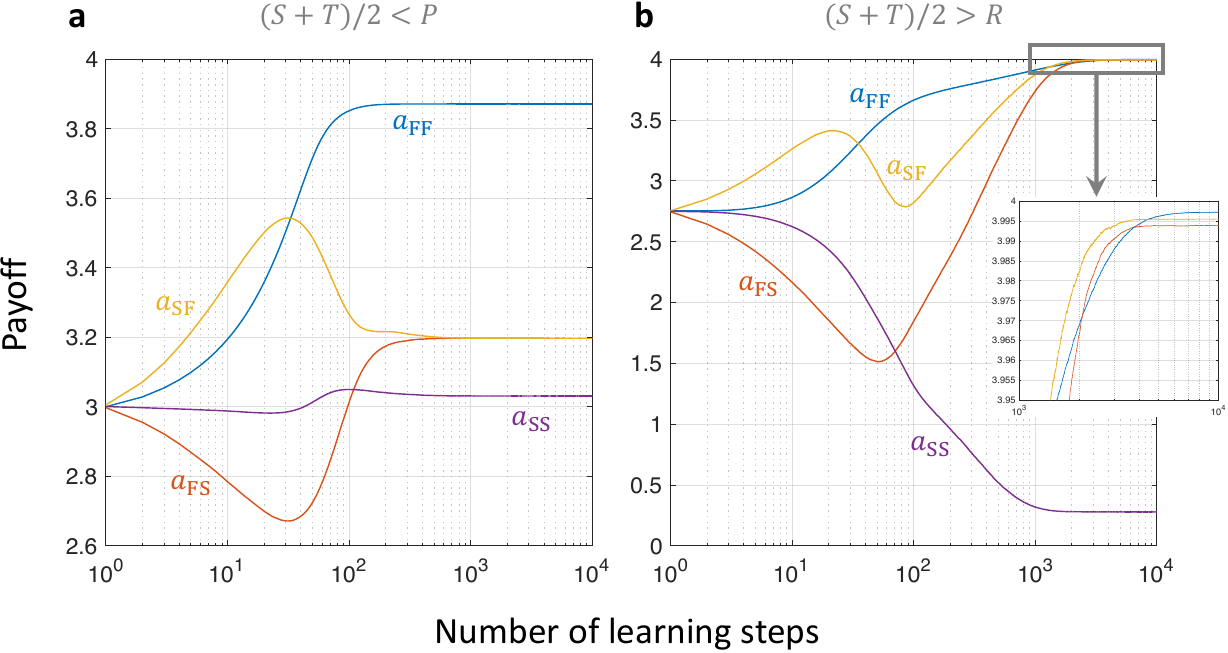}
	\caption{\textbf{Supergame payoffs in non-standard prisoner's dilemma interactions.} The donation game is a prisoner's dilemma satisfying $P<\left(S + T\right)/2<R$. Here, we instead consider prisoner's dilemmas in which one of these two inequalities does not hold. (There is no prisoner's dilemma in which both inequalities fail to hold.) In \textbf{a}, alternating cooperation is the most inefficient outcome, whereas in \textbf{b}, alternating cooperation is most efficient outcome. In both cases, the learning process eventually turns the resulting supergame into a harmony game, which means that \FMTL{} can invade and replace a population of selfish learners. In {\bf a,} this happens after $\approx 100$ learning steps (when $a_\textrm{FS}$ exceeds $a_\textrm{SS}$). In {\bf b,} it happens slightly earlier. Game parameters: \textbf{a,} $R=4$, $S=0$, $T=5$, and $P=3$; \textbf{b,} $R=3$, $S=-1$, $T=9$, and $P=0$.\label{fig:other_pds}}
\end{figure}

\clearpage
\newpage

\begin{figure}[p]
	\centering
	\includegraphics[width=1.0\linewidth]{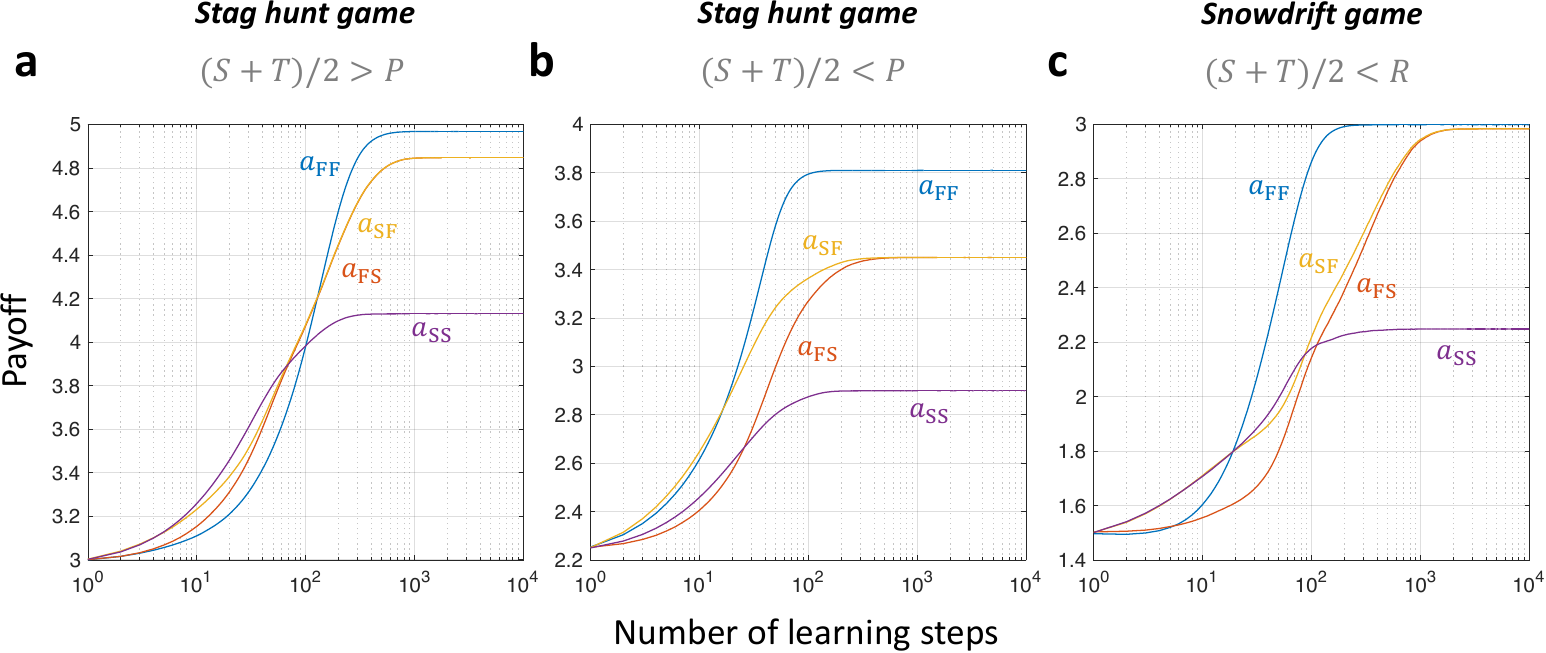}
	\caption{\textbf{Supergame payoffs in conflicts for which selfish learning is suboptimal.} In each of these games, mutual selfish learning leads to relatively low payoffs (purple). \FMTL{} is able to greatly improve these payoffs, on average. As a result, after a reasonably small number of learning steps, \FMTL{} globally dominates selfish learning in the resulting evolutionary dynamics. Game parameters: \textbf{a} $R=5$, $S=1$, $T=4$, and $P=2$; \textbf{b} $R=4$, $S=0$, $T=3$, and $P=2$; \textbf{c} $R=3$, $S=0$, $T=4$, and $P=-1$.\label{fig:supergame_others}}
\end{figure}

\clearpage
\newpage

\begin{figure}[p]
	\centering
	\includegraphics[width=0.9\linewidth]{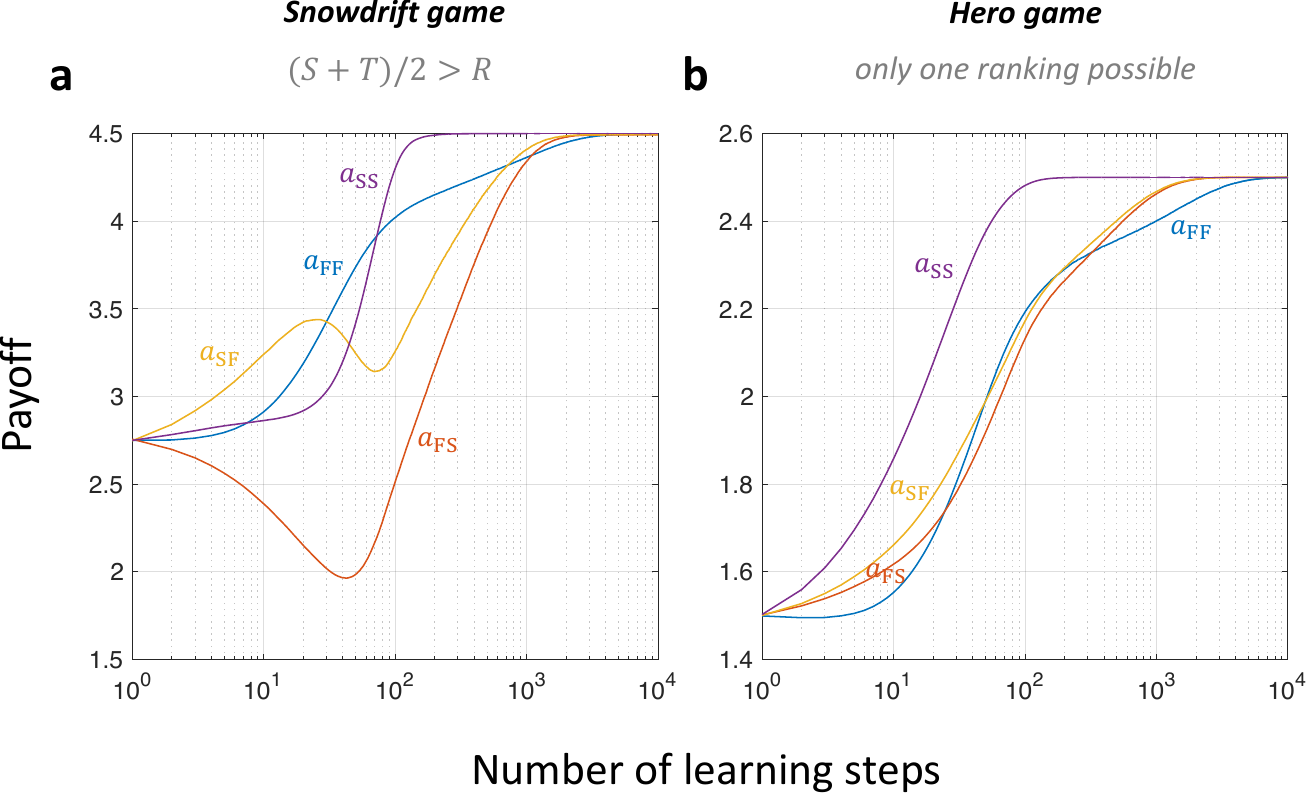}
	\caption{\textbf{Supergame payoffs in conflicts for which selfish learning is nearly optimal.} Since selfish learning is already nearly optimal (on average) in the snowdrift and hero games, \FMTL{} is approximately neutral relative to selfish learning after sufficiently many learning steps. Since \FMTL{} also values fairness (which in these cases does not affect evolutionary dynamics), it is less efficient in converging to a final outcome. As a result, on shorter learning timescales, selfish learning is significantly favored relative to \FMTL{}. Moreover, since \FMTL{} picks up a small number of additional suboptimal points at mutual cooperation in the snowdrift game (\sfig{others_neutral}\textbf{b,c}), after a sufficiently long learning horizon we find that there is bistable competition between \FMTL{} and selfish learning. However, since the long-run payoffs of the four curves in each of \textbf{a} and \textbf{b} all converge to approximately the same value, any competition between the two learning rules is nearly neutral after enough learning steps. Moreover, unlike in the previous conflicts we have studied, competition between the two learners in these two games does not significantly change the payoffs in a population. Game parameters: \textbf{a} $R=3$, $S=0$, $T=9$, and $P=-1$; \textbf{b} $R=1$, $S=3$, $T=2$, and $P=0$.\label{fig:supergame_others_neutral}}
\end{figure}

\clearpage
\newpage

\begin{figure}[p]
	\centering
	\includegraphics[width=1.0\linewidth]{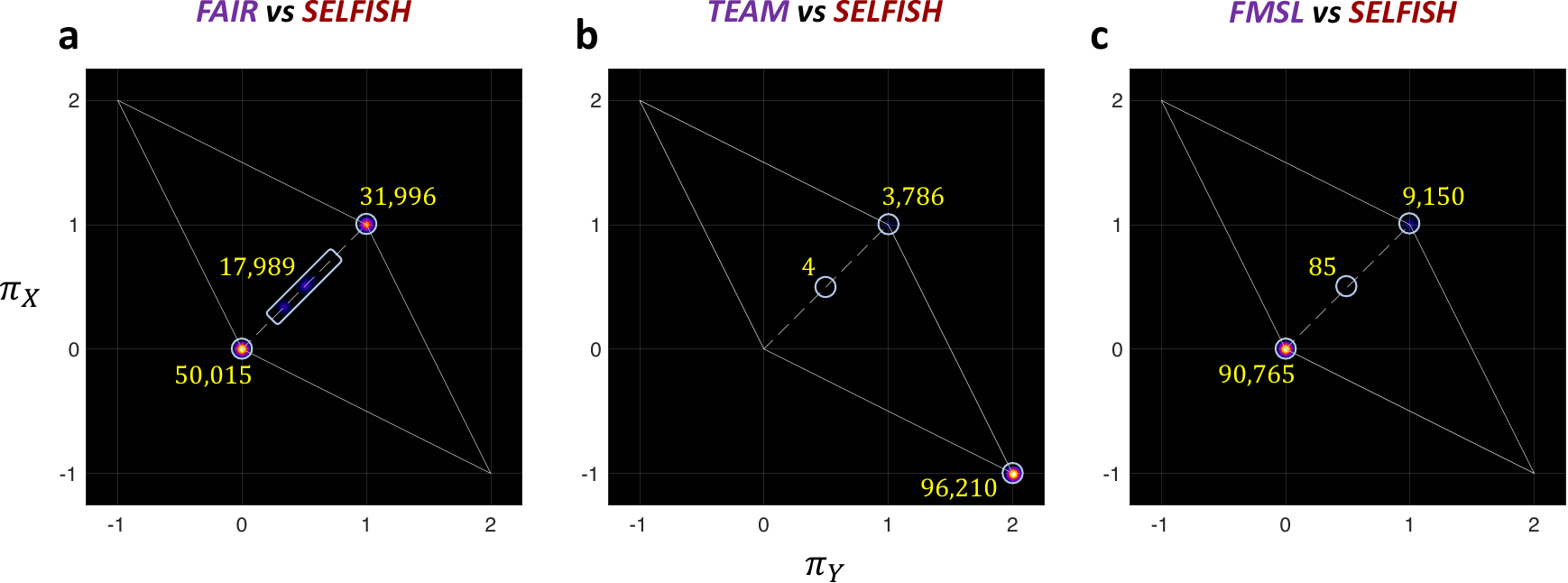}
	\caption{\textbf{Performance of related learning rules.} Since \FMTL{} balances both fairness and team-based learning, a natural question is whether either of these two objectives alone is sufficient when paired with a selfish learner. We consider the donation game with $b=2$ and $c=1$ as an example. \textbf{a} and \textbf{b} demonstrate that neither one alone can reliably lead to good outcomes. The final outcomes in \textbf{a} are fair but often involve a substantial amount of defection. In \textbf{b}, a selfish learner exploits a team-based learner in most runs. Finally, \textbf{c} addresses the question of whether good outcomes are possible with fairness-mediated \textit{selfish} learning (\textit{FMSL}). The results are reminiscent of when a selfish learner is paired with another selfish learner; in particular, the majority of outcomes from \textit{FMSL} versus a selfish learner involve a substantial amount of defection. The endpoints in all panels are based on $10^{5}$ random initial strategy pairs.\label{fig:alternative_rules}}
\end{figure}

\clearpage
\newpage

\begin{figure}[p]
	\centering
	\includegraphics[width=0.9\linewidth]{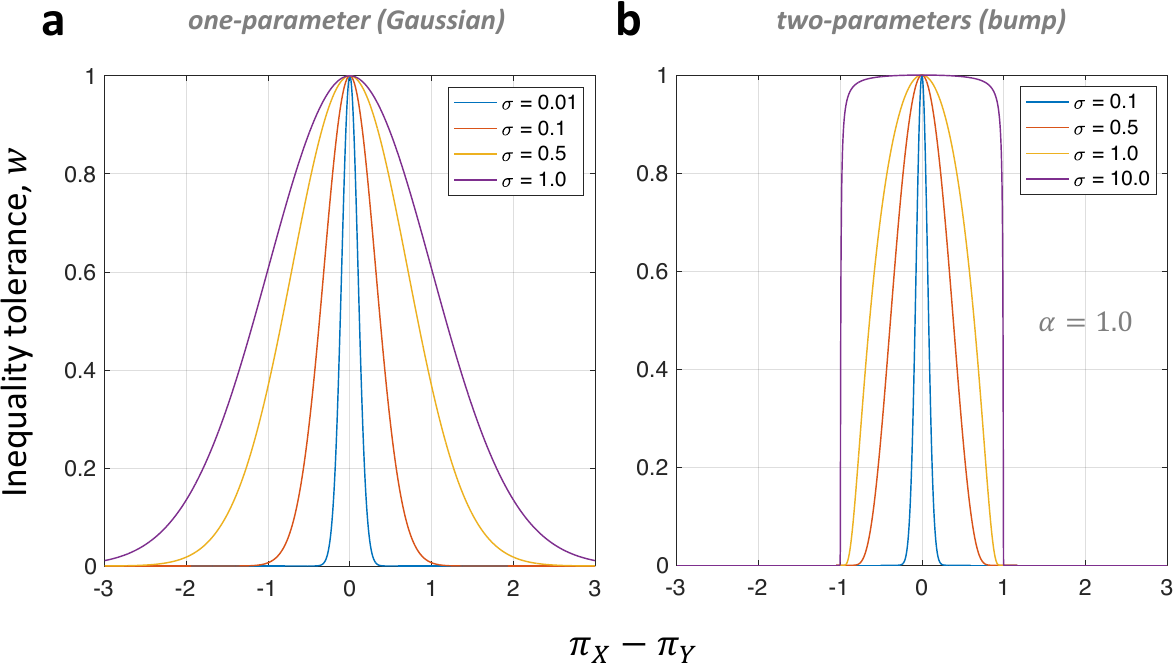}
	\caption{\textbf{Inequality tolerance functions for \FMTL{}.} Suppose that $X$ and $Y$ use strategies $\p$ and $\q$, respectively. In the \FMTL{} rule, $\omega$ represents the probability that the learner maximizes the sum of the players' payoffs, $\pi_{X}\left(\p,\q\right) +\pi_{Y}\left(\p,\q\right)$, as opposed to minimizing the magnitude of the difference, $\left|\pi_{X}\left(\p,\q\right) -\pi_{Y}\left(\p,\q\right)\right|$. This probability is a function of the current value of $\left|\pi_{X}\left(\p,\q\right) -\pi_{Y}\left(\p,\q\right)\right|$. \textbf{a} depicts a Gaussian tolerance function, which involves only one parameter (see \eq{gaussian_function}). \textbf{b} shows another option, a bump function, which allows for more control over both the rate of decay and the support of the function. This flexibility comes at the added cost of an additional parameter (see \eq{bump_function}). In our examples, we do not see significant qualitative differences between the two functions when implemented in \FMTL{}, so we use the Gaussian function for simplicity.\label{fig:w_function}}
\end{figure}

\clearpage
\newpage

\begin{figure}[p]
	\centering
	\includegraphics[width=1.0\linewidth]{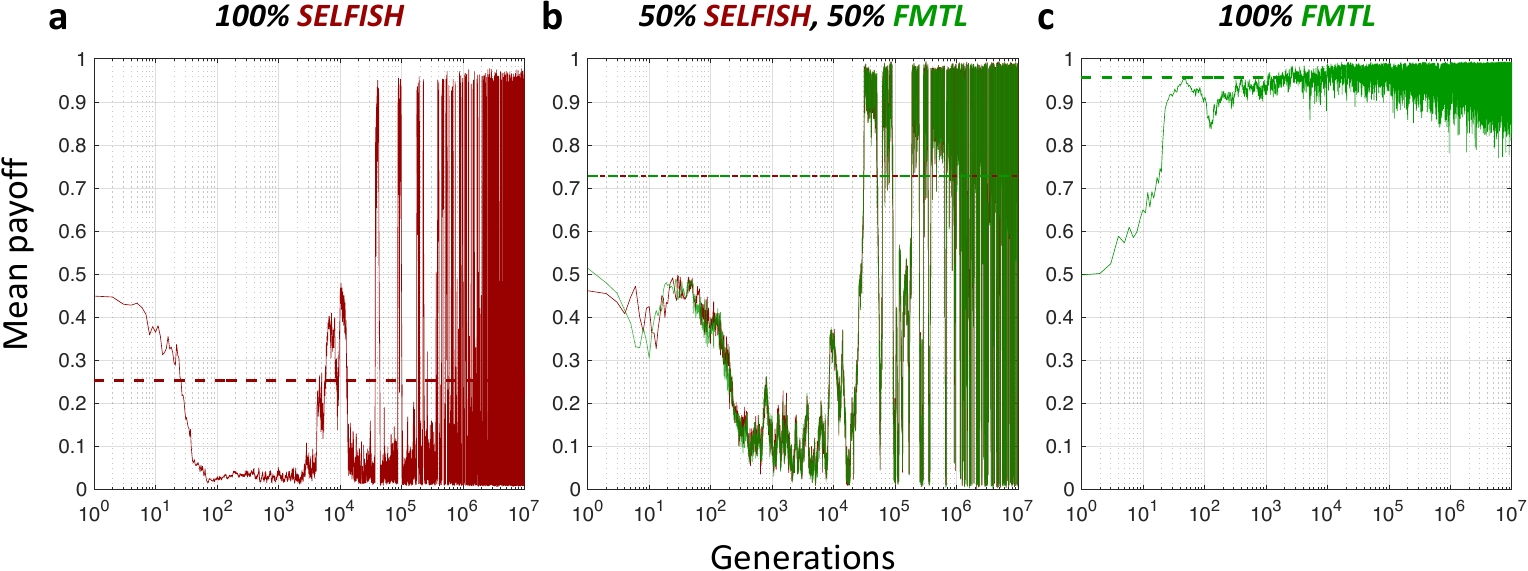}
	\caption{\textbf{Imitation-based learning dynamics.} In a population of size $N=100$, we consider synchronous imitation dynamics based on different social preferences. At the start, each individual chooses a random memory-one strategy, with each coordinate chosen independently from an arcsine distribution. In \textbf{a}, all individuals imitate based on a desire to improve their own payoffs. Panels \textbf{b} and \textbf{c} illustrate the presence of \FMTL{} imitators, who imitate based on either efficiency or fairness, with an equal mixture of selfish and \FMTL{} in \textbf{b} and only \FMTL{} in \textbf{c}. In each, mean payoffs over $10^{7}$ generations are depicted using dashed lines. Notably, these mean payoffs increase with the presence of \FMTL{}.\label{fig:imitation}}
\end{figure}

\clearpage
\newpage

\begin{figure}[p]
	\centering
	\includegraphics[width=0.5\linewidth]{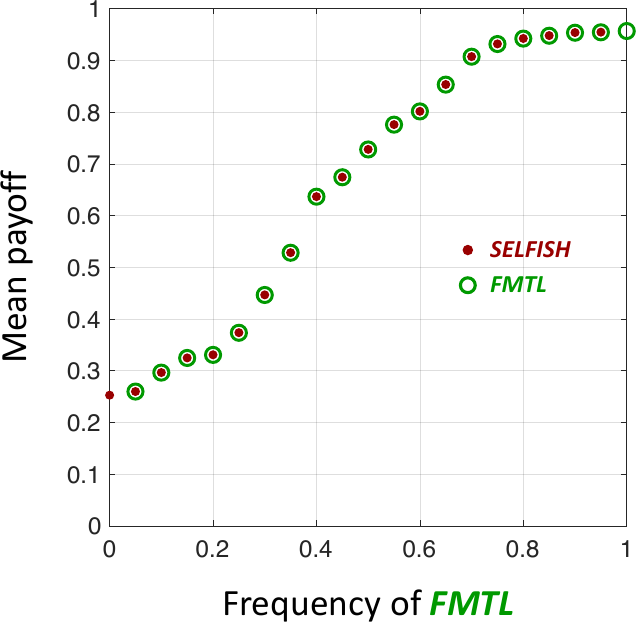}
	\caption{\textbf{Mean payoffs as functions of \FMTL{} frequency under imitation dynamics.} In a population of size $N=100$, mean payoffs over $10^{7}$ generations are shown in red for selfish imitators and in green for \FMTL{} imitators (corresponding to the dashed lines in \fig{imitation}). The number of \FMTL{} present in the population ranges from $0$ to $N$ in increments of $5$, and the trend is that \FMTL{} increases the mean payoffs of both kinds of imitators. In particular, a selfish imitator would always benefit from switching to \FMTL{}, which is qualitatively similar to the findings for introspection dynamics in the main text. The stochasticity in these mean payoffs is due to finite population effects and the choice of a finite number of generations ($10^{7}$) over which to average.\label{fig:imitation_trend}}
\end{figure}

\clearpage
\newpage

\begin{figure}[p]
	\centering
	\includegraphics[width=0.9\linewidth]{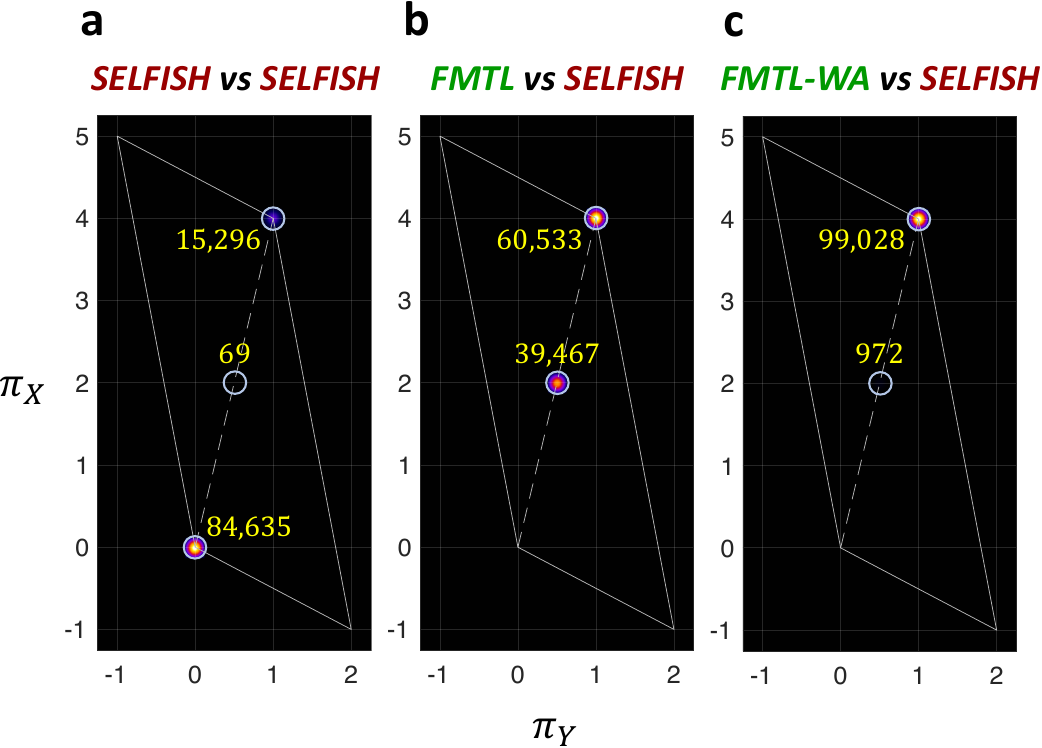}
	\caption{\textbf{Fairness-mediated team learning in a weakly-asymmetric game.} The learning rule \FMTL{}, if not properly modified for asymmetric games, may be thought of as ``equality-mediated team learning.'' Although this learning rule results in better outcomes against a selfish learner relative to selfish learning (panels \textbf{a} and \textbf{b}), it does not properly take into account fairness since equality no longer serves as a proxy for fairness in asymmetric settings. Instead, since this asymmetric donation game is weakly asymmetric, there is a natural correspondence between the strategies of the two players (which, in this case, is just $C$ to $C$ and $D$ to $D$). This correspondence allows the players to compare payoffs when strategies are swapped (\eq{VF_wa}), giving rise to a weakly-asymmetric version of \FMTL{}, which we denote \emph{FMTL-WA} (panel \textbf{c}). This rule elicits fair, optimal outcomes against a selfish learner in over 99\% of runs, reflecting the qualitative findings discussed for symmetric games in the main text. The dashed line represents the outcomes that are perfectly fair, i.e. which maximize the objective function in \eq{VF_wa}. The endpoints in each panel are based on $10^{5}$ randomly-sampled initial strategies, with $b_{X}=2$, $b_{Y}=5$, and $c=1$.\label{fig:FMTL_asymmetric}}
\end{figure}

\clearpage
\newpage

\renewcommand\tablename{Tab.}

\begin{table}
	\centering
	\includegraphics[width=0.9\textwidth]{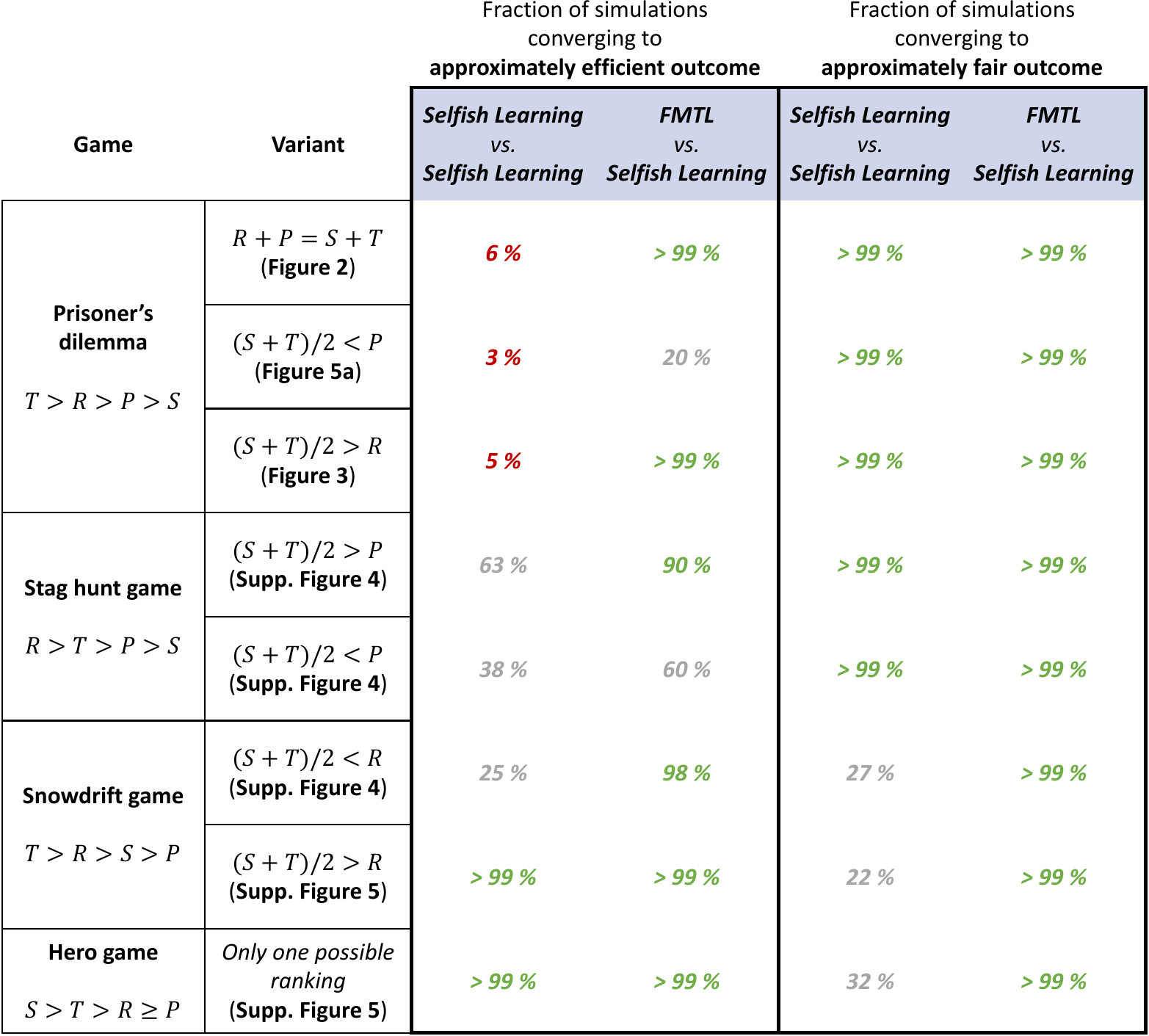}
	\caption{\textbf{Performance of learning rules across different repeated games.} When facing a selfish learner, we ask, for each learning rule, whether the resulting dynamics eventually lead to efficiency and fairness. For efficiency, we consider the degree to which a learner can evade all inefficient equilibria. For fairness, we ask whether players are expected to achieve an equitable outcome. We note that in games in which alternating cooperation is particularly detrimental, such that $\left(S+T\right) /2<P$, no combination of learning rules can escape from mutual defection. However, if at least one player adopts \FMTL{}, players are much less likely to settle at mutual defection in the first place (\fig{other_pds}a, \sfig{others}d,e, and \sfig{supergame_others}b).\label{tab:SummaryFixedLearningRules}}
\end{table}

\clearpage
\newpage

\begin{table}
	\centering
	\includegraphics[width=0.9\textwidth]{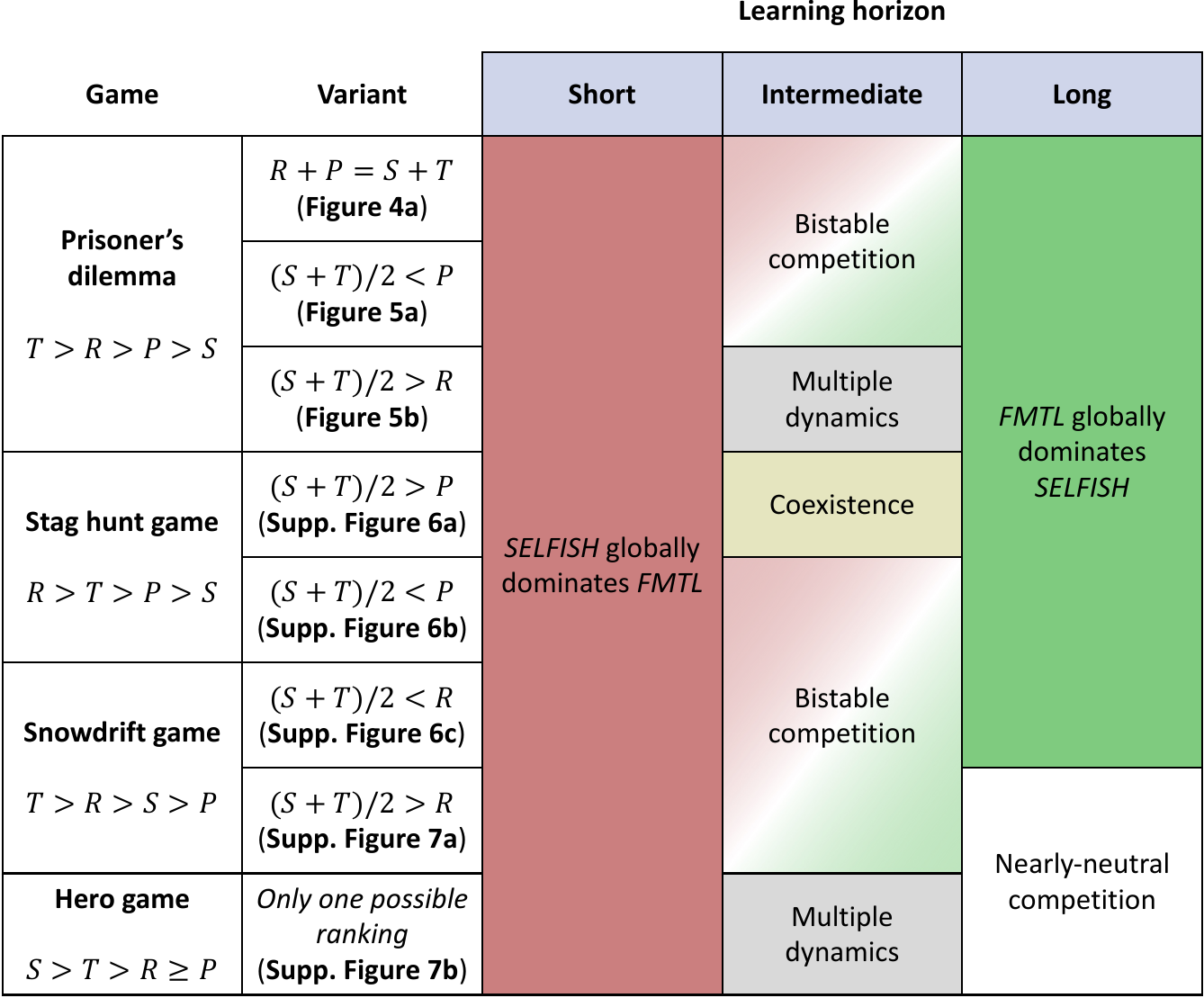}
	\caption{\textbf{Evolutionary dynamics among learning rules across different repeated games.} Here, we summarize the results shown in \fig{replicatorDynamics}, \fig{other_pds}, \sfig{supergame_others}, and \sfig{supergame_others_neutral}. For each of the game types we consider, the table shows the qualitative dynamics between selfish learning and \FMTL{}. The results depend on the players' learning horizon. The table broadly distinguishes between three cases. A short learning horizon refers to the limiting case in which players assess the performance of their learning rule based on a minimum of learning steps, for $n=1$. A long learning horizon refers to the other limiting case in which players assess their learning rule based on which outcome they yield eventually, for $n\rightarrow\infty$. We use the intermediate class to summarize all phase transitions that we observe between those extremes. For two games, we observe multiple phase transitions. The first such game is the prisoner's dilemma with $\left(S+T\right) /2 > R$. Here, both bistable competition and coexistence are possible for intermediate learning horizons (\fig{other_pds}b). In the hero game, we observe both bistable competition and selfish learning dominating \FMTL{} (\sfig{supergame_others_neutral}b). We note that while the short and the long class are formally defined as a limiting case, the behaviors described here are robust. In particular, selfish learning remains globally stable for $n>1$, as long as the number of learning steps is below a threshold that depends on the game. Similarly, \FMTL{} can be globally stable long before the learning dynamics converge to an equilibrium.\label{tab:SummaryEvolution}}
\end{table}

\clearpage
\newpage

\renewcommand\figurename{Vid.}
\setcounter{figure}{0}

\begin{figure}
	\caption{\textbf{Escape from ALLD.} Both players start with strategies of ALLD, mutual unconditional defection, in the donation game with $b=2$ and $c=1$. $X$ uses \FMTL{} while $Y$ is a selfish learner. After each learning step, the current payoff is depicted in purple, while the feasible region of the strategy used by $X$ is shown in light grey. This region represents the space of all possible payoffs against this particular strategy of $X$. As the number of learning steps (shown in the upper-right corner) increases, $X$ is successfully able to draw $Y$ out of defection, eventually settling on strategies that support mutual cooperation. The feasible region for the final strategy used by $X$ makes it clear why this strategy cannot be exploited: any deviation in payoffs must decrease the score of $Y$. (Video file available at \texttt{https://github.com/alexmcavoy/fmtl/blob/main/paper\_data/VidS1.mp4}.)\label{vid:trajectory}}
\end{figure}

\clearpage
\newpage

\begin{center}
	\Huge\textbf{Supporting Information}
\end{center}

\setcounter{equation}{0}
\setcounter{section}{0}
\renewcommand{\thesection}{S\arabic{section}}
\renewcommand{\thesubsection}{S\arabic{section}.\arabic{subsection}}
\renewcommand{\theequation}{S\arabic{equation}}

\section{Related literature}
Our study explores how different learning rules affect the dynamics of two-player interactions. It is motivated by previous work in three fields, the literature on evolution in repeated games, the literature on social preferences, and the literature on multi-agent learning. In the following, we discuss each of these three fields in more detail.

\subsection{Related work on evolution in repeated games}
There is a rich literature on evolution in repeated games. A general summary of this literature can be found in the books by \citet{axelrod:BB:1984}, \citet{nowak:BP:2006}, and \citet{sigmund:PUP:2010}, or in recent review articles \citep{hilbe:NHB:2018,dalbo:JEL:2018}. Research in this area explores three broad themes. \emph{(i)} It aims to identify strategies that can sustain cooperation \citep{axelrod:Science:1981,molander:jcr:1985,Kraines:TaD:1989,nowak:Nature:1992a,nowak:Nature:1993,hauert:PRSB:1997,fischer:PNAS:2013,Murase:SciRep:2020}. \emph{(ii)}~It describes how likely such cooperative strategies evolve \citep{lindgren:bookchapter:1997,szabo:PRE:2000b,nowak:Nature:2004,imhof:PRSB:2010,martinez-vaquero:plosone:2012,szolnoki:scirep:2014,baek:SR:2016} and persist \citep{boyd:Nature:1987,boyd:JTB:1989,garcia:jet:2016,Garcia:FRAI:2018} in a population. \emph{(iii)} It evaluates to which extent the predicted strategies can be observed in humans \citep{milinski:PNAS:1998,dalbo:AER:2011,fudenberg:aer:2012,hilbe:ncomms:2014} or other animal species \citep{Raihani:JEB:2011}. While our study speaks to all three aspects mentioned above, we are most concerned with a more general question. We ask how individuals themselves can learn to adopt beneficial strategies more effectively, when given the choice between different learning rules. 

Previous research on evolution in repeated games presumes that individuals engage in what we have termed ``selfish learning.'' Selfish learners evaluate the performance of their strategies by the immediate payoffs they yield. Herein, we argue that selfish learning may be replaced by alternative learning rules if selection operates on a learning rule's long-run performance. The selection pressure for an alternative learning rule may be particularly strong in games with multiple equilibria. In such games, selfish learning can lead individuals to settle at inefficient outcomes although more profitable equilibria are available. Such inefficiencies can occur in the repeated prisoner's dilemma \citep{hilbe:NHB:2018} but also in any other social dilemma that requires coordination \citep{pacheco:PRSB:2009,santos:PNAS:2011,Iyer:PLosCB:2016,Couto:JTB:2020}.

Within the literature on repeated games, our work is most closely related to studies that explore properties of memory-one strategies \citep{press:PNAS:2012,stewart:PNAS:2013,stewart:pnas:2014,hilbe:GEB:2015,akin:Games:2015,Akin:chapter:2016,McAvoy:PNAS:2016,Ichinose:JTB:2018,Mamiya:PRE:2020,mcavoy:PRSA:2019}. For the repeated prisoner's dilemma without discounting, \citet{akin:Games:2015,Akin:chapter:2016} was first to describe all memory-one strategies that stabilize full cooperation in a Nash equilibrium. This work was extended by \citet{stewart:pnas:2014} to describe all symmetric Nash equilibria. In equilibrium, only four outcomes are feasible: mutual cooperation, mutual defection, alternating cooperation, and equilibria that are enforced by so-called equalizer strategies \citep{boerlijst:AMM:1997}. Because Nash equilibria result in outcomes where no player can unilaterally increase their payoff, any such equilibrium is a rest point of the learning dynamics among two selfish learners. When players can switch to any other memory-one strategy, the converse is also true: any rest point of the learning dynamics among two selfish learners must be a Nash equilibrium.

However, similar to the literature on adaptive dynamics \citep{geritz:PRL:1997,dieckmann:TREE:1997}, we focus on players who adapt locally. This means that in each updating step, learners make only minor modifications to their existing strategies. In that case, two selfish learners may get stuck in rest points that are locally stable without being Nash equilibria. Local adaptation leads to ``smoother'' trajectories of the learning dynamics. However, as shown by \citet{stewart:games:2015}, the qualitative outcomes are comparable to those of global updating. Again, eventually players either mutually cooperate, defect, alternate, or they employ equalizer strategies. However, simulations suggest that they become even more likely to settle at mutual defection \citep{stewart:games:2015}. Selfish learning is therefore particularly prone to result in inefficient outcomes under local adaptation.\\ 

\noindent
Overall, we contribute to this literature as follows:
\begin{enumerate} \setlength{\itemsep}{0cm}
\item We introduce a framework to formalize learning in repeated games. While we focus on one learning rule in particular (\FMTL{}), our notion of learning rules is general, and it can be used to describe many alternative modes of learning. 
\item We explore the learning dynamics that transpire when players have objectives that are different from immediate payoff maximization. With \FMTL{}, we identify a learning rule that shows a surprisingly strong performance across many different types of games. In these games, {\it one} \FMTL{} learner is usually sufficient for {\it both} players to be better off eventually.
\item Our research clarifies under which conditions selfish learning can be expected to persist in a population. We find that the evolutionary robustness of selfish learning depends on the game being played, and on the relative timescale of evolution. When learning rules evolve at a similar rate as the learners' strategies, selfish learning is prevalent. However, when learning rules evolve at a much slower timescale, selfish learning is dominated in most games we consider. The only exceptions arise when selfish learning generally results in efficient outcomes, as in the hero game and in certain snowdrift games.
\end{enumerate}

\subsection{Related work on social preferences}
In addition to the evolutionary game theory literature, our work is related to the branch of behavioral economics that explores social (or other-regarding) preferences. Research in this area is centered around the following three themes. 
\emph{(i)}~It explores experimentally to which extent humans take the payoffs of other group members into account when making decisions. \emph{(ii)}~It aims to integrate these empirically estimated preferences into standard models of economic decision making. \emph{(iii)}~It describes how humans may have developed social preferences over an evolutionary timescale. For a general overview of this literature and a critical evaluation, we refer to the reviews by \citet{fehr:Handbook:2006}, \citet{List:ARE:2009}, and \citet{Alger:ARE:2019}.

Social preferences can manifest themselves in various forms. Decision makers may aim to increase the payoff of the group member with the lowest payoff (``maximin preferences''); they may strive to have the highest payoff within their group (``competitive preferences''); they may want to maximize the joint payoff of all group members (``efficiency preferences''); or they may wish to reduce economic inequality within their group (``inequity aversion,'' which is usually interpreted as a preference for fairness). To the extent that decision-makers are willing to pay a personal cost to achieve these objectives, their preferences represent deviations from the classical {\it homo economicus}. 

In the context of our study, the last two of the above-mentioned preferences are particularly relevant--those related to fairness and efficiency. In the following, we first discuss some of the experimental evidence for the existence of such preferences. In a next step, we briefly review the existing theoretical literature on preference evolution. 

\subsubsection{Experimental evidence for fairness and efficiency preferences}
There is a rather extensive literature suggesting that humans value egalitarian outcomes \citep{fehr:nature:2003,mcauliffe:NHB:2017}. One piece of evidence comes from experiments on ultimatum-bargaining \citep{guth:JEBO:1982}. Here, participants typically reject mutually beneficial offers if these offers put them at a notable disadvantage. Another piece of evidence comes from experiments on public good games with punishment \citep{fehr:AER:2000}. Here, a non-negligible fraction of participants are willing to pay a cost to reduce the payoff of free-riders, even if any long-run advantage of punishment is excluded by the experimental design. While there is considerable cross-cultural variation in the extent to which humans reject unfair offers and retaliate against free-riders \citep{henrich:AER:2001,herrmann:Science:2008}, an aversion against disadvantageous inequity seems to be universal \citep{Blake:Nature:2015} and it is developed by children at a relatively young age \citep{fehr:Nature:2008}. The brain regions that are activated when rejecting unfair offers include the ventral striatum and the ventromedial prefrontal cortex \citep{tricomi:Nature:2010}. The former is associated with social comparisons, whereas the latter integrates costs and benefits into a decision value \citep{mcauliffe:NHB:2017}. 

Preferences for inequity aversion can be integrated into economic models by extending the individual's utility functions. While the precise specifications differ between studies \citep{fehr:QJE:1999,bolton:AER:2000}, these models usually represent utilities as a linear combination of the individual's own payoff and some measure of group inequality. 
By calculating the Nash equilibria for the modified utility functions, these models can be used to make experimental predictions.

There is similar research on the role of efficiency in human decision making. For example, \citet{andreoni:E:2002} study different variants of a dictator game. In their experiments, a randomly assigned participant (the ``dictator'') decides how to divide a fixed number of tokens between herself and a recipient. Different treatments vary the relative worth of tokens. In some treatments, tokens are more valuable to the dictator; in others they are more valuable to the recipient. The study finds that there is considerable heterogeneity among dictators. However, a substantial number of dictators give more tokens to the player to whom tokens are worth the most (even if this happens to be the other player). This may be taken as evidence that such dictators value efficiency. Similar qualitative findings are reported by \citet{charness:QJE:2002} and by \citet{engelmann:AER:2004}. 
As with inequity aversion, preferences for efficiency can be incorporated into standard models of game theory by considering appropriately parameterized utility functions \citep{andreoni:E:2002,charness:QJE:2002}.

\subsubsection{Indirect evolutionary approach}
Individuals with social preferences may adopt strategies that do not maximize their material payoff. If one assumes that evolutionary processes select for payoff-maximizing behavior, the question arises how social preferences can evolve. To address this question, researchers have adopted the ``indirect evolutionary approach'' \citep{guth:IJGT:1995,guth:RS:1998}. Similar to our evolutionary model, this literature considers dynamics on two timescales. In the short run, the players' social preferences are fixed. Given these preferences, individuals coordinate on an equilibrium that maximizes their subjective utilities. In the long run, the players' preferences evolve. For the evolution of preferences, material payoffs are relevant. Only those preferences that yield high material payoffs in equilibrium proliferate. For a detailed review of this literature, see \citet{Alger:ARE:2019}. In the following, we describe some of the key insights.

The indirect evolutionary approach has identified two major mechanisms that allow social preferences to be evolutionarily stable. The first mechanism is based on the observation that social preferences can serve as a commitment device (e.g. Refs.~\citen{guth:IJGT:1995, guth:RS:1998, heifetz:ET:2007, akcay:PNAS:2009}). The respective models assume (sometimes implicitly) that a player's preference type is publicly observable. In this ``complete information'' scenario, preferences affect a player's payoff in two ways. \emph{(i)}~They have a direct effect as they influence which action the player will choose. \emph{(ii)}~They have an indirect effect as they influence how the player's opponent will react. While the direct effect has negative consequences on the player's payoff, the indirect effect can be positive. Because the direct payoff effect can be expected to be vanishingly small for sufficiently small deviations from payoff-maximizing behavior, social preferences can evolve \citep{heifetz:ET:2007}. 

As an example of such a setup, \citet{akcay:PNAS:2009} consider a model where two individuals share resources with each other. Each individual's material payoff is monotonically decreasing in the amount of resource given away, and monotonically increasing in the amount of resource received. Players with other-regarding preferences share more than they would in a Nash equilibrium. Because of the assumed multiplicative form of the players' preference functions, this also induces the second player to share more of their resource. As long as the overall effect is positive, population members have an incentive to extend the magnitude of their other-regarding preferences. Interestingly, the eventual equilibrium can only be sustained {\it because} the players have evolved other-regarding preferences. This is somewhat different from what happens in our model. In all examples we have studied, the eventual equilibrium payoffs of two \FMTL{} learners are also viable equilibrium payoffs among two selfish players. However, \FMTL{} makes it more likely that two learners would coordinate on the respective fair and efficient equilibria.

The second (and more recently explored) mechanism that can give rise to social preferences is population structure. 
When players with similar preferences are more likely to interact with each other, evolution may give rise to behavior consistent with {\it Homo moralis} \citep{alger:ECO:2013}. For pairwise interactions, this type of player maximizes a linear combination of the material payoff and the payoff the player would obtain if the co-player adopted the same strategy. While we focus on well-mixed populations throughout most of the main text, we describe in the {\bf Methods} section that population structure can also have a positive effect within our learning framework.\\

\noindent
Our work adds to this literature as follows:
\begin{enumerate} \setlength{\itemsep}{0cm}
\item Existing theoretical models on social preferences use a static perspective to describe the strategic interaction (i.e. the short run dynamics). They introduce modified utility functions to explore the effect on the game's Nash equilibria. To make clear predictions, this approach often requires the respective equilibria to be unique.\newline
Instead, we are interested in a dynamic theory of learning. In our framework, players do not play equilibrium strategies from the outset, but they adapt their strategies over time. Learning rules do not only affect whether a given payoff allocation can be achieved in equilibrium. Instead, they also affect how likely players are to coordinate on each possible equilibrium outcome. 
\item Because the existing literature on the indirect evolutionary approach focuses on equilibrium behavior, they typically assume a complete separation of timescales; by the time preferences evolve, the players' strategies have reached an equilibrium. Instead, we add to this literature by exploring how the players' learning horizon affects the long-run evolutionary dynamics. 
\item The experimental literature on social preferences highlights the role of fairness and efficiency for human decision making. We show that these two objectives can be combined to create a learning rule that exhibits a remarkable performance in the context of multi-agent learning.
\end{enumerate}

\subsection{Related work on multi-agent learning}
In addition to its relationship to the social sciences, our work touches upon topics of interest in the field of multi-agent learning. The literature on single- and multi-agent reinforcement literature is vast; here, we briefly mention several studies that are closely related. For general overviews of the basic goals of multi-agent learning, as well as its relationship to evolutionary game theory, we refer the reader to Refs.~\citen{shoham:AI:2007,tuyls:AI:2007,bloembergen:JAIR:2015,zhang:HRLC:2021}.

Optimization problems involving a single agent in a stationary environment are usually approached using the well-studied theory of Markov decision processes, where algorithms have strong performance guarantees. Multi-agent learning introduces two key complications \citep{tuyls:AI:2012,bloembergen:JAIR:2015}, which are captured by our simplified model. First, the presence of other agents, all of which can perform actions influencing each other, and all of which are actively learning and changing their own behavior, introduces a ``moving target'' aspect to the environment, where the outcome of the learning algorithm of one agent is interdependent with the learning trajectory of the other agent \citep{tuyls:AAMAS:2005}. The environment is no longer stationary or Markovian. In the multiplied learning setting considered in this paper (as opposed to divided or active learning) each agent acts on their own and is not necessarily aware of or aligned with the goals and algorithms of other agents. Since increased awareness allows for simpler solutions and direct incorporation of other agents' objective functions \citep{foerster2018learning}, which may not be possible in real-life settings, we assume agents to be mutually unaware.

Second, the behavior of the other agents cannot necessarily be predicted based on simple assumptions. Unlike fully cooperative scenarios where all agents have aligned goals, or fully competitive zero-sum games, where one can reason unambiguously about the best response of adversaries \citep{silver2016mastering,silver2017mastering,vinyals2019grandmaster} and make progress with methods such as tree-search or self-play, most real world scenarios combine both cooperative and adversarial components and incentive alignment is unclear or variable. As such, a degree of coordination with the other player and their learning trajectory is necessary to achieve outcomes that maximize both payoffs while limiting exploitation. In this new regime, the optimal choice of objective function for the learner is unclear. How to align incentives and find globally optimal outcomes in this general/unaligned case remains poorly understood \citep{shoham:AI:2007}.

Many approaches to multi-agent learning in the context of repeated social dilemmas assume multiple self-interested agents as a standard starting point \citep{leibo2017multi}. An example is infinitesimal gradient ascent (including variants such as ``win or learn fast,'' where learning rate is high when algorithm is losing, otherwise low \citep{bowling:AI:2002}), which follows the gradient of expected payoffs and is guaranteed to converge to the Nash equilibrium of two-player, two-action games. (In our study, even when discussing selfish learning, our primary focus is on optimization via random search, as opposed to gradient-based techniques.) Another example is Nash-Q, an exemplary algorithm for multi-agent learning, which aims to find Nash equilibria \citep{hu:JMLR:2003}. While we note that a high $b/c$ ratio can give rise to increasingly favorable outcomes for self-interested agents (see \fig{donationGame}), self-interested agents and Nash equilibria can result in globally poor outcomes in the general case, especially in social dilemmas.

With that being said, in the same way that the idea of social preferences in behavioral economics has been around for decades, considering the opponent's score in multi-agent optimization is not new. A simple example is a learning rule called ``cooperative reward shaping'' \citep{ivanov:AAMAS:2021}, which blends selfish objectives with a desire to increase efficiency. Simple convex combinations of the players' payoffs has also proven effective in coordination games such as the stag hunt \citep{peysakhovich:AAMAS:2018}. Inequity aversion, based on the model of \citet{fehr:QJE:1999}, has been shown to promote cooperation in temporally-extended social dilemmas \citep{hughes:NEURIPS:2018}. A slightly different approach is to incorporate the opponent's learning process directly into one's own (``learning with opponent-learning awareness'' \citep{foerster2018learning}), provided this information is available. In fact, one could readily conjecture that other-regarding preferences do (and will) play an important role in algorithms based on the fact that humans, a species subject to evolutionary pressures, face similar challenges to those encountered in multi-agent learning and are documented to have social preferences.

The authors of Ref.~\citen{shoham:AI:2007} describe five agendas in the field of multi-agent learning: computational, descriptive, normative, prescriptive (cooperative), and prescriptive (non-cooperative). Our work would be best described as descriptive and prescriptive (non-cooperative). It is descriptive, at least partially, in that the learning rules we consider are inspired by behavioral experiments, in which humans are known to care about (and even balance) fairness and efficiency. In that respect, we are concerned with whether such behaviors can drive selfishness out of a population due their abilities to shape the incentives of a selfish learner. Our work is prescriptive, in the non-cooperative sense, because it addresses the question of how a learner can/ought to behave in order to overcome the drawbacks of selfish learning (since selfish learning is inefficient in non-cooperative games).

As it relates to the multi-agent learning literature, our contribution can be summarized as follows:
\begin{enumerate} \setlength{\itemsep}{0cm}
\item We present a novel learning rule, which balances objectives of fairness and efficiency in a dynamic fashion and achieves excellent outcomes against selfish learning across a variety of games.
\item We study ``supergames'' of learning rules. When trying to understand competition between learning rules in a population, one needs to know the outcomes of all possible combinations of learning rules. This kind of analysis can allow one to make deductions about evolutionary dynamics without studying a specific kind of update rule (e.g. for dominance games). We study this supergame and its qualitative transitions (e.g. bistable competition, coexistence, and dominance games) as a function of the timescale of learning.
\item We show that evolutionary models do not take into account aspects of learning rules that might be nonetheless important in settings of multi-agent learning. While fairness is important for eventually arriving at outcomes with high payoffs, on average, evolutionary dynamics often do not take into account inequity. The hero game, as well as a variant of the snowdrift game, illustrate this point clearly: \FMTL{} is effectively neutral relative to selfish learning (\fig{supergame_others_neutral}), yet \FMTL{} is able to elicit fair outcomes against an opponent with near certainty, while two selfish learners often end up with unequal payoffs (\fig{others_neutral}).
\end{enumerate}

\section{An analysis of local strategy revisions}
In our study, we consider learning processes among players with memory-one strategies in repeated two-player interactions. To this end, we consider players who adapt {\it locally}. In each learning step, the respective player compares its current strategy $\mathbf{p}$ with an alternative strategy $\mathbf{p}'$ that is sufficiently nearby. The alternative strategy is adopted if it leads to an improvement, given the player's objectives. In the following, we ask under which conditions such local improvements are possible. This analysis will help us to characterize which strategy pairs $\left(\mathbf{p},\mathbf{q}\right)$ are rest points of the learning dynamics. Rest points are those strategy profiles for which both players are unable to make any further local improvements (given their objectives). We begin our analysis by deriving some useful properties of memory-one strategies in repeated games.

\subsection{Game setup and some useful lemmas} 
We consider two players, $X$ and $Y$, with memory-one strategies $\mathbf{p}=\left(p_{0},p_{CC},p_{CD},p_{DC},p_{DD}\right)$ and $\mathbf{q}=\left(q_{0},q_{CC},q_{CD},q_{DC},q_{DD}\right)$, respectively. They repeatedly interact in a game with one-shot payoffs $\mathbf{u}_{X}=\left(R,S,T,P\right)$ and $\mathbf{u}_{Y}=\left(R,T,S,P\right)$. Future payoffs are discounted by $\lambda <1$. The players' payoffs for the repeated game can thus be written as in \eq{discountedPayoff} of the main text,
\begin{subequations}\label{eq:discountedpayoff1}
\begin{align}
\pi_{X}\left(\mathbf{p},\mathbf{q}\right) &= \left< \mathbf{v}, \mathbf{u}_{X} \right> ; \\
\pi_{Y}\left(\mathbf{p},\mathbf{q}\right) &= \left< \mathbf{v}, \mathbf{u}_{Y} \right> .
\end{align}
\end{subequations}
Here, the vector $\mathbf{v}$ is given by
\begin{align} \label{eq:invariant_distribution}
\mathbf{v} = \left(v_{CC},v_{CD},v_{DC},v_{DD}\right) \coloneqq \left(1-\lambda\right) \nu_{0}\left(\p ,\q\right)\big(I-\lambda M\left(\p ,\q\right)\big)^{-1},
\end{align}
where 
\begin{align}\label{eq:initial_distribution}
\nu_{0}\left(\p ,\q\right) &= \Big(p_{0}q_{0},~p_{0}\left(1-q_{0}\right),~\left(1-p_{0}\right) q_{0},~\left(1-p_{0}\right)\left(1-q_{0}\right)\Big)
\end{align}
and 
\begin{align}
M\left(\p ,\q\right) &= 
\begin{pmatrix}
p_{CC}q_{CC} & p_{CC}\left(1-q_{CC}\right) & \left(1-p_{CC}\right) q_{CC} & \left(1-p_{CC}\right)\left(1-q_{CC}\right) \\
p_{CD}q_{DC} & p_{CD}\left(1-q_{DC}\right) & \left(1-p_{CD}\right) q_{DC} & \left(1-p_{CD}\right)\left(1-q_{DC}\right) \\
p_{DC}q_{CD} & p_{DC}\left(1-q_{CD}\right) & \left(1-p_{DC}\right) q_{CD} & \left(1-p_{DC}\right)\left(1-q_{CD}\right) \\
p_{DD}q_{DD} & p_{DD}\left(1-q_{DD}\right) & \left(1-p_{DD}\right) q_{DD} & \left(1-p_{DD}\right)\left(1-q_{DD}\right)
\end{pmatrix} .
\end{align}
For $\lambda <1$, these payoffs, $\pi_{X}\left(\mathbf{p},\mathbf{q}\right)$ and $\pi_{Y}\left(\mathbf{p},\mathbf{q}\right)$, are well-defined for all $\mathbf{p},\mathbf{q}\in\left[0,1\right]^{5}$, and they fall within the {\it feasible region} of the game, which is defined as the convex hull of the players' one-shot payoffs, namely $\left(R,R\right)$, $\left(S,T\right)$, $\left(T,S\right)$, and $\left(P,P\right)$. In the following, it will be useful to have an alternative expression for the repeated game payoffs. This alternative expression is somewhat more explicit, as it does not require computing the inverse of a matrix. The following result is by Mamiya and Ichinose \citep{Mamiya:PRE:2020}, who themselves use the framework by Press and Dyson \citep{press:PNAS:2012}.

\begin{lemma}[An alternative representation of payoffs] \label{Lem:AlternativePayoff}
For $\mathbf{u}=\left(u_{1},u_{2},u_{3},u_{4}\right)\in\mathbb{R}^{4}$ and $\mathbf{p},\mathbf{q}\in\left[0,1\right]^{5}$, let $A\left(\mathbf{p},\mathbf{q},\mathbf{u}\right)$ denote the matrix
\begin{align} \label{eq:PressDysonMatrix}
\begin{pmatrix}
\left(1-\lambda\right) p_0q_0 + \lambda p_{CC}q_{CC} - 1  &\left(1-\lambda\right) p_0 + \lambda p_{CC}-1  &\left(1-\lambda\right) q_0 + \lambda q_{CC}-1    &u_{1}\\
\left(1-\lambda\right) p_0q_0 + \lambda p_{CD}q_{DC}   &\left(1-\lambda\right) p_0 + \lambda p_{CD} - 1  &\left(1-\lambda\right) q_0 + \lambda q_{DC}    &u_{2}\\
\left(1-\lambda\right) p_0q_0 + \lambda p_{DC}q_{CD}   &\left(1-\lambda\right) p_0 + \lambda p_{DC}  &\left(1-\lambda\right) q_0 + \lambda q_{CD}- 1     &u_{3}\\
\left(1-\lambda\right) p_0q_0 + \lambda p_{DD}q_{DD}   &\left(1-\lambda\right) p_0 + \lambda p_{DD}  &\left(1-\lambda\right) q_0 + \lambda q_{DD}     &u_{4}\\
\end{pmatrix} .
\end{align}
The payoffs to players $X$ and $Y$, according to \eq{discountedpayoff1}, can then be written as
\begin{subequations}\label{eq:discountedpayoff2}
\begin{align}
\pi_{X}\left(\mathbf{p},\mathbf{q}\right) &= \frac{\det A\left(\mathbf{p},\mathbf{q},\mathbf{u}_{X}\right)}{\det A\left(\mathbf{p},\mathbf{q},\mathbf{1}\right)} ; \\
\pi_{Y}\left(\mathbf{p},\mathbf{q}\right) &= \frac{\det A\left(\mathbf{p},\mathbf{q},\mathbf{u}_{Y}\right)}{\det A\left(\mathbf{p},\mathbf{q},\mathbf{1}\right)} ,
\end{align}
\end{subequations}
where $\mathbf{1}\in\mathbb{R}^{4}$ is the vector of ones.
\end{lemma}

\noindent
As shown by \citet{mcavoy:PRSA:2019} for games without discounting, a certain monotonicity property holds if only one component of player $X$'s strategy is varied at a time. To explore local strategy revisions, here we first generalize their result to discounted games. To this end, it will be useful to introduce the basis vectors
\begin{subequations}
\begin{align}
\mathbf{e}_{0} &= \left(1,0,0,0,0\right) ; \\ 
\mathbf{e}_{CC} &= \left(0,1,0,0,0\right) ; \\ 
\mathbf{e}_{CD} &= \left(0,0,1,0,0\right) ; \\ 
\mathbf{e}_{DC} &= \left(0,0,0,1,0\right) ; \\ 
\mathbf{e}_{DD} &= \left(0,0,0,0,1\right) .
\end{align}	
\end{subequations}
These vectors allow us to introduce the notion of marginal functions, which track how the payoffs to $X$ and $Y$ change as $X$ varies just one parameter of his or her strategy, $\mathbf{p}$. If $X$ varies only $p_{0}$, then we have the marginal functions $\varphi_{0}\left(z\right)\coloneqq\pi_{X}\left(\mathbf{p}+z\mathbf{e}_{0},\mathbf{q}\right)$ for $X$ and $\psi_{0}\left(z\right)\coloneqq\pi_{Y}\left(\mathbf{p}+z\mathbf{e}_{0},\mathbf{q}\right)$ for $Y$. If $X$ varies only $p_{xy}$ for $x,y\in\left\{C,D\right\}$, then we have the marginal functions $\varphi_{xy}\left(z\right)\coloneqq\pi_{X}\left(\mathbf{p}+z\mathbf{e}_{xy},\mathbf{q}\right)$ for~$X$ and $\psi_{xy}\left(z\right)\coloneqq\pi_{Y}\left(\mathbf{p}+z\mathbf{e}_{xy},\mathbf{q}\right)$ for~$Y$. We note that $\varphi_0$ and $\psi_0$ are well-defined for all $z\in I_0\coloneqq\left[-p_0,1-p_0\right]$. Similarly, for $x,y\in\left\{C,D\right\}$, both $\varphi_{xy}$ and $\psi_{xy}$ are well-defined for all $z\in I_{xy}\coloneqq\left[-p_{xy},1-p_{xy}\right]$.

\begin{lemma}\label{lem:line}
Consider a game with arbitrary one-shot payoffs $\mathbf{u}_X$ and $\mathbf{u}_Y, respectively$. 
Then, for $x,y\in\left\{C,D\right\}$, the sets $\left\{\big(\psi_{0}\left(z\right) ,\varphi_{0}\left(z\right)\big)\right\}_{z\in I_0}$ and $\left\{\left(\psi_{xy}\left(z\right) ,\varphi_{xy}\left(z\right)\right)\right\}_{z\in I_{xy}}$ are line segments within the feasible region. Furthermore, the functions $\varphi_{0}\left(z\right)$, $\psi_{0}\left(z\right)$, $\varphi_{xy}\left(z\right)$, and $\psi_{xy}\left(z\right)$ vary monotonically in $z$. In particular, as $z$ increases, movement is possible in at most one direction along these line segments.
\end{lemma}
\begin{proof}
The fact that $\left\{\left(\psi_{0}\left(z\right) ,\varphi_{0}\left(z\right)\right)\right\}_{z\in I_0}$ forms a line segment in the feasible region is immediate from \eqs{discountedpayoff1}{initial_distribution} since $p_{0}$ (and therefore $z$) enters only into the initial distribution over the states. Because the respective functions are linear in $z$, the monotonicity property also follows immediately. 

For $\left\{\left(\psi_{xy}\left(z\right) ,\varphi_{xy}\left(z\right)\right)\right\}_{z\in I_{xy}}$, we note that by \eq{discountedpayoff2} (and the fact that the denominators for $\pi_{X}$ and $\pi_{Y}$ are identical), there exist $\alpha_{1},\alpha_{2},\beta_{1},\beta_{2},\gamma_{1},\gamma_{2}$, all independent of $z$ (but depending on $x,y\in\left\{C,D\right\}$), such that
\begin{subequations}
\begin{align}
\varphi_{xy}\left(z\right) &= \frac{\alpha_{1}z+\alpha_{2}}{\beta_{1}z+\beta_{2}} ; \\
\psi_{xy}\left(z\right) &= \frac{\gamma_{1}z+\gamma_{2}}{\beta_{1}z+\beta_{2}} .
\end{align}
\end{subequations}
A straightforward calculation then gives
\begin{align}
\left(\beta_{1}\gamma_{2}-\beta_{2}\gamma_{1}\right)\varphi_{xy}\left(z\right) &= \left(\alpha_{2}\beta_{1}-\alpha_{1}\beta_{2}\right)\psi_{xy}\left(z\right) + \alpha_{1}\gamma_{2}-\alpha_{2}\gamma_{1} ,
\end{align}
which establishes that $\left\{\left(\psi_{xy}\left(z\right) ,\varphi_{xy}\left(z\right)\right)\right\}_{z\in I_{xy}}$ is also a line segment. Finally, to see that one cannot move in two directions along these segments as $z$ increases, we note that $\varphi_{xy}'\left(z\right)$ and $\psi_{xy}'\left(z\right)$ have the same signs as $\alpha_{1}\beta_{2}-\alpha_{2}\beta_{1}$ and $\gamma_{1}\beta_{2}-\gamma_{2}\beta_{1}$, respectively, which are independent of $z$.
\end{proof}

\subsection{Local strategy revisions}
In the following, we ask under which conditions a player is able to locally find profitable deviations from its current strategy (where ``profitable'' needs to be interpreted in terms of the player's objectives). To this end, consider an arbitrary but fixed base game with payoff vectors $\mathbf{u}_X$ and $\mathbf{u}_Y$. As usual, we denote the player's current strategies by $\mathbf{p}$ (for player $X$) and $\mathbf{q}$ (for player $Y$). Without loss of generality, we assume that it is player $X$ who considers revising his or her strategy.

\subsubsection{Local improvements}

\begin{definition}[Local improvements]
We say that player $X$ can locally improve its objective function $V$ if for all distances $d>0$ there is a strategy $\mathbf{p'}$ with $\left\|\mathbf{p}' - \mathbf{p}\right\| <d$ such that $V\left(\mathbf{p'},\mathbf{q}\right) >V\left(\mathbf{p},\mathbf{q}\right)$.
\end{definition}
In this definition, it does not matter which exact norm $\left\|\cdot\right\|$ we use. However, to be consistent with the simulations, we will usually interpret it as the uniform norm. As a consequence, the relation $\left\|\mathbf{p}' - \mathbf{p}\right\| <d$ holds if and only if the difference in each entry is at most $d$ in absolute value.

Trivially, local improvements are only possible if the current value of $V\left(\mathbf{p},\mathbf{q}\right)$ is not optimal already. To formalize this observation, we define player $X$'s {\it maximum attainable value with respect to objective} $V$ as
\begin{equation}
V^{\ast}\left(\mathbf{q}\right) = \max_{\mathbf{p}\in\left[0,1\right]^{5}} V\left(\mathbf{p},\mathbf{q}\right) .
\end{equation}
Similarly, we say $\mathbf{p^{\ast}}$ is a \emph{best response with respect to objective} $V$ if it is an element of the set
\begin{equation}
\textrm{BR}_{V}\left(\mathbf{q}\right) = \left\{ \mathbf{p}\in\left[0,1\right]^{5}\ \mid\ V\left(\mathbf{p},\mathbf{q}\right) = V^{\ast}\left(\mathbf{q}\right)~\right\} .
\end{equation}
We note that for $\mathbf{p}\not\in\textrm{BR}_{V}\left(\mathbf{q}\right)$, the task of improving player $X$'s objective function would be simple if global mutations were considered. In this case the player could always switch to some best response $\mathbf{p^{\ast}}\in\textrm{BR}_{V}\left(\mathbf{q}\right)$ (or by continuity, to any other strategy that is sufficiently close to~$\mathbf{p^{\ast}}$). 

To identify a local improvement, a natural approach may then be to study interpolations of the form $\mathbf{p'}\left(z\right) =\left(1-z\right)\mathbf{p}+z\mathbf{p^{\ast}}$ for $z\in\left[0,1\right]$. These interpolations have two useful properties: \emph{(i)}~they approach a best response for $z \rightarrow 1$ and \emph{(ii)}~they are arbitrarily close to the player's current strategy $\mathbf{p}$ for $z\rightarrow 0$. For such an interpolation approach to work, however, one would need to prove that $V\left(\mathbf{p'}\left(z\right),\mathbf{q}\right)$ is a monotonic function of $z$. Unfortunately, this turns out to be not quite how $V\left(\mathbf{p'}\left(z\right),\mathbf{q}\right)$ behaves in general: even if player $X$ can improve its objective function $V$ by taking a large step into the direction of $\mathbf{p^{\ast}}$, the value of $V$ may drop when steps are too small.

However, as shown in {\bf Lemma~\ref{lem:line}}, such a monotonicity property may hold if only one component of player $X$'s strategy is varied at a time (i.e. if we consider the respective marginal functions). To do so, we first consider a special class of objectives.

\subsubsection{Weighted payoff objectives}
In the following, we restrict ourselves to objectives that are related to weighted sums of the players' payoffs. For given payoff weights $\pmb{\kappa} = \left(\kappa_X, \kappa_Y\right)$, we define the associated objective by 
\begin{equation}
V_{\pmb{\kappa}}\left(\mathbf{p},\mathbf{q}\right) = \kappa_X\pi_X\left(\mathbf{p},\mathbf{q}\right) + \kappa_Y\pi_Y\left(\mathbf{p},\mathbf{q}\right) . 
\end{equation}
As special cases, we recover the objective of a selfish learner by setting $\pmb{\kappa}=\left(1,0\right)$ and the objective of an efficiency-minded player by setting $\pmb{\kappa}=\left(1,1\right)$. To capture the objective function of a fairness-minded player, we consider $-\big|V_{\pmb{\kappa}}\left(\mathbf{p},\mathbf{q}\right)\big|$, where $\pmb{\kappa}=\left(1,-1\right)$. 

Because $V_{\pmb{\kappa}}$ is linear in the players' payoffs, and because of the basic properties of scalar products, we can use \eq{discountedpayoff1} to derive more explicit expression for these objective functions,
\begin{align}\label{eq:ObjectiveKappa1}
V_{\pmb{\kappa}}\left(\mathbf{p},\mathbf{q}\right) &= \kappa_X \left< \mathbf{v}, \mathbf{u}_{X} \right> + \kappa_Y\left< \mathbf{v}, \mathbf{u}_{Y} \right> \nonumber \\
&= \left< \mathbf{v}, \kappa_X \mathbf{u}_{X} +\kappa_Y\mathbf{u}_Y\right> ,
\end{align}
where $\mathbf{v}$ is as defined by \eq{invariant_distribution}. Equivalently, due to \eq{discountedpayoff2} and the fact that the determinant is a linear function of each column of a matrix, we can write
\begin{align} \label{eq:ObjectiveKappa2}
V_{\pmb{\kappa}}\left(\mathbf{p},\mathbf{q}\right) &= \kappa_X\frac{\det A\left(\mathbf{p},\mathbf{q},\mathbf{u}_{X}\right)}{\det A\left(\mathbf{p},\mathbf{q},\mathbf{1}\right)} + \kappa_Y\frac{\det A\left(\mathbf{p},\mathbf{q},\mathbf{u}_{Y}\right)}{\det A\left(\mathbf{p},\mathbf{q},\mathbf{1}\right)} \nonumber \\
&= \frac{\det A\left(\mathbf{p},\mathbf{q},\kappa_X\mathbf{u}_{X}+\kappa_Y\mathbf{u}_Y\right)}{\det A\left(\mathbf{p},\mathbf{q},\mathbf{1}\right)} .
\end{align}
where $A\left(\mathbf{p},\mathbf{q},\mathbf{u}\right)$ is the matrix defined by \eq{PressDysonMatrix}.

\begin{remark} \label{remark:equivalence}
By comparing \eq{ObjectiveKappa1} and \eq{ObjectiveKappa2} for player $X$'s objective function with \eq{discountedpayoff1} and \eq{discountedpayoff2} for the player's payoff, we identify a useful equivalence: a player who wishes to maximize $V_{\pmb{\kappa}}$ given one-shot payoffs $\mathbf{u}_X$ is equivalent to a player who wishes to maximize $\pi_X$ for a game with modified one-shot payoffs $\mathbf{u}_{\pmb{\kappa}}\coloneqq\kappa_X \mathbf{u}_X + \kappa_Y \mathbf{u}_Y$.
\end{remark}

\subsubsection{Marginal functions}
\begin{definition}[Marginal functions with respect to a player's objective $V$]
Consider two players with strategies $\mathbf{p}$ and $\mathbf{q}$, and suppose player $X$'s objective function is~$V$. Let $I_{0}\coloneqq\left[-p_{0},1-p_{0}\right]$ and, for $x,y\in\left\{C,D\right\}$, let $I_{xy}\coloneqq\left[-p_{xy},1-p_{xy}\right]$.
\begin{enumerate}
\item The marginal function $F_0:I_{0}\rightarrow \mathbb{R}$ with respect to player $X$'s initial cooperation probability is defined as $F_0\left(z\right) \coloneqq V\left(\mathbf{p}+z\mathbf{e}_0,\mathbf{q}\right)$. 
\item Similarly, for each action profile ($x,y$) with $x,y\in\{C,D\}$ we define a respective marginal function $F_{xy}:I_{xy}\rightarrow \mathbb{R}$ by $F_{xy}\left(z\right) \coloneqq V\left(\mathbf{p}+z\mathbf{e}_{xy},\mathbf{q}\right)$.
\end{enumerate}
\end{definition}

Marginal functions reflect how payoff differences change as player $X$ varies a single entry of his or her strategy. By definition $F_0\left(0\right) =F_{xy}\left(0\right) =V\left(\mathbf{p},\mathbf{q}\right)$ corresponds to the status quo. In the special case of weighted payoff objectives, we indeed observe that the associated marginal functions satisfy a monotonicity property:
\begin{lemma}[Monotonicity of marginal functions with respect to weighted payoff objectives] \label{Lemma:MonotonicityMarginal}
For given payoffs weights $\pmb{\kappa}=\left(\kappa_X,\kappa_Y\right)$, each marginal function with respect to the weighted payoff objective $V_{\pmb{\kappa}}$ is either strictly monotonically increasing, strictly monotonically decreasing, or constant on its entire domain.
\end{lemma}
\begin{proof}
The result follows from {\bf Remark~\ref{remark:equivalence}}, by applying {\bf \lem{line}} to the game in which player $X$'s one-shot payoffs are given by $\mathbf{u}_{\pmb{\kappa}}=\kappa_X \mathbf{u}_X + \kappa_Y \mathbf{u}_Y$.
\end{proof}

The following two Lemmas make use of the above monotonicity property. They show that quite stringent conditions need to hold for player $X$ not to be able to locally improve a weighted payoff objective function:
\begin{lemma} \label{Lemma:MonotonicityMarginal2}
Suppose player $X$'s current objective function is $V_{\pmb{\kappa}}$ for some given $\pmb{\kappa}=\left(\kappa_X,\kappa_Y\right)$, and consider players with strategies $\mathbf{p}$ and $\mathbf{q}$ such that $\mathbf{p}\not\in\textrm{BR}_{V_{\pmb \kappa}}\left(\mathbf{q}\right)$. Moreover, suppose player $X$ is unable to locally improve its objective function.
\begin{enumerate}
\item If $p_0=0$, then $F_0\left(z\right)$ is either constant or monotonically decreasing in $z$. 
If $p_0=1$, then $F_0\left(z\right)$ is either constant or monotonically increasing. If~~$0<p_0<1$, then $F_0\left(z\right)$ is constant on its entire domain.
\item Analogously, if $p_{xy}=0$ for $x,y\in\left\{C,D\right\}$, then $F_{xy}\left(z\right)$ is either constant or monotonically decreasing in~$z$. If $p_{xy}=1$, then $F_{xy}\left(z\right)$ is either constant or monotonically increasing. If $0<p_{xy}<1$, then $F_{xy}\left(z\right)$ is constant on its entire domain.
\end{enumerate}
\end{lemma}
\begin{proof}
Suppose $F_0\left(z\right)$ is not constant. Because of {\bf Lemma~\ref{Lemma:MonotonicityMarginal}}, $F_0\left(z\right)$ then either needs to be strictly monotonically increasing or strictly monotonically decreasing. If it is increasing, it follows for any $z\in\left(0,1-p_0\right]$ that $V_{\pmb \kappa}\left(\mathbf{p}+z\mathbf{e}_0,\mathbf{q}\right) > V_{\pmb \kappa}\left(\mathbf{p},\mathbf{q}\right)$. For $p_0 < 1$, this yields a contradiction because $X$ was assumed to be unable to locally improve its objective function. A similar contradiction arises if $F_0\left(z\right)$ is strictly decreasing and $p_0>0$. In that case, $V_{\pmb \kappa}\left(\mathbf{p}+z\mathbf{e}_0,\mathbf{q}\right) > V_{\pmb \kappa}\left(\mathbf{p},\mathbf{q}\right)$ for all $z\in\left[-p_0,0\right)$. An analogous proof shows the result for all other marginal functions $F_{xy}\left(z\right)$.
\end{proof}

\begin{lemma} \label{Lemma:LocalInvariance}
Suppose player $X$'s current objective function is $V_{\pmb{\kappa}}$ for some given $\pmb{\kappa}=\left(\kappa_X,\kappa_Y\right)$, and consider players with strategies $\mathbf{p}\in\left(0,1\right)^5$ and $\mathbf{q}\in\left[0,1\right]^5$ such that $\mathbf{p}\not \in\textrm{BR}_{V_{\pmb \kappa}}\left(\mathbf{q}\right)$. Moreover, suppose player $X$ is unable to locally improve its objective function. Then, there is a $d>0$ such that $V_{\pmb{\kappa}}\left(\mathbf{p'},\mathbf{q}\right) =V_{\pmb{\kappa}}\left(\mathbf{p},\mathbf{q}\right)$ for all $\mathbf{p'}$ with $\left\|\mathbf{p'}-\mathbf{p}\right\|<d$.
\end{lemma}
\begin{proof}
Because player $X$ is unable to increase fairness and $\mathbf{p}\in\left(0,1\right)^5$, we can find a $d>0$ such that all $\mathbf{p}'$ with $\left\|\mathbf{p'}-\mathbf{p}\right\| <d$ satisfy $\mathbf{p'}\in\left(0,1\right)^5$ and $V_{\pmb{\kappa}}\left(\mathbf{p'},\mathbf{q}\right) \leqslant V_{\pmb{\kappa}}(\mathbf{p},\mathbf{q})$. Now assume that there is at least one such $\mathbf{p'}$ for which the inequality is strict, $V_{\pmb{\kappa}}\left(\mathbf{p'},\mathbf{q}\right) < V_{\pmb{\kappa}}\left(\mathbf{p},\mathbf{q}\right)$. Consider the sequence of strategies, 
\begin{subequations}
\begin{align}
\mathbf{p}^{0} &\coloneqq \left(p_0,p_{CC},p_{CD},p_{DC},p_{DD}\right) ; \\
\mathbf{p}^{1} &\coloneqq \left(p_0',p_{CC},p_{CD},p_{DC},p_{DD}\right) ; \\
\mathbf{p}^{2} &\coloneqq \left(p_0',p_{CC}',p_{CD},p_{DC},p_{DD}\right) ; \\
\mathbf{p}^{3} &\coloneqq \left(p_0',p_{CC}',p_{CD}',p_{DC},p_{DD}\right) ; \\
\mathbf{p}^{4} &\coloneqq \left(p_0',p_{CC}',p_{CD}',p_{DC}',p_{DD}\right) ; \\
\mathbf{p}^{5} &\coloneqq \left(p_0',p_{CC}',p_{CD}',p_{DC}',p_{DD}'\right) .
\end{align}
\end{subequations}
In particular, $\mathbf{p}^0=\mathbf{p}$,~$\mathbf{p}^5=\mathbf{p'}$, and in each step we only vary one coordinate. 
Moreover, $\left\|\mathbf{p}^{i}-\mathbf{p}\right\|\le\left\|\mathbf{p}'-\mathbf{p}\right\| <d$ for all $i$, and therefore $V_{\pmb{\kappa}}\left(\mathbf{p}^{i},\mathbf{q}\right) \le V_{\pmb{\kappa}}\left(\mathbf{p},\mathbf{q}\right)$ for all $i$. Let $j$ be the largest number such that $V_{\pmb{\kappa}}\left(\mathbf{p}^{j},\mathbf{q}\right) = V_{\pmb{\kappa}}\left(\mathbf{p},\mathbf{q}\right)$. By construction, we must have $j<5$. Since $\mathbf{p}^j\in\left(0,1\right)^5$ and player $X$ also needs to be unable to locally increase fairness around the new point $\left(\mathbf{p}^j,\mathbf{q}\right)$, it follows from {\bf Lemma~\ref{Lemma:MonotonicityMarginal2}} that any marginal function centered at the new point $\left(\mathbf{p}^j,\mathbf{q}\right)$ has to be constant. Since $\mathbf{p}^{j+1}$ and $\mathbf{p}^j$ only differ in one coordinate, it follows that $V_{\pmb{\kappa}}\left(\mathbf{p}^{j+1},\mathbf{q}\right) = V_{\pmb{\kappa}}\left(\mathbf{p}^j,\mathbf{q}\right) = V_{\pmb{\kappa}}\left(\mathbf{p},\mathbf{q}\right)$, which contradicts the definition of $j$ as the largest number for which equality holds.
\end{proof}

\subsubsection{Local improvements in the interior of the strategy space}
We can now show that, for generic strategies in the interior of the state space, player~$X$ can always locally increase her objective function $V_{\pmb{\kappa}}$ (unless $X$ uses a best response already). Note that the following result does not make any assumptions on the exact payoffs of the game:
\begin{proposition}[Existence of local improvements in the interior of the strategy space] \label{Prop:LocalImprovementInterior}
Consider a player $X$ with weighted payoff objectives $V_{\pmb{\kappa}}$ for some given $\pmb{\kappa}=\left(\kappa_X,\kappa_Y\right)$, and suppose the players' strategies are $\mathbf{p}\in\left(0,1\right)^5$ and $\mathbf{q}\in\left[0,1\right]^5$ such that $\mathbf{p}\not \in\textrm{BR}_{V_{\pmb \kappa}}\left(\mathbf{q}\right)$. Then player $X$ can locally improve his or her objective function.
\end{proposition}
\begin{proof}
Given the co-player's strategy, let $\mathbf{p^{\ast}}\in \textrm{BR}_{V_{\pmb \kappa}}\left(\mathbf{q}\right)$ be an arbitrary best response. Then, $V_{\pmb \kappa}\left(\mathbf{p},\mathbf{q}\right)<V_{\pmb \kappa}\left(\mathbf{p^{\ast}},\mathbf{q}\right)$ since $\mathbf{p}\not \in\textrm{BR}_{V_{\pmb \kappa}}\left(\mathbf{q}\right)$. We now define a function $h:\left[0,1\right]\rightarrow \mathbb{R}$ by
\begin{equation} \label{eq:h_function}
h\left(z\right) \coloneqq V_{\pmb{\kappa}}\big(\left(1-z\right)\mathbf{p}+z\mathbf{p^{\ast}},\mathbf{q}\big)-V_{\pmb{\kappa}}\left(\mathbf{p},\mathbf{q}\right) .
\end{equation}
By \eq{ObjectiveKappa2}, it follows that $h\left(z\right)$ is a rational function of $z$. Suppose to the contrary that player $X$ cannot locally improve $V_{\pmb \kappa}$. By {\bf Lemma~\ref{Lemma:LocalInvariance}}, $h\left(z\right)$ must be identically zero for all $z$ sufficiently small. But a rational function has either finitely many zeros or is identically zero, so $h\left(z\right) =0$ for all $z\in\left[0,1\right]$. However, by \eq{h_function}, $h\left(1\right) =V_{\pmb \kappa}\big(\mathbf{p^{\ast}},\mathbf{q}\big) -V_{\pmb \kappa}\left(\mathbf{p},\mathbf{q}\right) >0$.
\end{proof}
As a result, if both players adopt strategies in the interior of the strategy space, the restriction that players need to adapt locally does not introduce any new equilibria. If both players are unable to locally improve their objective function, then their respective strategies need to be (global) best replies to each other (relative to their respective objective functions).

Next, we apply {\bf{Proposition~\ref{Prop:LocalImprovementInterior}}} to discuss the special cases that player $X$ has objectives based on selfishness, efficiency, or fairness:
\begin{proposition}[Existence of local improvements for some important objective functions]
Consider a player $X$ with objective function $V$ in a symmetric base game with continuation probability $\lambda >0$ and payoffs $\mathbf{u}_X$, $\mathbf{u}_Y$, such that $R\ge P$. Moreover, suppose the player's strategies are $\mathbf{p}\in\left(0,1\right)^5$ and $\mathbf{q}\in\left[0,1\right]^5$, and that player $X$ cannot locally improve its objective function. 
\begin{enumerate}
\item If $X$ has selfish objectives, meaning $V\left(\mathbf{p},\mathbf{q}\right) = \pi_X\left(\mathbf{p},\mathbf{q}\right)$, then $\mathbf{q}$ must be an equalizer strategy. That is, for $\lambda <1$ the entries of $\mathbf{q}$ need to satisfy the equations \citep{hilbe:GEB:2015,Ichinose:JTB:2018}
\begin{subequations} \label{eq:Equalizers}
\begin{align}
\lambda\left(R-P\right) q_{CD} &= \lambda\left(T-P\right) q_{CC}-\left(T-R\right)\left(1+\lambda q_{DD}\right) ; \\
\lambda\left(R-P\right) q_{DC} &= \left(P-S\right)\left(1-\lambda q_{CC}\right) +\left(R-S\right)\lambda q_{DD} .
\end{align}
\end{subequations}
For $\lambda =1$, the strategy $\mathbf{q}$ additionally needs to be different from the degenerate strategy \emph{Repeat} defined by $q_{CC}=q_{CD}=1$ and $q_{DC}=q_{DD}=0$.
\item If $X$ has efficiency objectives, i.e. $V\left(\mathbf{p},\mathbf{q}\right) = \pi_X\left(\mathbf{p},\mathbf{q}\right) +\pi_Y\left(\mathbf{p},\mathbf{q}\right)$, then the game needs to be constant-sum ($2R=2P=S+T$). In that case, any $\mathbf{p}$ and $\mathbf{q}$ satisfy the assumptions.
\item If $X$ has fairness objectives, i.e. $V\left(\mathbf{p},\mathbf{q}\right)  =  -\big| \pi_X\left(\mathbf{p},\mathbf{q}\right) - \pi_Y\left(\mathbf{p},\mathbf{q}\right)\big|$, then the payoffs need to be equal ($\pi_X\left(\mathbf{p},\mathbf{q}\right) =\pi_Y\left(\mathbf{p},\mathbf{q}\right)$).
\end{enumerate}
\end{proposition}
\begin{proof}
\begin{enumerate}
\item Because player $X$ is unable to locally improve its objective function, it follows by an argument similar to the proof of {\bf Proposition~\ref{Prop:LocalImprovementInterior}} that $V\left(\mathbf{p},\mathbf{q}\right) =V\left(\mathbf{p^*},\mathbf{q}\right)$ for all alternative strategies $\mathbf{p^*}$. In other words, player $X$'s payoff is independent of its strategy. The strategies $\mathbf{q}$ that have this property have been called {\it equalizers} \citep{boerlijst:AMM:1997}. For discounted games, the equalizers are exactly the strategies satisfying \eq{Equalizers} \citep{hilbe:GEB:2015,Ichinose:JTB:2018}. For games without discounting, it is well-known that \emph{Repeat} needs to be excluded explicitly \citep{press:PNAS:2012}.
\item By {\bf Remark~\ref{remark:equivalence}}, players with efficiency objectives can interpreted as players with selfish objective in a game with modified one-shot payoffs $\mathbf{u}_E\coloneqq\mathbf{u}_X+\mathbf{u}_Y=\left(2R,S+T,S+T,2P\right)$. We can thus use \eq{Equalizers} to obtain a necessary condition on player $Y$'s strategy: 
\begin{subequations} \label{eq:EqualizersEfficiency}
\begin{align}
2\lambda \left(R-P\right) q_{CD} &= \lambda \left(S+T-2P\right) q_{CC}-\left(S+T-2R\right)\left(1+\lambda q_{DD}\right) ; \\
2\lambda \left(R-P\right) q_{DC} &= \left(2P-T-S\right)\left(1-\lambda q_{CC}\right) +\left(2R-T-S\right)\lambda q_{DD} .
\end{align}
\end{subequations}
In the special case that $R=P$, the sum of these two equations simplifies to
$0 = 2R-T-S$,
and therefore $2R=2P=S+T$. 
In the other case, $R>P$, taking the difference between these two equations gives $\lambda\left(q_{CD}-q_{DC}\right) =1$. This equation cannot be satisfied when $\lambda <1$, and when $\lambda =1$ it requires $q_{CD}=1$ and $q_{DC}=0$. Plugging these values into \eq{EqualizersEfficiency}, we find that $q_{CC}=1$ and $q_{DD}=0$, so $\mathbf{q}$ must be the strategy \emph{Repeat}, which is explicitly excluded.
\item Assume to the contrary that payoffs are unequal. Without loss of generality, we may assume that $\pi_X\left(\mathbf{p},\mathbf{q}\right) < \pi_Y\left(\mathbf{p},\mathbf{q}\right)$. It follows that locally, player $X$'s objective function has the form $V_{\pmb\kappa}$ with ${\pmb\kappa}=\left(1,-1\right)$. Since $V_{\pmb \kappa}\left(\mathbf{p},\mathbf{q}\right)  <  0 = V_{\pmb \kappa}\left(\mathbf{q},\mathbf{q}\right)$, we have $\mathbf{p}\not\in \textrm{BR}_{V_{\pmb \kappa}}\left(\mathbf{q}\right)$. By {\bf Proposition~\ref{Prop:LocalImprovementInterior}}, it then follows that player $X$ can locally improve its objective function. That is, for any distance $d>0$, one can find a strategy $\mathbf{p}_d$ with $\left\| \mathbf{p}_d - \mathbf{p} \right\| <d$ such that $V_{\pmb \kappa}\left(\mathbf{p}_d,\mathbf{q}\right) > V_{\pmb \kappa}\left(\mathbf{p},\mathbf{q}\right)$. Moreover, because $V_{\pmb \kappa}$ is a continuous function of its inputs, $V_{\pmb \kappa}\left(\mathbf{p}_d,\mathbf{q}\right) <0$ for all $d$ that are sufficiently small. For any such $d$, we have $-\left|V_{\pmb \kappa}\left(\mathbf{p},\mathbf{q}\right)\right| < -\left|V_{\pmb \kappa}\left(\mathbf{p}_d,\mathbf{q}\right)\right|$. Thus, local improvements are possible for all $d$, a contradiction. \qedhere
\end{enumerate}
\end{proof}

\subsubsection{Local improvements on the boundary of the strategy space}
In the following, we briefly discuss under which condition player $X$ is guaranteed to find a local improvement when its current strategy is on the boundary, $\mathbf{p} \in \textrm{bd}\left(\left[0,1\right]^5\right)$. 

Obviously, any local improvement is impossible when both players already use a best response to each other (given their objectives). However, even in the standard case of two players with selfish objectives, very little is known about the corresponding set of all Nash equilibria among the memory-one strategies. Most available results address games without discounting, $\lambda =1$. For such games, Akin \citep{akin:Games:2015} described all Nash equilibria profiles $\left(\mathbf{p},\mathbf{q}\right)$ that sustain full cooperation in the prisoner's dilemma. Stewart and Plotkin \citep{stewart:pnas:2014} describe all symmetric Nash equilibria $\left(\mathbf{p},\mathbf{p}\right)$ among the memory-one strategies for all $2\times 2$ games. For generic cases (with $R\neq P$ and $T\neq S$) they find that in addition to the equalizer strategies of the form~\eq{Equalizers}, there are three more classes of equilibrium strategies $\mathbf{p}$:
\begin{enumerate}
\item {\bf Strategies that result in mutual cooperation} are a Nash equilibrium if and only if 
\begin{subequations}
\begin{align}
p_{CC} &= 1 ; \\
\left(T-R\right) p_{DC} &\leqslant \left(R-S\right)\left(1-p_{CD}\right) ; \\
\left(T-R\right) p_{DD} &\leqslant \left(R-P\right)\left(1-p_{CD}\right) .
\end{align}
\end{subequations}
\item {\bf Strategies that result in alternating cooperation} are a Nash equilibrium if and only if 
\begin{subequations}
\begin{align}
p_{CD} &= 0 ; \\
p_{DC} &= 1 ; \\
\left(T-S\right) p_{CC} &\leqslant 2\left(T-R\right) ; \\
\left(T-S\right) p_{DD} &\leqslant \left(S+T-2P\right) .
\end{align}
\end{subequations}
\item {\bf Strategies that result in mutual defection} are a Nash equilibrium if and only if 
\begin{subequations}
\begin{align}
p_{DD} &= 0 ; \\
\left(R-P\right) p_{DC} &\leqslant \left(P-S\right)\left(1-p_{CC}\right) ; \\
\left(T-P\right) p_{DC} &\leqslant \left(P-S\right)\left(1-p_{CD}\right) . 
\end{align}
\end{subequations}
\end{enumerate}

In the following, we ask whether these three qualitative outcomes are still feasible if (at least) one of the players is motivated by efficiency:
\begin{proposition}
Consider a player $X$ whose objective is to maximize efficiency, $V\left(\mathbf{p},\mathbf{q}\right) =\pi_X\left(\mathbf{p},\mathbf{q}\right) +\pi_Y\left(\mathbf{p},\mathbf{q}\right)$ in a symmetric base game with $R>P$ and continuation probability $\lambda >0$. 
\begin{enumerate}
\item Suppose that $R \geqslant (S+T)/2$ and that both players mutually cooperate, giving $\pi_{X}=\pi_Y=R$. Then, player $X$ is unable to locally (or globally) improve its objective function. 
\item Suppose $R < (S+T)/2$, and both players cooperate in an alternating fashion, giving $\pi_{X}=\pi_Y=(S+T)/2$. Then, player $X$ is unable to locally (or globally) improve its objective function. 
\item Suppose both players mutually defect, giving $\pi_{X}=\pi_Y=P$. Then there are strategies~$\mathbf{q}$ for player~$Y$ such that player $X$ is unable to locally (or globally) improve its objective function if and only if $\left(S+T\right) /2  \leqslant P$.
\end{enumerate}
\end{proposition}
\begin{proof}
\begin{enumerate}
\item By assumption, the current efficiency is $\pi_X\left(\mathbf{p},\mathbf{q}\right)+\pi_Y\left(\mathbf{p},\mathbf{q}\right)=2R$. After any deviation of player $X$ to some other strategy $\mathbf{p^{\ast}}$, this efficiency takes the form 
\begin{equation}
\pi_X\left(\mathbf{p^{\ast}},\mathbf{q}\right)+\pi_Y\left(\mathbf{p^{\ast}},\mathbf{q}\right) = a_1\left(2R\right) + a_2\left(S+T\right) + a_3\left(2P\right) , \label{eq:efficiency_deviation}
\end{equation}
where $a_{1},a_{2},a_{3}\geqslant 0$ and $a_{1}+a_{2}+a_{3}=1$. Because $R \geqslant \left(S+T\right) /2$ and $R>P$, this efficiency is at most $2R$.
\item By assumption, the current efficiency is $\pi_X\left(\mathbf{p},\mathbf{q}\right) +\pi_Y\left(\mathbf{p},\mathbf{q}\right) =S+T$. After a deviation, the efficiency again takes the form of \eq{efficiency_deviation}. Because now $\left(S+T\right) /2 \ge R \ge P$, again it follows that this efficiency is at most $S+T$.
\item First, assume that $\left(S+T\right)/2  > P$, and suppose player $X$ deviates by choosing a strategy $\mathbf{p^{\ast}}=\mathbf{p}+\varepsilon\mathbf{e}_{DD}$, with $\varepsilon >0$ being arbitrarily small. Again, the efficiency after the deviation takes the form of \eq{efficiency_deviation}, with either $a_1$ or $a_2$ strictly positive. Because $R>P$ and $\left(S+T\right) /2  >P$, this efficiency is larger than the current efficiency of $2P$. Hence, player $X$ is able to locally improve its objective function.

For the other direction, assume now that $\left(S+T\right)/2  \leqslant P$ and that player $Y$ adopts the strategy ALLD. Then any deviation $\mathbf{p^{\ast}}$ by player $X$ leads to an efficiency of the form
\begin{equation}
\pi_X\left(\mathbf{p^{\ast}},\mathbf{q}\right) +\pi_Y\left(\mathbf{p^{\ast}},\mathbf{q}\right) =  a\left(S+T\right) +\left (1-a\right)\left(2P\right) ,
\end{equation}
which is at most $2P$. Thus player $X$ is unable to locally (or globally) improve its objective function.\qedhere
\end{enumerate}
\end{proof}

\noindent
In particular, this result implies that in donation games, players cannot settle at mutual defection if at least one of them applies \FMTL{}. On the other hand, in the special case of a prisoner's dilemma with $\left(S+T\right) /2  \leqslant P$, mutual defection can be stable even if both players value efficiency. This effect arises because in our framework (and in evolutionary game theory more generally), learners adapt their strategies independently from each other. In games in which one player alone is unable to escape from inefficient strategy profiles, players may thus end up in a rest point that they both wish to avoid. 

\section{Evolutionary dynamics of learning rules}
Here, we briefly outline several qualitative observations about the evolutionary dynamics of learning rules. Recall that the dynamics we consider depend on mean payoffs only. Suppose that $\mathcal{R}$ and $\mathcal{S}$ are learning rules. For $\mathcal{X},\mathcal{Y}\in\left\{\mathcal{R},\mathcal{S}\right\}$, we let $a_{\mathcal{X}\mathcal{Y}}\left(n\right)$ denote the mean payoff to $\mathcal{X}$ against $\mathcal{Y}$ after $n$ learning steps, averaged over initial conditions and learning trajectories.

These mean payoffs motivate a qualitative comparison among learning rules based on how they perform against one another, on average. We consider three broad conditions:
\begin{itemize}

\item $\mathcal{R}$ \emph{exploits} $\mathcal{S}$ after $n$ steps if $a_{\mathcal{R}\mathcal{S}}\left(n\right) >a_{\mathcal{S}\mathcal{R}}\left(n\right)$;

\item $\mathcal{R}$ is \emph{fair relative to} $\mathcal{S}$ after $n$ steps if $a_{\mathcal{R}\mathcal{S}}\left(n\right) =a_{\mathcal{S}\mathcal{R}}\left(n\right)$;

\item $\mathcal{R}$ is \emph{exploited by} $\mathcal{S}$ after $n$ steps if $a_{\mathcal{R}\mathcal{S}}\left(n\right) <a_{\mathcal{S}\mathcal{R}}\left(n\right)$.

\end{itemize}
Note that each of these conditions disregards the performance of a learning rule against itself, i.e. $a_{\mathcal{R}\mathcal{R}}\left(n\right)$ and $a_{\mathcal{S}\mathcal{S}}\left(n\right)$. As we will see below, these conditions do not alone allow one to conclude anything about evolutionary stability, but they do provide reasonable classifications for a first comparison of two learning rules.

\begin{example}
In \fig{alternative_rules}b, team-based learning (maximizing $\pi_{X}+\pi_{Y}$) is exploited by selfish learning after sufficiently many learning steps. In all of our examples, \FMTL{} is fair relative to selfish learning after long enough learning horizons, but it can be exploited by selfish learning on shorter timescales (e.g. \fig{replicatorDynamics}a). However, as the example of \fig{replicatorDynamics}a shows, \FMTL{} can still be evolutionary stable relative to selfish learning even when it is being exploited, due to the superior performance of \FMTL{} against itself relative to selfish learning against itself. By symmetry, any learning rule is fair relative to itself.
\end{example}

\begin{example}
Selfish learning, i.e. caring about one's own payoff only, can actually still be exploited by some learning rules. To see why, we can formulate a learning rule that takes advantage of selfish learners. Suppose that $\chi >1$. In the donation game with benefit $b$ and cost $c$, consider the following learning rule: if $\pi_{X}\geqslant\chi\pi_{Y}$, then maximize $\pi_{X}+\pi_{Y}$; otherwise, minimize $\left|\pi_{X}-\chi\pi_{Y}\right|$. When $b=2$, $c=1$, $\chi =2$, and this learning rule is paired with a selfish learner, the mean payoffs to the two learners over $10^{5}$ runs are $1.1788$ and $0.5894$, respectively. We do not wish to dwell on this particular learning rule; rather, this example is intended simply to illustrate that even selfish learners can be exploited by learners with sufficiently coercive preferences.
\end{example}

The following result says that if $\mathcal{S}$ does not exploit an opposing learning rule, $\mathcal{R}$, then $\mathcal{S}$ cannot be evolutionary stable relative to $\mathcal{R}$ whenever $\mathcal{R}$ is optimal against itself. (To simplify the notation, we suppress the number of learning steps, $n$.)
\begin{proposition}\label{Prop:ESS}
Suppose that $\mathcal{R}$ either exploits $\mathcal{S}$ or is fair relative to $\mathcal{S}$, and suppose furthermore that $a_{\mathcal{R}\mathcal{R}}$ is maximal (which can be determined by looking at the feasible region for the repeated game). Then, either \textit{(i)} $\mathcal{R}$ is neutral relative to $\mathcal{S}$ or \textit{(ii)} $\mathcal{R}$ is evolutionarily stable relative to $\mathcal{S}$.
\end{proposition}
\begin{proof}
Since $\mathcal{R}$ exploits $\mathcal{S}$ or is fair relative to $\mathcal{S}$, we have $a_{\mathcal{R}\mathcal{S}}\geqslant a_{\mathcal{S}\mathcal{R}}$. Moreover, since $a_{\mathcal{R}\mathcal{R}}$ is maximal and the feasible region is convex, we must have $a_{\mathcal{R}\mathcal{R}}\geqslant a_{\mathcal{S}\mathcal{R}}$. If $a_{\mathcal{R}\mathcal{R}}>a_{\mathcal{S}\mathcal{R}}$, then $\mathcal{R}$ is automatically evolutionarily stable relative to $\mathcal{S}$. But if $a_{\mathcal{R}\mathcal{R}}=a_{\mathcal{S}\mathcal{R}}$, then $a_{\mathcal{R}\mathcal{S}}\geqslant a_{\mathcal{S}\mathcal{R}}=a_{\mathcal{R}\mathcal{R}}\geqslant a_{\mathcal{S}\mathcal{S}}$. If these inequalities are all equalities, then $\mathcal{R}$ is neutral relative to $\mathcal{S}$; otherwise, $\mathcal{R}$ satisfies Maynard Smith's second condition \citep{maynardsmith:Nature:1973} and is thus evolutionarily stable relative to $\mathcal{S}$.
\end{proof}
For example, in the hero game, we know that \FMTL{} is both fair and optimal against itself after sufficiently many learning steps (\sfig{supergame_others_neutral}b). Thus, selfish learning cannot be evolutionary stable relative to \FMTL{}. In this case, since selfish learning is also optimal against itself, \FMTL{} is neutral relative to selfish learning. Of course, one can always deduce evolutionary stability by looking at the mean payoffs $a_{\mathcal{X}\mathcal{Y}}$ themselves. \textbf{Proposition~\ref{Prop:ESS}} simply says that a learner who performs optimally against itself is \emph{at least} neutral against any nonexploitative opponent.

\section{Beyond introspection: imitation as a form of learning}\label{sec:imitation}
The ``evolutionary'' aspect of our model enters in describing competition between learning rules (the ``supergame'' dynamics), not in the implementation of learning rules themselves. Rather, the learning rules we consider are implemented between two players (a trivial case of a population) using introspection. Each player samples a new strategy nearby and then determines whether it wants to adopt this strategy or retain the old one. The difference between a selfish learning and \FMTL{} amounts to the objective(s) used in determining whether to adopt the new strategy.

Introspection is a natural choice for learners because each learner knows their own strategy and can thus sample something nearby to test against an opponent. Whether the player accepts this strategy or not depends on observed payoffs; a player does not specifically need to know the strategy used by the opponent. In contrast, imitation dynamics require that a player can copy an opponent's strategy, which may or may not be a reasonable assumption. In classical models involving a small number of strategic types, a player simply needs to see the action taken by the opponent and then copy that action \citep{traulsen:PNAS:2009}. But repeated games involve more complicated conditional behavior, in general, and a player might not be able to faithfully infer such a strategy by interacting with an opponent. This observation is not to discount the value of imitation but rather to say that it likely entails a degree of error or mutation in strategy transmission.

Nonetheless, we can study what happens when a player's objectives are evaluated through imitation rather than introspection. Suppose that $s\in\left[0,1\right]$. If $X$ imitates the strategy $\q$ of $Y$, then we assume that the $i$th coordinate of $X$'s strategy is $\min\left\{ \max\left\{ q_{i} + z_{i} , 0 \right\} , 1 \right\}$, where $z_{i}$ is uniformly distributed on $\left[-s,s\right]$, and $z_{i}$ is independent of $z_{j}$ for $j\neq i$. This mutation procedure is identical to the one used for introspection dynamics in the main text, although here we use $s=10^{-2}$; there are always errors in copying strategies, but these errors are not too large.

The baseline for comparison here, analogous to a selfish learner in the main text, is a selfish imitator. Suppose that $X$ is a selfish imitator using $\p$, and let $Y$ be a model player who uses $\q$. $X$ chooses another individual, $Z$, using $\mathbf{r}$, and asks how $X$ fares against $Z$ relative to how $Y$ fares against $Z$. Since our focus is on symmetric games, we can simplify notation and write $\pi\left(\p ,\mathbf{r}\right)$ for $\pi_{X}\left(\p ,\mathbf{r}\right)$ and $\pi\left(\mathbf{r},\p\right)$ for $\pi_{Z}\left(\p ,\mathbf{r}\right)$. $X$ then copies the strategy of $Y$ with probability
\begin{align}
\frac{1}{1+e^{-\delta\left(\pi\left(\q ,\textbf{r}\right) -\pi\left(\p ,\textbf{r}\right)\right)}} ,
\end{align}
where $\delta\geqslant 0$ represents the intensity of selection. This is known as Fermi updating \citep{szabo:PRE:1998,traulsen:JTB:2007b}.

The analogue of \FMTL{} first chooses an objective function to optimize, based on the current value of $\left|\pi\left(\p ,\mathbf{r}\right) -\pi\left(\mathbf{r},\p\right)\right|$, exactly as prescribed in the main text. The key distinction between this implementation of \FMTL{} and the one in the main text lies in the next step. If efficiency is the objective, then $X$ imitates $Y$ based on whether $Y$ and $Z$ have a larger total payoff than $X$ and $Z$. If the objective is fairness, then $X$ imitates $Y$ based on whether $Y$'s payoff is closer to $Z$'s than $X$'s is to $Z$'s. More specifically, the probability that $X$ imitates $Y$'s strategy, relative to $Z$, is
\begin{align}
\begin{cases}
\displaystyle\frac{1}{1+e^{-\delta\left(\left(\pi\left(\q ,\mathbf{r}\right) +\pi\left(\mathbf{r},\q\right)\right) -\left(\pi\left(\p ,\mathbf{r}\right) +\pi\left(\mathbf{r},\p\right)\right)\right)}} & \displaystyle\textrm{with probability }\omega\left(\p ,\textbf{r}\right) , \\
& \\
\displaystyle\frac{1}{1+e^{-\delta\left(\left|\pi\left(\p ,\mathbf{r}\right) -\pi\left(\mathbf{r},\p\right)\right| -\left|\pi\left(\q ,\mathbf{r}\right) -\pi\left(\mathbf{r},\q\right)\right|\right)}} & \displaystyle\textrm{with probability }1-\omega\left(\p ,\textbf{r}\right) ,
\end{cases}
\end{align}
where $\omega$ is the function of \eq{gaussian_function} (or \eq{bump_function}) of the main text. For both selfish and \FMTL{} imitators, we use a selection intensity of $\delta =1.0$.

Using the donation game as an example, our findings in \fig{donationGame} and \fig{replicatorDynamics}\textbf{a} say that \FMTL{} improves the outcomes for all learners and cannot be exploited by selfish learners. To explore whether this behavior is preserved under imitation dynamics, we consider three settings: \emph{(i)} populations consisting of only selfish imitators; \emph{(ii)} populations consisting of a mixture of selfish imitators and \FMTL{} imitators; and \emph{(iii)} populations consisting of only \FMTL{} imitators. In unstructured populations of finite size $N$, there are $N-1$ possibilities for option \emph{(ii)}, one for each number of \FMTL{} imitators between $1$ and $N-1$. \fig{imitation} illustrates imitation dynamics in a population of size $N=100$ when \textbf{a} all imitators are selfish, \textbf{b} there are as many \FMTL{} imitators as selfish imitators, and \textbf{c} all imitators use \FMTL{}. Qualitatively, we find the same outcomes reported in the main text: selfish imitators lead to relatively poor outcomes overall, but the presence of \FMTL{} leads to substantial improvements while avoiding exploitation. \fig{imitation_trend} demonstrates this trend beyond populations with an equal number of each kind of imitator. Worthy of note is the fact that a selfish imitator would always be better off if they switched to \FMTL{}.

The payoffs of selfish imitators and \FMTL{} imitators in \fig{imitation}\textbf{b} are highly correlated. However, this is not necessarily attributed entirely to the fact that \FMTL{} emphasizes fairness. Because changes in strategies are based on imitation, a player who imitates based on \FMTL{} can copy a strategy that was shaped by selfish objectives and vice versa. What was learned by one player can thus be directly transferred to another player, even when the two players have different preferences over how to imitate. In this sense, imitation dynamics result in blended learning within a population, which partly contributes to the correlations between payoffs across types. Actually, even team-based imitation alone can improve outcomes for all, relative to selfish imitation, which is a finding that does not hold in the introspective setting (\fig{alternative_rules}\textbf{b}).

Beyond an individual's ability to infer and imitate a complicated strategy, other issues arise when considering social preferences in classical models, especially when strategies are inherited instead of imitated. Under birth-death dynamics, individuals reproduce and pass down their strategies for a repeated game, subject to mutation. More successful strategies tend to produce more offspring, so these dynamics may be thought of as encoding a kind of selfish learning. However, unlike in imitation dynamics, individuals do not necessarily have the agency to choose their reproductive rates based on preferences. Just because an individual values efficiency or fairness does not mean that their birth rate increases as those objectives get closer to their optima.

In summary, \FMTL{} is most natural when implemented in settings for which social preferences can be effectively expressed, including introspection and imitation. Our main focus is on introspection-based learning in pairs, which is quite different from the population-based model of imitation presented here. However, the latter illustrates an important point: the key beneficial properties of \FMTL{} are not due to introspection but rather to the specific preferences such learners use when making choices.

\section{Asymmetric games}\label{sec:asymmetric_games}

\subsection{Weakly-asymmetric games}
Our focus in this study is on symmetric games, but it is natural to ask how the results extend to settings with asymmetric interactions. Consider a general two-action bimatrix game,
\begin{align}
\bordermatrix{%
& L & R \cr
U &\ a_{1},b_{1} & \ a_{2},b_{2} \cr
D &\ a_{3},b_{3} & \ a_{4},b_{4} \cr
}\ . \label{eq:bimatrix}
\end{align}
Here, $X$ is the row player and $Y$ is the column player. The respective action spaces are $S_{X}=\left\{U,D\right\}$ (``up'' and ``down'') for $X$ and $S_{Y}=\left\{L,R\right\}$ (``left'' and ``right'') for $Y$, with $a_{i}$ denoting $X$'s payoff and $b_{i}$ denoting $Y$'s payoff. If $A\coloneqq\begin{pmatrix}a_{1} & a_{2} \\ a_{3} & a_{4}\end{pmatrix}$ and $B\coloneqq\begin{pmatrix}b_{1} & b_{2} \\ b_{3} & b_{4}\end{pmatrix}$, then this game is symmetric if $S_{X}=S_{Y}$ and $B^{\intercal}=A$. Note that this definition is based on a comparison of the exact values in $A$ and $B$, and it implicitly identifies $U$ with $L$ and $D$ with $R$. We will consider a weaker version of this definition, in which we use relative rankings of the values in each matrix. More specifically, let $\textrm{rank}\left(a_{i},A\right) =k$ mean that $a_{i}$ is the $k$th largest value in $A$ (ties are allowed when two values are equal). Now, consider the two ranked matrices $A^{\textrm{ranked}} \coloneqq \begin{pmatrix}\textrm{rank}\left(a_{1},A\right) & \textrm{rank}\left(a_{2},A\right) \\ \textrm{rank}\left(a_{3},A\right) & \textrm{rank}\left(a_{4},A\right)\end{pmatrix}$ and $B^{\textrm{ranked}} \coloneqq \begin{pmatrix}\textrm{rank}\left(b_{1},B\right) & \textrm{rank}\left(b_{2},B\right) \\ \textrm{rank}\left(b_{3},B\right) & \textrm{rank}\left(b_{4},B\right)\end{pmatrix}$. We say that this game is \emph{weakly-asymmetric} if $\left(B^{\textrm{ranked}}\right)^{\intercal}=A^{\textrm{ranked}}$. More generally, we say that:
\begin{definition}[Weakly-asymmetric game]\label{def:weakly_asymmetric}
An asymmetric $N$-player game with action spaces $S_{1},\dots ,S_{N}$ and utility function $u:\prod_{i=1}^{N}S_{i}\rightarrow\mathbb{R}^{N}$ is \emph{weakly asymmetric} if there exists a symmetric game with action space $S$ and utility function $u':S^{N}\rightarrow\mathbb{R}^{N}$, together with bijections $\left\{\phi_{i}:S_{i}\rightarrow S\right\}_{i=1}^{N}$ such that for all $i=1,\dots ,N$ and $s,s'\in\prod_{i=1}^{N}S_{i}$, $\textrm{sgn}\left(u_{i}\left(s\right) -u_{i}\left(s'\right)\right) =\textrm{sgn}\left(u_{i}'\left(\phi\left(s\right)\right) -u_{i}'\left(\phi\left(s'\right)\right)\right)$, where $\phi =\phi_{1}\times\cdots\times\phi_{N}$ and $\textrm{sgn}\left(x\right)$ is $-1$ if $x<0$, $0$ if $x=0$, and $1$ if $x>0$. In other words, for all $s,s'\in\prod_{i=1}^{N}S_{i}$,
\begin{align}
u_{i}\left(s\right)\geqslant u_{i}\left(s'\right) &\iff u_{i}'\left(\phi\left(s\right)\right)\geqslant u_{i}'\left(\phi\left(s'\right)\right) ,
\end{align}
with equality on the left-hand side if and only if there is equality on the right-hand side.
\end{definition}

\begin{example}
Consider the two-player ``battle of the sexes'' game \citep{luce:D:1989}, with utility $u$ given by
\begin{align}
\bordermatrix{%
& F & B \cr
F &\ 3,2 & \ 1,1 \cr
B &\ 0,0 & \ 2,3 \cr
}\ . 
\end{align}
The interpretation of this game is that two individuals (row and column) have different preferences and choose between two options: attend a fight ($F$) and attend a ballet ($B$). The row player prefers seeing a fight, while the column player prefers seeing a ballet. They prefer, however, to be together rather than to be apart, which gives the payoff ranking depicted in the matrix.

Let $\phi_{1},\phi_{2}:\left\{F,B\right\}\rightarrow\left\{C,D\right\}$ be the bijections with $\phi_{1}\left(F\right) =C$ and $\phi_{2}\left(F\right) =D$. With these correspondences on the strategy spaces, this battle of the sexes game maps to the symmetric ``hero'' game described in the main text, with utility function $u'$ given by the payoff matrix
\begin{align}
\bordermatrix{%
& C & D \cr
C &\ 1 & \ 3 \cr
D &\ 2 & \ 0 \cr
}\ . 
\end{align}
This mapping preserves the value of $u_{i}\left(s\right) -u_{i}\left(s'\right)$, so the battle of the sexes game is weakly asymmetric. However, note that there is no such mapping to a symmetric game if $\phi_{1}\left(F\right) =\phi_{2}\left(F\right) =C$. The reason is that if we have a symmetric matrix game with parameters $R$, $S$, $T$, and $P$, then the fact that $u_{1}\left(F,F\right) =3>2=u_{1}\left(B,B\right)$ requires that $R>P$. At the same time, the fact that $u_{2}\left(F,F\right) =2<3=u_{2}\left(B,B\right)$ requires that $R<P$, and we cannot have both of these inequalities. Thus, in mapping the battle of the sexes game to a symmetric game, it was necessary to map one player's choice of $F$ to $C$ and the other player's choice of $F$ to $D$.

This property of the battle of the sexes game partially motivates the definition of weakly-asymmetric games. The two players view the same option, $B$, differently. So, when considering the action $F$ for one player, the comparable choice for the second player is not necessarily the same nominal action. Rather, it is the action that, relative to player two, is most comparable to the action $F$, relative to player one. In this case, we can think of $C$ as meaning ``choose one's own preference'' and $D$ as ``choose the partner's preference.'' This aspect of the battle of the sexes game will be important for understanding fairness because we will need to talk about what happens when two players swap actions (or imitate) one another; and if player one chooses $F$, then when player two imitates this action, she actually chooses $B$ rather than $F$.
\end{example}

Another, slightly simpler example is the asymmetric donation game, in which player $X$ can pay a cost $c$ to donate $b_{X}$ and player $Y$ can pay $c$ to donate $b_{Y}$. Taking such an action is considered ``cooperation'' for each player, whereas ``defection'' for either $X$ or $Y$ amounts to paying nothing and donating nothing. In this case, the action $C$ of $X$ corresponds to the action $C$ of $Y$, so we do not have to deal with the same subtlety that arises in the battle of the sexes game. In fact, we will use this game as our main illustration. However, before describing results for the asymmetric donation game, we note that the approach outlined above assumes that players have comparable actions, so that they can swap strategies in the first place. Within the space of weakly-asymmetric games, there must exist a correspondence $\phi_{Y}^{-1}\circ\phi_{X}:S_{X}\rightarrow S_{Y}$, so we know that there is at least one way of relating the strategies of $X$ to those of $Y$. Generically, this correspondence is unique:
\begin{proposition}
In a weakly-asymmetric bimatrix game of the form depicted in \eq{bimatrix}, any two bijections $\phi_{X}$ and $\phi_{Y}$ satisfying the conditions in Definition~\ref{def:weakly_asymmetric} give rise to a unique bijection between actions, $\phi_{Y}^{-1}\circ\phi_{X}:\left\{U,D\right\}\rightarrow\left\{L,R\right\}$, unless $a_{1}=a_{4}$, $a_{2}=a_{3}$, $b_{1}=b_{4}$, and $b_{2}=b_{3}$.
\end{proposition}
\begin{proof}
By relabeling the strategies in \eq{bimatrix} if necessary, we may assume that the bijections $\phi_{X}$ and $\phi_{Y}$ satisfying Definition~\ref{def:weakly_asymmetric} are such that the bijection $\phi_{Y}^{-1}\circ\phi_{X}:\left\{U,D\right\}\rightarrow\left\{L,R\right\}$ sends $U$ to $L$. By the definition of a weakly-asymmetric game, we then have the equation
\begin{align}
\begin{pmatrix}
\textrm{rank}\left(a_{1},A\right) & \textrm{rank}\left(a_{2},A\right) \\
\textrm{rank}\left(a_{3},A\right) & \textrm{rank}\left(a_{4},A\right)
\end{pmatrix}
&= 
\begin{pmatrix}
\textrm{rank}\left(b_{1},B\right) & \textrm{rank}\left(b_{3},B\right) \\
\textrm{rank}\left(b_{2},B\right) & \textrm{rank}\left(b_{4},B\right)
\end{pmatrix} . \label{eq:ABt}
\end{align}
If there were another bijection $\widetilde{\phi}_{Y}^{-1}\circ\widetilde{\phi}_{X}:\left\{U,D\right\}\rightarrow\left\{L,R\right\}$ sending $U$ to $R$ instead, such that $\widetilde{\phi}_{X}$ and $\widetilde{\phi}_{Y}$ also satisfy Definition~\ref{def:weakly_asymmetric}, then by swapping the columns of \eq{bimatrix} we have
\begin{align}
\begin{pmatrix}
\textrm{rank}\left(a_{2},A\right) & \textrm{rank}\left(a_{1},A\right) \\
\textrm{rank}\left(a_{4},A\right) & \textrm{rank}\left(a_{3},A\right)
\end{pmatrix}
&= 
\begin{pmatrix}
\textrm{rank}\left(b_{2},B\right) & \textrm{rank}\left(b_{4},B\right) \\
\textrm{rank}\left(b_{1},B\right) & \textrm{rank}\left(b_{3},B\right)
\end{pmatrix} . \label{eq:ABt_swapped}
\end{align}
Combining \eqs{ABt}{ABt_swapped}, we see that $a_{1}=a_{4}$, $a_{2}=a_{3}$, $b_{1}=b_{4}$, and $b_{2}=b_{3}$.
\end{proof}

\subsection{\FMTL{} for weakly-asymmetric games}
Clearly, not all asymmetric games are weakly asymmetric because players can either have incomparable actions or even a different number of actions to choose from. An analysis of ``fairness'' in general asymmetric games is complicated and beyond the scope of this paper. However, owing to the unique correspondence between the players' actions in weakly asymmetric games, there is a natural way to extend \FMTL{} beyond symmetric games. This is important because, in an asymmetric game, it is no longer necessarily true that equal outcomes are necessarily ``fair.''

Another way of thinking about the objective $V_{F}\left(\p ,\q\right) =-\left|\pi_{X}\left(\p ,\q\right) -\pi_{Y}\left(\p ,\q\right)\right|$ is in terms of the taxicab distance between $\left(\pi_{Y}\left(\p ,\q\right) ,\pi_{X}\left(\p ,\q\right)\right)$ and $\left(\pi_{Y}\left(\q ,\p\right) ,\pi_{X}\left(\q ,\p\right)\right)$ in $\mathbb{R}^{2}$. Note that when $X$ and $Y$ swap strategies, they implicitly take into account the correspondence $\phi_{Y}^{-1}\circ\phi_{X}$. In other words, if $\q$ says that $Y$ plays $y$ with probability $q$, then when $X$ plays $\q$ she plays $\phi_{X}^{-1}\circ\phi_{Y}\left(y\right)$ with probability $q$. When the game is weakly asymmetric, we can let
\begin{align}
V_{F}\left(\p ,\q\right) &= -\frac{1}{2}\left|\pi_{X}\left(\p ,\q\right) -\pi_{X}\left(\q ,\p\right)\right| - \frac{1}{2}\left|\pi_{Y}\left(\p ,\q\right) -\pi_{Y}\left(\q ,\p\right)\right| \label{eq:VF_wa}
\end{align}
serve as a proxy for measuring fairness. The idea is to allow $X$ and $Y$ to swap behaviors and then measure how far away the resulting points in payoff space are from one another. The taxicab metric is not special; we simply choose it here (and in the definition of \FMTL{} in general) for convenience. In symmetric games, this objective $V_{F}$ reduces to the one studied in the main text.

Returning to the asymmetric donation game, $X$ receives $b_{Y}-c$ and $Y$ receives $b_{X}-c$ when both players cooperate in the iterated game. For two such strategies, $\p$ for $X$ and $\q$ for $Y$, their payoffs do not change if they swap strategies with one another, so the unequal outcome of $b_{Y}-c$ for $X$ and $b_{X}-c$ for $Y$ is considered fair. This makes intuitive sense because both players are taking the same action (``donate'') and making the same sacrifice (paying $c$). More generally, $V_{F}\left(\p ,\q\right) =0$ if and only if $\pi_{X}\left(\p ,\q\right) =\pi_{X}\left(\q ,\p\right)$ and $\pi_{Y}\left(\p ,\q\right) =\pi_{Y}\left(\q ,\p\right)$. For this asymmetric donation game, these equations hold if and only if there exists $K$ such that $\pi_{X}\left(\p ,\q\right) =K\left(b_{Y}-c\right)$ and $\pi_{Y}\left(\p ,\q\right) =K\left(b_{X}-c\right)$. As a result, the line connecting the payoffs for mutual defection and mutual cooperation represent the fairest possible outcomes.

For measuring efficiency, we retain the team payoff, $V_{E}\left(\p ,\q\right) =\pi_{X}\left(\p ,\q\right) +\pi_{Y}\left(\p ,\q\right)$. The associated learning rule, which we denote \emph{FMTL-WA} (``WA'' for ``weakly asymmetric'') is defined analogously to \FMTL{} in the main text: first decide which of $V_{F}$ and $V_{E}$ to use, based (stochastically) on the value of $V_{F}$; then sample in order to optimize the chosen objective function. \fig{FMTL_asymmetric} shows the results of three learning rules against a selfish learner in an asymmetric donation game. The first learner is itself a selfish learner; the second uses the (symmetric) version of \FMTL{} described in the main text; and the third is \emph{FMTL-WA}. Although \FMTL{} itself performs much better against a selfish learner than does another selfish learner, it is \emph{FMTL-WA} that is able to bring out optimal payoffs in more than 99\% of runs. Note that the dashed line in \fig{FMTL_asymmetric} no longer represents equality but rather fairness as defined by $V_{F}=0$ (see above).

\end{document}